\documentclass[aps, prd, superscriptaddress, preprintnumbers, showkeys, showpacs]{revtex4}
\usepackage[english]{babel}
\usepackage{graphicx,amsfonts,amsmath,amssymb,amstext}
\usepackage{float,wrapfig}
\usepackage{subfigure, psfrag}
\usepackage{dsfont}
\usepackage{epsfig}
\usepackage{color}
\usepackage{bm}
\usepackage{multirow,textcomp}
\usepackage{natbib}
\usepackage[normalem]{ulem}
\usepackage{hyperref}

\definecolor{purple}{rgb}{0.8,0,0.6}

\makeatletter
\newcommand{\vast}{\bBigg@{2}}
\newcommand{\Vast}{\bBigg@{3}}
\makeatother

\begin{document}

\title{Electron states in the field of charged impurities in two-dimensional Dirac systems}
\date{\today}

\author{E. V. Gorbar}
\affiliation{Department of Physics, Taras Shevchenko National University of Kiev, Kiev, 03680, Ukraine}
\affiliation{Bogolyubov Institute for Theoretical Physics, Kiev, 03680, Ukraine}

\author{V. P. Gusynin}
\affiliation{Bogolyubov Institute for Theoretical Physics, Kiev, 03680, Ukraine}

\author{O. O. Sobol}
\affiliation{Department of Physics, Taras Shevchenko National University of Kiev, Kiev, 03680, Ukraine}

\begin{abstract}
We review the theoretical and experimental results connected with the electron states in two-dimensional Dirac systems paying
a special attention to the atomic collapse in graphene. Two-electron bound states of a Coulomb impurity are considered too.
A rather subtle role of a magnetic field in the supercritical charge problem in graphene is discussed. The electron states in
the field of two equally charged impurities are studied and the conditions for supercritical instability to occur are determined.
It is shown that the supercriticality of novel type is realized in gapped graphene with two unlikely charged impurities. For
sufficiently large charges of impurities, it is found that the wave function of the occupied electron bound state of the highest
energy changes its localization from the negatively charged impurity to the positively charged one as the distance between the
impurities increases. The specifics of the atomic collapse in bilayer graphene is considered and it is shown that the atomic
collapse in this material is not related to the phenomenon of the fall-to-center.
\end{abstract}


\maketitle


\tableofcontents

\section{Introduction}
\label{section:introduction}

The phenomenon of the fall-to-center is deeply rooted in the history of physics. The Rutherford's discovery of the planetary model of the atom
immediately brought to the light the problem of the stability of the atom. Indeed, classically, the electron rotating around the nucleus should
emit electromagnetic radiation, lose its energy, and fall to the nucleus. We know that the atoms are stable and the atomic collapse is avoided
due to the uncertainty principle of quantum mechanics. While the Coulomb interaction scales like $-Ze/r$, where $r$ is the distance to the
nucleus and $Ze$ is its charge, the positive electron kinetic energy diverges more strongly $\mathbf{p}^2/(2m_e)\sim \hbar^2/(2m_er^2)$ as
$r \to 0$. Therefore, the fall to the nucleus is energetically forbidden.

This qualitative argument shows that the fall-to-center may still be possible in quantum mechanics for more singular potentials $V(r)\sim 1/r^n$
with $n \ge 2$. In fact, the Schr\"{o}dinger equation with the potential $V(r)=-\beta/r^2$ provides the canonical textbook  example of the
fall-to-center in quantum mechanics \cite{Landau:t3}, which takes place for $\beta > \hbar^2/(8m_e)$ when the  energy spectrum is not bounded
from below. If the interaction potential is regularized at some distance $r_0$, then the electron wave function of the ground state is localized
in the region of the radius $r_0$ which shrinks to the origin as $r_0 \to 0$.

Still physically as $|\mathbf{p}|$ attains the value of order $m_ec$, where $c$ is the speed of light, the relativistic effects become relevant.
Since the kinetic term in the Dirac equation depends linearly on momentum, the kinetic energy of the electron in the relativistic regime scales
like $\hbar/r$ as $r \to 0$. This means that already the Coulomb interaction could lead to the atomic collapse. In quantum electrodynamics (QED)
for the regularized Coulomb potential, the atomic collapse takes place for $Z \gtrsim 170$ [\onlinecite{Pomeranchuk1945,Zeldovich1972,Greiner1985}]
when the lowest energy electron bound state dives into the lower continuum transforming into a narrow resonance. This leads to the spontaneous
creation of electron-positron pairs with
the electrons screening the positively charged nucleus and the positrons emitted to infinity. Since supercritically charged nuclei are not encountered
in nature, this phenomenon was never observed in QED. It was suggested in the 70-ties \cite{Gershtein1970,Rafelski1971,Muller1972,Zeldovich1972} that
the supercritical instability in QED can nevertheless be experimentally tested in a collision of two heavy nuclei. Although subsequent experiments
confirmed the existence of supercritical fields in collisions of very heavy nuclei and the gross features of positron emission \cite{Greiner1985},
the analysis of the supercritical regime turned out to be a difficult problem mainly due to the transient nature of supercritical fields generated
during collisions.

It is an interesting question whether the supercritical instability could be observed in the condensed matter systems. The first natural place to
look is the narrow gap semiconductors whose conductance and valence bands are separated by a small gap. There exist also condensed matter systems
with the relativistic-like energy spectrum of quasiparticles. Bismuth, whose quasiparticles are described by the massive Dirac equation, provides the
historically first example of such a system (for a review, see Refs.\cite{Falkovsky1968,Edelman1976}). Long time ago Herring argued \cite{Herring1937}
that the conductance and valence bands in solids could, in general, meet at discrete touching points. Remarkably, the energy dispersion in the vicinity
of these bands touching points is linear and resembles the Weyl equation. The recently discovered Dirac and Weyl semimetals whose itinerant electrons
are described by the 3D Dirac and Weyl equations, respectively, experimentally realize the Herring's prediction (for a review, see, e.g.,
Ref.\cite{Armitage2017}). However, the corresponding materials are characterized by the large dielectric constants. The small value of the effective
coupling constant makes it practically impossible to realize the supercritical instability in these materials.

The situation is different in graphene whose effective coupling constant $\alpha_{g}=e^2/(v_{\rm F}\hbar)\approx 2.2$, where $v_{\rm F} \approx c/300$
is the Fermi velocity, exceeds unity. This drastically decreases the value of the critical charge in graphene [\onlinecite{Shytov2007a,Shytov2007b,Pereira2007,Fogler2007}]. Although, according to the theory, the supercritical instability should be easily
realized for charged impurities in graphene, its experimental observation remained elusive until recently. The problem is that it is difficult to
produce highly charged impurities because of their fast recombination. Still one can reach the supercritical regime by collecting a
large enough number of charged impurities in a certain region of graphene. Such an approach was recently successfully realized [\onlinecite{Wang2013}]
by using the tip of a scanning tunneling microscope in order to create clusters of charged calcium dimers.

In addition, the external charge in the realistic experimental set-up should be smeared over a finite region of the graphene plane because,
otherwise, the Dirac equation is no longer applicable and other nearest $\sigma$-bands should be included in the analysis \cite{Fogler2007}.
Thus, the potential of charged impurities should be necessarily regularized at small distances in order that the continuum
problem be well posed physically. For instance, the charged impurities displaced from the graphene plane provide such a natural
regularization and help to avoid the reconstruction of the band spectrum which takes place if they are placed directly
into the graphene plane or a disorder is present \cite{Feher2009,Loktev2012,Skrypnik2007,Dora2008}.

An interesting aspect of the electron physics in graphene is its two-dimensional character. Therefore, the supercritical instability in the field
of a charged impurity in graphene is, in fact, the atomic collapse in a Flatland. Of course, this does not mean that the theory governing the
electron-electron interactions in graphene is $\mbox{QED}$ in (2+1) dimensions. Although the electrons are confined in the plane of graphene, the electromagnetic
force lines spread beyond the graphene's plane resulting in the standard Coulomb interaction potential $V_C(r)=e^{2}/r$. The crucial advantage
of graphene compared to QED is its experimental accessibility where atomic collapse can be investigated in table-top experiments varying such
parameters as doping and gate voltage.

The supercritical charge instability is closely related to the excitonic instability in graphene in the strong coupling regime $\alpha>\alpha_c\sim1$
(see, Refs.[\onlinecite{Gamayun2009,Wang2010,Sabio2010}]) and possible gap opening, which may transform graphene into an insulator \cite{Khveshchenko2001,Gorbar2002,Gorbar2003,Khveshchenko2004,Gamayun2010,Gonzalez2012,Drut2009a,Drut2009b,Armour2010,Buividovich2012}. Indeed,
the excitonic instability can be viewed as a many-body analog of the supercritical instability in the field of a charged impurity and the critical
coupling $\alpha_{c}$ is an analog of the critical coupling constant $Z_{c}\alpha$ in the problem of the Coulomb center. In the strong coupling
regime $\alpha>\alpha_{c}$ the electron can spontaneously create from the vacuum the electron-hole pair (in the same way as the supercritical charge
creates electron-hole pairs). The initial electron attracts the hole and forms a bound state (an exciton) and the emitted electron (which also has
the supercritical charge) can spontaneously create another pair, etc. The process of creating pairs continues leading to the formation of excitonic
condensate and, as a result, the quasiparticles acquire a gap. The semimetal-insulator transition in graphene  is similar to the chiral symmetry
breaking phase transition  in strongly coupled QED studied in the 1970s and 1980s (for a review see Ref.\cite{FominReview1983}). The latter QED
transition induced by strong electromagnetic fields was searched in experiments in heavy-ion collisions \cite{Peccei1988}.

To stay closer to the experimental situation, one should make a further step by considering electron states in the field of two Coulomb centers,
both like and unlike charged. The electron states in the field of charged impurities in graphene and in the presence of a magnetic field are
also of considerable interest from the experimental point of view. It was shown in Refs.\cite{Luican-Mayer2014,Mao2016} that the
strength of a charged impurity can be tuned by controlling the occupation of Landau-level states with a gate voltage.

All these topics are considered in the present review paper which is organized as follows. The phenomenon of the supercritical charge
instability is briefly discussed in Sec.~\ref{section:introduction}. In Sec.~\ref{section:monolayer-graphene}, we analyse the electron states
in gapless and gapped monolayer graphene. The experimental data of the observation of the atomic collapse in graphene are provided in
Sec.~\ref{section:experiment}. The impact of
a magnetic field on the supercritical charge problem in graphene is studied in Sec.~\ref{section:magnetic}. Two-electron bound states of a
Coulomb impurity are considered in Sec.~\ref{section:bound-states}. The atomic collapse in the field of two charged impurities is investigated in
Sec.~\ref{section:two-centers}. The dipole problem is studied in Sec.~\ref{section:dipole}. The specifics of the atomic collapse in bilayer
graphene is considered in Sec.~\ref{section:bilayer}. The results are summarized and conclusions are given in Sec.~\ref{section:Summary}.

\section{Atomic collapse in monolayer graphene}
\label{section:monolayer-graphene}

The electron quasiparticle states in the vicinity of the $K_{\pm}$ points of graphene
in the potential $V(\mathbf{r})$  of charged impurities are described by the following Dirac Hamiltonian in $2+1$
dimensions:
\begin{equation}
H(\mathbf{p},\xi)=v_{\rm F}\boldsymbol\sigma\boldsymbol{p}+\xi\Delta\sigma_z+V(\mathbf{r}),
\label{Master-Hamiltonian}
\end{equation}
where $v_{\rm F}$ is the Fermi velocity of graphene, $\boldsymbol{p}=-i\boldsymbol{\nabla}$ is the canonical momentum, $\sigma_{i}$ are the Pauli matrices, $\Delta$
is a quasiparticle gap, and $\xi$ is an index, which corresponds to the valley $K_{+}$ ($\xi=+1$) or  $K_{-}$ ($\xi=-1$). Although the pristine
graphene is gapless, a quasiparticle gap $\Delta$ can be generated if graphene sheet is placed on a substrate and two carbon sublattices
become inequivalent because of interaction with the substrate  (for band structure calculation of such a configuration see, for instance,
Ref.\cite{Giovannetti2007}). The gap can arise also in graphene ribbons due to geometrical quantization\cite{Son2006} or due to many-body electron
correlations \cite{Khveshchenko2001,Gorbar2002,Gorbar2003,Khveshchenko2004,Gamayun2010,Gonzalez2012,Drut2009a,Drut2009b,Armour2010,Buividovich2012}.

The Hamiltonian (\ref{Master-Hamiltonian}) acts on two component spinor $\Psi_{\xi s}$ which carries the valley ($\xi=\pm)$ and spin ($s=\pm$)
indices. We will use the standard convention: $\Psi^{T}_{+s}=(\psi_{A},\psi_{B})_{K_{+}s}$, whereas $\Psi^{T}_{-s}=(\psi_{B},\psi_{A})_{K_{-}s}$,
and $A,B$ refer to two sublattices of hexagonal graphene lattice. Since the interaction potential does not depend on spin, we will omit the spin
index $s$ in what follows. Further, for the sake of definiteness, we will consider electrons in the $K_+$ valley. The Hamiltonians at two
valleys are related by means of the time reversal operator $\Theta=is_2\sigma_1K$:
\begin{equation}
\Theta H(\mathbf{p},\xi=+1)\Theta^{-1}=H(-\mathbf{p},\xi=-1),
\end{equation}
where $s_2$ is the Pauli spin matrix and K is the complex conjugation.

The supercritical instability in the field of a single charged impurity was studied quite in detail in the literature
\cite{Shytov2007a,Shytov2007b,Pereira2007,Fogler2007,Gamayun2009,Novikov2007,Pereira2008,Terekhov2008,Shytov2009,CastroNeto2009a}. In this section
we will summarize its main features.

\subsection{Resonance states in gapless graphene in quasiclassical approach}
\label{monolayer-resonances}

Let us start our analysis with the case of gapless graphene. Since massless particles cannot form bound states, the atomic collapse is revealed
for massless particles through resonance states which appear when the Coulomb potential strength exceeds a certain critical value $\alpha_c=1/2$.
In order to demonstrate the presence of these states, it is instructive to begin with the semiclassical analysis. We follow in this subsection
the derivation in Ref.\cite{Shytov2007a}.

In relativistic classical theory, the electron trajectories can spiral around the charged center and eventually fall down on it\,\cite{Darwin1913}
if the electron angular momentum is small enough $M<M_c=Ze^2/c$.

These states can be constructed quasiclassically from relativistic dynamics described by the Hamiltonian $H=v_{\rm F}|\mathbf{p}|+V(r)$, where
$V(r)=-Ze^2/(\kappa r)$ and $\kappa$ is a dielectric constant. The collapsing trajectories with angular momenta $M< M_c=Ze^2/(\kappa v_{\rm F})$
are separated from non-falling trajectories by a centrifugal barrier. This is manifested in the expression for the radial momentum square
\begin{equation}
\label{eq:p_r}
p_r^2  = v_{\rm F}^{-2}\left(E + \frac{Ze^2}{\kappa r}\right)^2 - \frac{M^2}{r^2}.
\end{equation}
Clearly, there is
a classically forbidden region, the annulus $r_1<r<r_2$,
$r_{1,2}=(\frac{Ze^2}{\kappa} \mp Mv_{\rm F})/|E|$, where the right-hand side of Eq.(\ref{eq:p_r}) is negative.
The quasi-stationary states trapped by this barrier are obtained from the Bohr-Sommerfeld quantization $\int_{r_0}^{r_1} p_r dr=\pi\hbar n$,
where $r_0$ is a regularization parameter, which is of order of lattice spacing. Evaluating the integral with logarithmic accuracy, we obtain
$ \gamma \ln \frac{Ze^2}{\kappa r_0|E|}= \pi\hbar n$, where $\gamma\equiv \left( M_c^2-M^2\right)^{1/2}$, which gives the quasi-Rydberg states
\begin{equation}
\label{eq:En_quasiclass}
E_n \approx
-\frac{Ze^2}{\kappa r_0}e^{-\pi \hbar n/\gamma} ,\quad n>0.
\end{equation}
The energies of these states converge to zero, $E_n\to 0$, at large $n$, whereas their radii diverge, similar to the Rydberg states in the
hydrogen atoms. To find the transparency of the barrier, we integrate ${\Im m\,} p_r$ and obtain the tunneling action
\begin{equation}\label{eq:S_coulomb}
S = \int_{r_1}^{r_2} dr \sqrt{\frac{M^2}{r^2} - \left(\frac{E}{v_{\rm F}} + \frac{M_c}{r}\right)^2}
= \pi \left(M_c - \gamma\right).
\end{equation}
Taken near the threshold $\gamma\approx 0$, the transparency $e^{-2S/\hbar}$ gives the width $\Gamma_n \sim |E_n|\exp (-2\pi Z\alpha)$, where
$\alpha=e^{2}/(\kappa \hbar v_{\rm F})$ is the effective coupling constant.
The quasi-Rydberg states manifest themselves in the local density of states that can be probed experimentally. Also, resonance scattering
on the quasi-bound states manifests itself in the dependence of transport properties on the carrier density. For supercritical potential strength
$|Z\alpha|>1/2$ there are oscillations of the Ohmic conductivity which have a characteristic form of Fano resonances centered at $E_n$ \cite{Shytov2007a}.
In this regime the conductivity exhibits peaks at the densities for which the Fermi energy $E_F$ equals $E_n$. The peak position is highly sensitive
to the potential strength $Z\alpha$, changing by an order of magnitude when $Z\alpha$ varies from $-1.0$ to $-1.3$.

It is instructive to compare these results to the exact solution of the Coulomb center problem that we do in the next section.

\subsection{Supercritical instability in graphene}
\label{monolayer-gapped}
\subsubsection{Gapped graphene, subcritical regime}
Now, let us include into consideration a quasiparticle gap that on the one side makes more transparent the derivation of the instability condition
(diving of the lowest energy level into the negative continuum), while on the other hand takes into account a possible presence of a gap due to the
interaction with a substrate. In this subsection, we follow the study performed in Ref.\cite{Gamayun2009}. The electron quasiparticle states in
graphene in the field of a single Coulomb impurity are described by Dirac Hamiltonian (\ref{Master-Hamiltonian}) with a regularized Coulomb potential
\begin{equation}
\label{potential-single-regularized}
V(r) = -\frac{Ze^{2}}{\kappa r},
\,\,(r>r_{0}),\quad V(r) = -\frac{Ze^{2}}{\kappa r_{0}}, \,\, (r<r_{0}).
\end{equation}
As we discussed in the Introduction,  to avoid the fall-to center problem we should regularize the Coulomb
potential at small distances. Potential (\ref{potential-single-regularized}) represents the simplest ``cutoff'' regularization. Since the Hamiltonian (\ref{Master-Hamiltonian}) with potential (\ref{potential-single-regularized}) commutes with the total angular momentum operator $J_z=L_z+S_z=-i\hbar\frac{\partial}{\partial \phi}+\frac{\hbar}{2}\sigma_z$, we seek eigenfunctions
in the following form:
\begin{equation}
\Psi = \frac{1}{r}\left(\begin{array}{c}
e^{i\phi (j-1/2)}\,a(r) \\
i\,e^{i\phi (j+1/2)}\,b(r)
\end{array}\right).
\label{spinor}
\end{equation}
Then we obtain a system of two coupled ordinary differential
equations of the first order
\begin{equation}
a' - (j+1/2)\frac{a}{r} + \frac{E+\Delta-V(r)}{\hbar v_{\rm F}}b =0,
\quad\quad\quad\quad\quad
b' + (j-1/2)\frac{b}{r} - \frac{E-\Delta-V(r)}{\hbar v_{\rm F}}a =0.
\quad\quad\quad\quad
\label{two-component-Dirac-Eq}
\end{equation}
It is convenient to define the quantities $\epsilon=
{E}/{\hbar v_{\rm F}},\, m= {\Delta}/{\hbar v_{\rm F}}$, and $\alpha
=\alpha_{g}/\kappa={e^2}/{\hbar v_{\rm F}\kappa}$.

The discrete spectrum of Eqs.~(\ref{two-component-Dirac-Eq}) exists for $|\epsilon| < m$. In this case it is convenient to
define
\begin{equation}
u
=\sqrt{m^2-\epsilon^2},\quad \rho =2 u r, \quad a =\frac{\sqrt{m+\epsilon}}{2}(g-f), \quad b=\frac{\sqrt{m-\epsilon}}{2}(g+f) \label{ab}
\end{equation}
and rewrite Eqs.~($\ref{two-component-Dirac-Eq}$) in the region $r>r_{0}$ as follows:
\begin{eqnarray}
\rho g' + g\left(\frac{\rho}{2}-\frac{1}{2}-Z\alpha\frac{\epsilon}{u}\right)+
f\left(j+Z\alpha\frac{m}{u}\right) =0,\nonumber\\
\rho f' -
f\left(\frac{\rho}{2}+\frac{1}{2}-Z\alpha\frac{\epsilon}{u}\right)+g\left(j-Z\alpha\frac{m}{u}\right)
=0.
\label{system-eqs-continuum}
\end{eqnarray}
Substituting $f$ from the first equation into the second
one, we obtain the equation for the $g$ component
\begin{equation}
\frac{d^2g}{d\rho^2} +
\left(-\frac{1}{4}+\frac{\frac{1}{2}+Z\alpha\frac{\epsilon}{u}}{\rho}+\frac{\frac{1}{4}-j^2
	+Z^2\alpha^2}{\rho^2}\right)g=0,
\end{equation}
which is the well-known Whittaker equation \cite{Gradshtein-book}. Its general solution is
\begin{equation}
g =C_1 W_{\mu,\nu}(\rho) + C_2
M_{\mu,\nu}(\rho),\quad \mu=\frac{1}{2}+\frac{Z\alpha\epsilon}{u}, \quad \nu =\sqrt{j^2-Z^2\alpha^2}.
\end{equation}
Taking into account the asymptotic of the Whittaker
functions $W_{k,\nu}(z),M_{k,\nu}(z)$ at infinity,
\begin{equation}
W_{\mu,\nu}(\rho)\simeq e^{-u r}(2u r)^{\mu},\quad M_{\mu,\nu}(\rho)\simeq\frac{\Gamma(1+\nu)}{\Gamma(\frac{1}{2}-\mu+\nu)}\,
e^{u r}(2u r)^{-\mu},\quad r\to\infty,
\end{equation}
we find that the regularity condition at infinity requires $C_2 =0$. Then the first equation in
(\ref{system-eqs-continuum}) gives the following solution for the $f$ component in the region II ($r > r_{0}$):
\begin{equation}
f_{II}= C_1 \left(j-Z\alpha\frac{ m}{u}\right)W_{-\frac{1}{2}+Z\alpha\frac{\epsilon}{u},\nu}(\rho).
\label{boundstate-wf}
\end{equation}
Solutions in the region I ($r<r_{0}$) are easily obtained
\begin{eqnarray}
b_I = A_1\,rJ_{|j+1/2|}\left(r\sqrt{\left(\epsilon+\frac{Z\alpha}{r_{0}}\right)^2-m^2}\right),\\
a_I = A_1\,\mbox{sgn}(j)\,
\sqrt{\frac{\epsilon+Z\alpha/r_{0}+m}{\epsilon+Z\alpha/r_{0}-m}}\,\,rJ_{|j-1/2|}\left(r\sqrt{\left(\epsilon+
	\frac{Z\alpha}{r_{0}}\right)^2-m^2}\right),
\label{solutions-region-I}
\end{eqnarray}
where $A_1$ is a constant and we took into account the infrared boundary condition
which selects only regular solution for $b_I$ and $a_I$. Energy levels are
determined through the continuity condition of the wave function at $r=r_{0}$,
\begin{equation}
\left.\frac{b_I}{a_I}\right|_{r=r_{0}}=\left.\frac{b_{II}}{a_{II}}\right|_{r=r_{0}},
\end{equation}
that gives the equation
\begin{equation}
\frac{W_{\frac{1}{2}+\frac{Z\alpha\epsilon}{u},\nu}(\rho)}{\left(j-\frac{Z\alpha
		m}{u}\right)
	W_{-\frac{1}{2}+\frac{Z\alpha\epsilon}{u},\nu}(\rho)}\Big|_{r=r_{0}} =
\frac{k+1}{k-1},\,\,\,\, k =
\mbox{sign}(j)\,\frac{m+\epsilon}{u}\sqrt{\frac{\epsilon+Z\alpha/r_{0}-m}{\epsilon+Z\alpha/r_{0}+m}}
\frac{J_{|j+1/2|}(\widetilde{\rho})}{J_{|j-1/2|}(\widetilde{\rho})},\,
\widetilde{\rho}=\sqrt{(Z\alpha+\epsilon r_{0})^{2}-m^{2}r_{0}^{2}}.
\label{matching}
\end{equation}
We analyze this equation in the limit $r_{0} \to 0$ where we can use the asymptotical behavior
of the Whittaker function at $\rho\rightarrow 0$,
\begin{equation}
W_{\mu,\nu}(\rho)\simeq\frac{\Gamma(2\nu)}{\Gamma(\frac{1}{2}-\mu+\nu)}\rho^{\frac{1}{2}-\nu}
+\frac{\Gamma(-2\nu)}{\Gamma(\frac{1}{2}-\mu-\nu)}\rho^{\frac{1}{2}+\nu}.
\end{equation}
In the limit $r_{0} \to 0$ Eq.(\ref{matching}) reduces to the
following one,
\begin{eqnarray}
\frac{\Gamma(-2\nu)}{\Gamma(2\nu)}\frac{\Gamma\left(1+\nu -
	Z\alpha\frac{\epsilon}{u}\right)}{\Gamma\left(1-\nu -
	Z\alpha\frac{\epsilon}{u}\right)}(2u r_{0})^{2\nu}=
-\frac{j+\nu-\frac{Z\alpha(m+\epsilon)}{u}+k_{0}\left(j-\nu-\frac{Z\alpha(m-\epsilon)}{u}\right)}
{j-\nu-\frac{Z\alpha(m+\epsilon)}{u}+k_{0}\left(j+\nu-\frac{Z\alpha(m-\epsilon)}{u}\right)}+O(r_{0}),
\label{eq:Rtozero}
\end{eqnarray}
where
\begin{equation}
k_{0}=\mbox{sign}(j)\,\frac{m+\epsilon}{u}\frac{J_{|j+1/2|}(Z\alpha)}{J_{|j-1/2|}(Z\alpha)}
\equiv \frac{m+\epsilon}{u}\,\sigma(Z\alpha,j).
\end{equation}

\begin{figure}[ht]
	\centering
	\includegraphics[width=6.5cm]{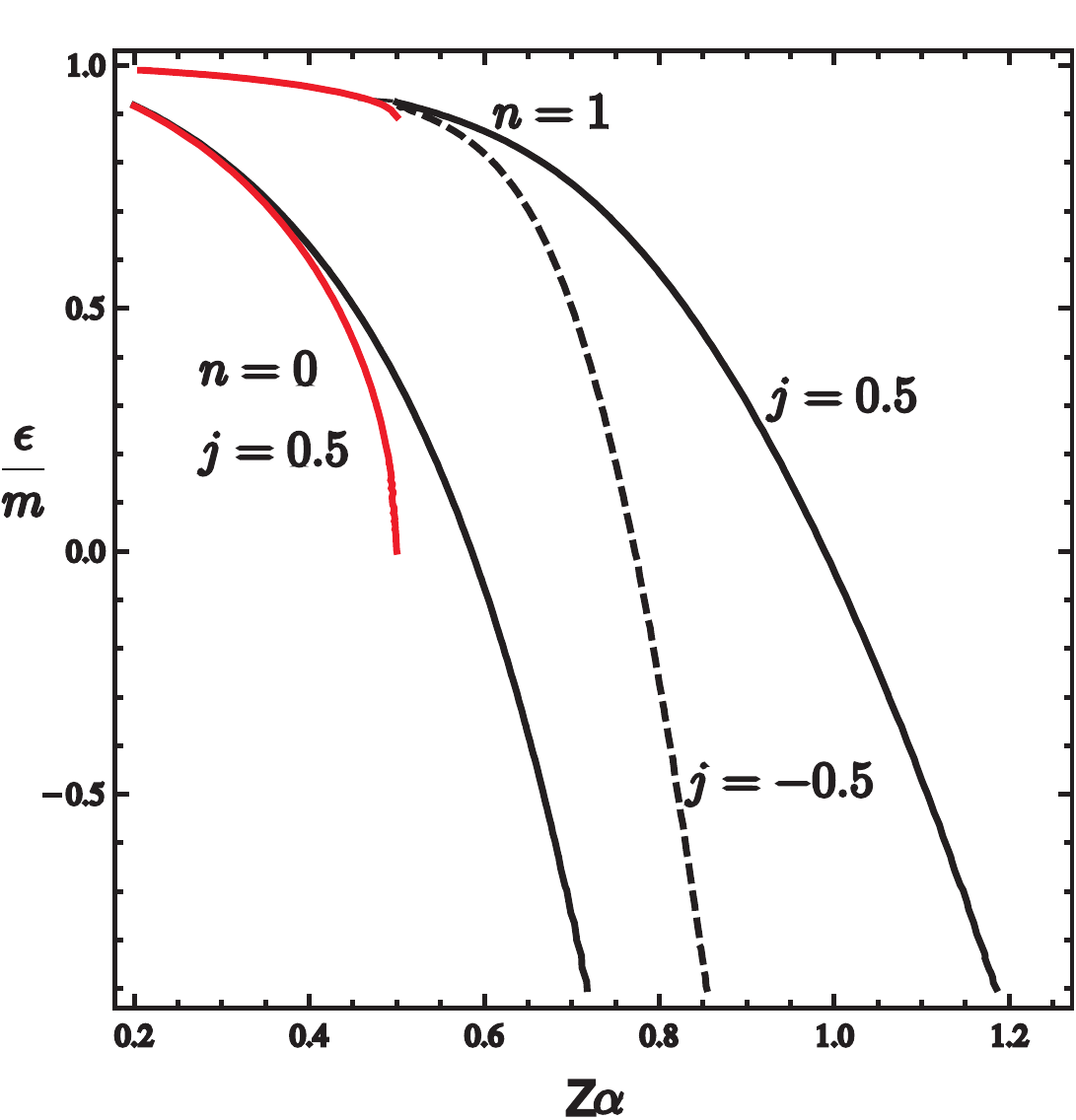}
    \includegraphics[width=9cm]{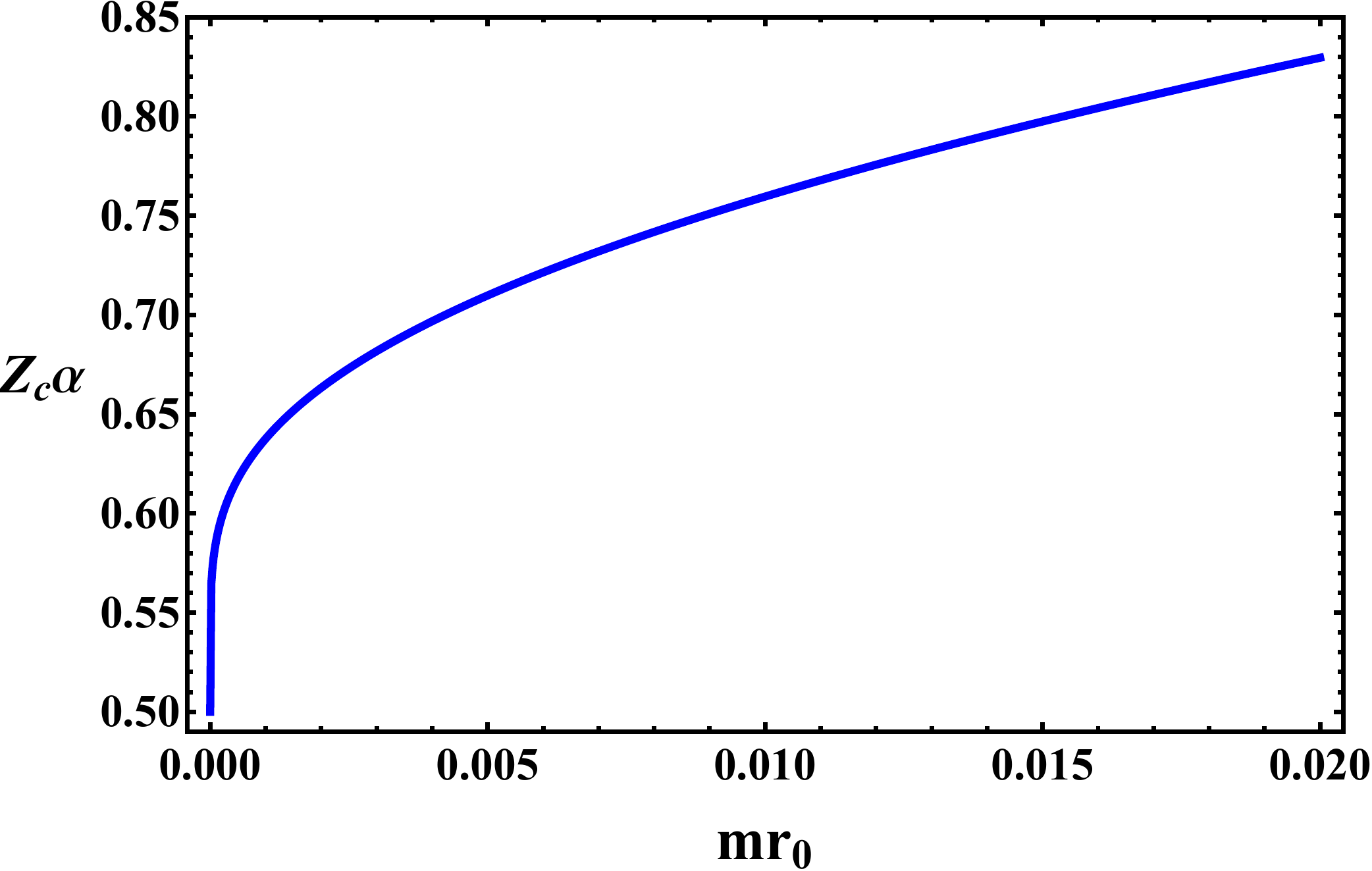}
	\caption{Left panel: The lowest energy levels as functions of $Z\alpha$. Red lines correspond to the pure Coulomb potential (they exist only
             for $Z\alpha < 1/2$); black solid lines are numerical solutions for $j=1/2,\, m r_{0}=0.01$; black dashed line is a numerical solution
             for $j=-1/2,\, m r_{0}=0.01$. Right panel: The critical coupling as a function of $m r_{0}$ for the $1S_{1/2}$ level.} \label{cr1}
\end{figure}

Equation~(\ref{eq:Rtozero}) can be rewritten in more convenient form
\begin{equation}
\frac{\Gamma(-2\nu)}{\Gamma(2\nu)}\frac{\Gamma\left(1+\nu -
	Z\alpha\frac{\epsilon}{u}\right)}{\Gamma\left(1-\nu -
	Z\alpha\frac{\epsilon}{u}\right)}(2u r_{0})^{2\nu}=
-\frac{j-\nu-\frac{Z\alpha(m-\epsilon)}{u}}{j+\nu-\frac{Z\alpha(m-\epsilon)}{u}}
\frac{j+\nu-Z\alpha\sigma(Z\alpha,j)}{j-\nu-Z\alpha\sigma(Z\alpha,j)}.
\label{eq:Rtozero2nd}
\end{equation}
In the limit $r_{0}\to 0$ the energy levels are determined by the poles of the gamma function $\Gamma\left(1+\nu -Z\alpha\frac{\epsilon}{u}\right)$
and by a zero of the right hand side of Eq.(\ref{eq:Rtozero2nd}), this leads to the familiar result (analogue of the Balmer's formula in
QED) \cite{Khalilov1998} (rederived also in \cite{Novikov2007}),
\begin{equation}
\epsilon_{n,j}=
m\left[1+\frac{Z^2\alpha^2}{(\nu+n)^2}\right]^{-1/2},\quad \left\{\begin{array}{c}n=0,1,2,3,...,\, j>0,\\
\hspace{-2.2mm}n=1,2,3,...,\, \hspace{3mm}j<0.\end{array}\right.
\label{energy-levels}
\end{equation}
The bound states for $n\ge1$ are doubly degenerate, $\epsilon_{n,j}=\epsilon_{n,-j}$. The lowest energy level is given by
\begin{equation}
\epsilon_{0,j=1/2} = m\sqrt{1-(2Z\alpha)^2}\,. \label{lowest}
\end{equation}
If $Z\alpha$ exceeds $1/2$, then the ground state energy (\ref{lowest}) becomes purely imaginary, i.e., the fall into the
center phenomenon occurs \cite{Shytov2007a,Shytov2007b,Shytov2009,CastroNeto2009a}. In fact, all energies $\epsilon_{n,1/2}$ become
complex for $Z\alpha>1/2$. The unphysical complex energies indicate that the Hamiltonian of the system is not a self-adjoint operator for
supercritical values $Z\alpha>1/2$ and should be extended to become a self-adjoint operator.
According to \cite{Pomeranchuk1945,Zeldovich1972}, nonzero $r_{0}$ resolves this problem. For $Z\alpha>j$, $\nu$  is imaginary for
certain $j$ and for such $j$ we denote $\nu=i\beta,\beta=\sqrt{Z^2\alpha^2-j^2}$. For finite $r_{0}$ discrete levels also exist for $Z\alpha>1/2$.
Their energy decreases with increasing of $Z\alpha$ until they reach the lower continuum. The behavior of lowest energy levels with $j=1/2$ as
functions of the coupling $Z\alpha$ is shown in Fig.\ref{cr1} (left panel).

The critical charge $Z_{c}$ that corresponds to diving into the continuum is obtained from Eq.(\ref{eq:Rtozero2nd}) setting $\epsilon = -m$ there and using
the corollary of the Stirling formula: $\frac{\Gamma(x+iy)}{\Gamma(x-iy)} \rightarrow e^{2iy \ln x} ,\,\,\,\, x \rightarrow +\infty$. We come at the
equation
\begin{equation}
\label{crc}
e^{-2i\beta\ln(2Z\alpha mr_{0})} = \frac{i\beta - j +
	Z\alpha\sigma(Z\alpha,j)}{-i\beta - j +
	Z\alpha\sigma(Z\alpha,j)}\frac{\Gamma(1-2i\beta)}{\Gamma(1+2i\beta)}\,,
\end{equation}
or,
\begin{equation}
-\beta\ln(2Z\alpha m r_{0})=\arg\,(Z\alpha\sigma({Z\alpha,j})-j+i\beta)
+\arg\,\Gamma(1-2i\beta)+\pi n, \quad n=0,\,1,\,... \label{critcoupling:eq}
\end{equation}
It is not difficult to check that for $j=1/2$ and $n=1$ the critical coupling $Z_c\alpha$ approaches the value $1/2$ for
$mr_{0} \rightarrow 0$. The dependence of the critical coupling $Z_c\alpha$ on $mr_{0}$ for $j=1/2$ is shown in
the right panel of Fig.\ref{cr1}.

\subsubsection{Gapped graphene, supercritical regime}

Let us analyze Eq.~(\ref{matching}) in the supercritical case $Z\alpha >1/2$ and show that there are resonant states for
$|\epsilon|> m$ (we define the gap $\Delta>0$). The Whittaker function $W_{\mu,\nu}(\rho)$ with $\mu=1/2+Z\alpha\epsilon/u$,
$\nu=\sqrt{j^{2}-Z^2\alpha^{2}}$ describes bound states for $|\epsilon|<m$ which are situated on the first physical sheet of
the variable $u$ and for which ${\Re e\,}u>0$ (see, Eq.(\ref{boundstate-wf})). The quasistationary states are described by the
same function $W_{\mu,\nu}(\rho)$ and are on the second unphysical sheet with ${\Re e\,}u<0$. We shall look for the solutions
corresponding to the quasistationary states which define outgoing hole waves at $r\to\infty$ with
\begin{equation}
{\Re e\,}\epsilon<0,\quad {\Im m\,}\epsilon<0, \quad {\Re e\,}u<0,\quad {\Im m\,}u<0.
\end{equation}
For solutions with $Z^2\alpha^2 > j^2$ resonance states are determined by Eq.(\ref{matching}) for bound states where $\nu$ is
replaced by $\nu=i\beta$. We will consider the states with $j=1/2$ which correspond to the $nS_{\frac{1}{2}}$-states, in particular,
the lowest energy state belongs to them. The corresponding equation then takes the form
\begin{equation}
\frac{W_{\frac{1}{2}+\frac{Z\alpha\epsilon}{u},i\beta}(\rho)}{\left(\frac{1}{2}
	-\frac{Z\alpha
		m}{u}\right)W_{-\frac{1}{2}+\frac{Z\alpha\epsilon}{u},i\beta}
	(\rho)}\Big|_{r=r_{0}} = \frac{k+1}{k-1},\,\,\,\, k =
\,\frac{m+\epsilon}{u}\sqrt{\frac{\epsilon+Z\alpha/r_{0}-m}{\epsilon+Z\alpha/r_{0}+m}}
\frac{J_{1}(\widetilde{\rho})}{J_{0}(\widetilde{\rho})},\,
\widetilde{\rho}=\sqrt{(Z\alpha+\epsilon r_{0})^2-m^2 r_{0}^2}.
\end{equation}

The analytical
results can be obtained for the near-critical values of $Z$ when $Z\alpha-1/2 \ll 1$. We assume that $|2u r_{0}| \ll 1$, then using
the asymptotic of the Whittaker function, we find
\begin{equation}
(2u r_{0})^{2i\beta}\frac{\Gamma(1-2i\beta)}{\Gamma(1+2i\beta)}\frac{\Gamma\left(1+i\beta
	-\frac{Z\alpha\epsilon}{u}\right)}
{\Gamma\left(1-i\beta-\frac{Z\alpha\epsilon}{u}\right)}=
\frac{\frac{1}{2}-i\beta -
	\frac{Z\alpha(m-\epsilon)}{u}}{\frac{1}{2}+i\beta
	-\frac{Z\alpha(m-\epsilon)}{u}}
\frac{\frac{1}{2}+i\beta-Z\alpha\frac{J_{1}(Z\alpha)}{J_{0}(Z\alpha)}}
{\frac{1}{2}-i\beta-Z\alpha\frac{J_{1}(Z\alpha)}{J_{0}(Z\alpha)}}\,.
\label{matching-resonance-final}
\end{equation}
Expanding Eq.(\ref{matching-resonance-final}) in the near critical region in powers of $\beta=\sqrt{Z^2\alpha^2-1/4}$, we find the
following equation:
\begin{equation}
(-2i\sqrt{\epsilon^2-m^2} r_{0})^{2i\beta} =
1+4i\beta\left[\frac{J_0(1/2)}{J_0(1/2)-J_1(1/2)}+\Psi(1)-\frac{1}{2}
\Psi\left(1-\frac{i}{2}\frac{\epsilon}{\sqrt{\epsilon^2-m^2}}\right)-
\frac{1}{1+i\sqrt{\frac{\epsilon-m}{\epsilon+m}}}\right].
\label{resonance-first-expansion}
\end{equation}
Here $\Psi(x)$ is the psi-function and we put $u=-i\sqrt{\epsilon^{2}-m^{2}}$ where ${\Im m}\sqrt{\epsilon^{2}-m^{2}}<0$ on the
second sheet.

It is instructive to consider
resonant states in the vicinity of the level $\epsilon = - m$ when bound states dive into the lower continuum and determine their
real and imaginary parts of energy. First of all, nonzero $m$ increases the value of the critical charge.
Indeed, using Eq.(\ref{critcoupling:eq}), we obtain that the critical value $Z_c\alpha$ for $j=1/2$ scales with $m$ like
(see Fig.1 (right panel))
\begin{equation}
Z_{c}\alpha\simeq \frac{1}{2}+\frac{\pi^{2}}{\ln^{2}(cm r_{0})}, \quad
c=\exp\left[-2\Psi(1)-\frac{2J_{0}(1/2)}{J_{0}(1/2)-J_1(1/2)} \right] \approx
0.21.
\label{crit-charge-on-r0}
\end{equation}
Note that the dependence of the critical coupling on $m r_{0}$ is quite similar to that in the strongly coupled QED
\cite{Fomin1978,FominReview1983}.

For $Z>Z_c$, using Eq.(\ref{resonance-first-expansion}), we find the following resonant states:
\begin{equation}
\epsilon=-m\left(1+b+i\frac{3\pi}{8}e^{-\pi/\sqrt{2b}}\right),\quad
b=\frac{3\pi}{8}\,\frac{\beta-\beta_c}{\beta\beta_c},
\label{massive-resonant-states}
\end{equation}
where $\beta_c=\sqrt{(Z_c\alpha)^{2}-1/4}$. Like in QED \cite{Popov1971} the imaginary part of energy of these resonant states vanishes
exponentially as $Z \to Z_c$. Such a behavior is connected with tunneling through the Coulomb barrier in the problem under consideration.
For the quasielectron in graphene in a central potential $V(r)$, expressing the lower component of the Dirac spinor (\ref{spinor}) through
the upper one and following \cite{Zeldovich1972,Popov1971}, we obtain an effective second order differential equation in the form of the
Schr\"{o}dinger equation
\begin{equation}
\chi^{\prime\prime}(r)+k^2(r)\chi(r)=0, \quad\quad\quad
a(r)=\exp\left[\frac{1}{2}\int\left(\frac{1}{r}-\frac{\tilde{V}^{\prime}}{\epsilon+m-\tilde{V}}\right)\,dr\right]\,\chi(r).
\label{Schrodinger-like:eq}
\end{equation}
Here
\begin{equation}
\label{quasiclassical-momentum}
k^2(r)=2(\mathcal{E}-U(r)),\quad \mathcal{E}=\frac{\epsilon^2-m^2}{2},\quad \tilde{V}=\frac{V}{\hbar v_{\rm F}},
\end{equation}
and we represent the effective potential as the sum of two terms $U=U_{1}+U_{2}$, where $U_1$ is the effective potential for the Klein--Gordon
equation and $U_2$ takes into account the spin dependent effects,
\begin{eqnarray}
U_{1}&=&\epsilon\tilde{V}-\frac{\tilde{V}^{2}}{2}+\frac{j(j-1)}{2r^{2}},
\label{U1-pot}\\
U_{2}&=&\frac{1}{4}\left[\frac{\tilde{V}^{\prime\prime}}{\epsilon+m-\tilde{V}}+\frac{3}{2}
\left(\frac{\tilde{V}^{\prime}}{\epsilon+m-\tilde{V}}\right)^{2}+\frac{2j\tilde{V}^{\prime}}
{r(\epsilon+m-\tilde{V})}\right]. \label{U2-pot}
\end{eqnarray}
Note that Eq.(\ref{Schrodinger-like:eq}) and the potentials (\ref{U1-pot}), (\ref{U2-pot}) coincide with the corresponding equations in QED
\cite{Zeldovich1972}. One can show that in the near-critical regime ($Z \to Z_c$, $j=1/2$, and $\epsilon=-m$) the effective potential
$U(r)$ has the Coulomb barrier (see Fig.\ref{pot} below), which prevents the delocalization of the wave function.

The tight-binding approach (solved exactly by using numerical techniques) was compared with the continuum approach based on the Dirac equation in Ref.\cite{Pereira2007}. It was shown that the latter provides a good qualitative description of the problem at low energies when properly regularized.
On the other hand, the Dirac description fails at moderate to high energies and at short distances when the lattice description should be used.

\subsubsection{Gapless graphene}

We consider now the case of gapless graphene, $m=0$. Writing $\epsilon=|\epsilon|e^{i\gamma}$ Eq.(\ref{resonance-first-expansion})
takes the form
\begin{equation}
\ln(2|\epsilon|r_{0})+i\left(\gamma-\frac{\pi}{2}\right)\simeq 2\left[\frac{J_0(1/2)}
{J_0(1/2)-J_1(1/2)}+\Psi(1)-\frac{1}{2}\Psi\left(1-\frac{i}{2}\right)-\frac{1}{1+i}\right]
-\frac{\pi n}{\beta}, \quad n=1,2,\dots.
\end{equation}
We find
\begin{equation}
\epsilon^{(0)}_{n}=a r_{0}^{-1}e^{i\gamma}\exp\left[-\frac{\pi
	n}{\sqrt{Z^2\alpha^{2}-1/4}}\right]= -(1.18+0.17i)r_{0}^{-1}\exp\left[-\frac{\pi
	n}{\sqrt{Z^2\alpha^{2}-1/4}}\right], \quad n=1,2,\dots,
\label{resonant-states}
\end{equation}
where
\begin{equation}
\gamma=\frac{\pi}{2}\left(1+\coth\frac{\pi}{2}\right)\approx 3.28, \quad a=\frac{1}{2}\exp\left[\frac{2J_0(1/2)}{J_0(1/2)-J_1(1/2)}
+2\Psi(1)-1-{\Re e\,}\Psi\left(1 -\frac{i}{2}\right)\right]\approx 1.19.
\end{equation}
These results are in agreement with Eq.~(\ref{eq:En_quasiclass}) and Refs.\cite{Shytov2007a,Shytov2007b}. The energy of quasistationary
states (\ref{resonant-states}) has a characteristic essential-singularity type dependence on the coupling constant reflecting the
scale invariance of the Coulomb potential. The infinite number of quasistationary levels is related to the long-range character of the
Coulomb potential. Note that a similar dependence takes place in the supercritical Coulomb center problem in QED \cite{FominMiransky1976}.

Since  the ``fine structure constant'' $e^{2}/\hbar v_{\rm F}\approx2.2$ in  graphene, an instability could potentially appear already
for the charge $Z=1$. However, in the analysis above we did not take into account the vacuum polarization effects. Considering these effects
and treating the electron-electron interaction in the Hartree approximation, it was shown in Ref.\cite{Terekhov2008} that the effective charge
of impurity $Z_{\rm eff}$ is such that the impurity with bare charge $Z=1$ remains subcritical, $Z_{\rm eff}e^{2}/(\kappa\hbar v_{\rm F})<1/2$,
for any coupling $e^{2}/(\kappa\hbar v_{\rm F})$, while impurities with higher $Z$ may become supercritical.

For finite $m$ and in the case $|\epsilon| \gg m,\,\,\Re e\,\epsilon < 0$, expanding Eq.(\ref{resonance-first-expansion}) in $m/\epsilon$ we
get up to the terms of order $m^2/\epsilon^2$,
\begin{equation}
\epsilon-\frac{m^2}{2\epsilon} = \epsilon^{(0)}_n
\left(1-\frac{m}{\epsilon}+\frac{m^2}{\epsilon^2}(0.29-0.23i)\right),\quad\quad\quad
n=1,2,\dots\, .
\label{energy-level}
\end{equation}
The resonant states with $\epsilon^{(0)}_n$ describe the spontaneous emission of positively charged holes when electron bound
states dive into the lower continuum in the case $m=0$. In order to find corrections to these energy levels due to nonzero $m$,
we seek solution of Eq.(\ref{energy-level}) as a series $\epsilon= \sum\limits_{k=0}^{\infty}\epsilon^{(k)}$ with $\epsilon^{(k)}$ of order $m^k$
and easily find the first two terms
\begin{equation}
\epsilon_{n}=\epsilon_{n}^{(0)}-m+ \frac{m^2}{|\epsilon^{(0)}_n|}(0.24+0.20i).
\end{equation}
Since ${\Im m\,}\epsilon_{n}^{(0)}<0$, the appearance of a gap results in decreasing the width $|{\rm Im}\epsilon|$ of quasistationary
states and, therefore, increases stability of the system. Also, as we showed above, the critical value $Z_c\alpha(mr_0)$,
determined by the condition of appearance of a nonzero imaginary part of the energy, increases with the increase of $m$. Thus there
are two possibilities for the system with supercritical charge to become stable: to create spontaneously electron-holes pairs and shield
the charge or to generate spontaneously the quasiparticle gap. In the problem of the supercritical Coulomb center only the first possibility
can be realized, which is already due to the formulation of the problem as the one-particle one. The second possibility - dynamical generation
of the gap - was studied in Refs.
\cite{Gamayun2009,Wang2010,Sabio2010,Khveshchenko2001,Gorbar2002,Gorbar2003,Khveshchenko2004,Gamayun2010,Gonzalez2012,Drut2009a,Drut2009b,Armour2010,Buividovich2012}.

Considering the many-body problem of strongly interacting gapless quasiparticles in graphene, it was shown
that the Bethe-Salpeter equation for an electron-hole bound state contains a tachyon in its spectrum in the supercritical regime $\alpha>\alpha_c$,
the critical constant $\alpha_c=1.62$ in the static random-phase approximation \cite{Gamayun2009} and $\alpha_c=0.92$ in the case of the
frequency-dependent polarization function \cite{Gamayun2010}. The tachyon states play the role of
quasistationary states in the problem of the supercritical Coulomb center and lead to the rearrangement of the ground state and the formation
of excitonic condensate. Thus, there is a close relation between the two instabilities, in fact, the tachyon instability can be viewed as the
field theory analog of the fall-to-center phenomenon and the critical coupling $\alpha_c$ is an analog of the critical coupling $Z_c\alpha$
in the problem of the Coulomb center. The physics of two instabilities is related to strong Coulomb interaction.

\subsection{Experimental observation of the atomic collapse in graphene}
\label{section:experiment}

Univalent charge impurities, such as K, Na, or ${\rm NH_3}$, all commonly used in graphene, are on the border of the supercritical regime.
To investigate this regime experimentally, one can use divalent or trivalent dopants such as alkaline-earth or rare-earth metals. However,
the observation of atomic collapse in the field of supercritical impurities has remained elusive for some time due to the difficulty of producing
highly charged impurities.

For the first time, the supercritical Coulomb behavior was observed in atomically-fabricated ``artificial nuclei'' assembled on the surface
of a gated graphene device in Ref.\cite{Wang2013}. Calcium atoms were deposited onto the graphene device at low temperature $T< 10\,K$.
Then graphene was warmed up before returning to lower temperature, thus causing the Ca adatoms to thermally diffuse and bind into dimers. Further,
as charges are transferred from a Ca dimer into graphene band states, the Ca dimer becomes positively charged. By making use of the density
functional theory calculations, it was found that Ca dimers acquire an effective positive charge $0.4\,e$.

\begin{figure}[ht]
	\centering
\includegraphics[width=5.3cm]{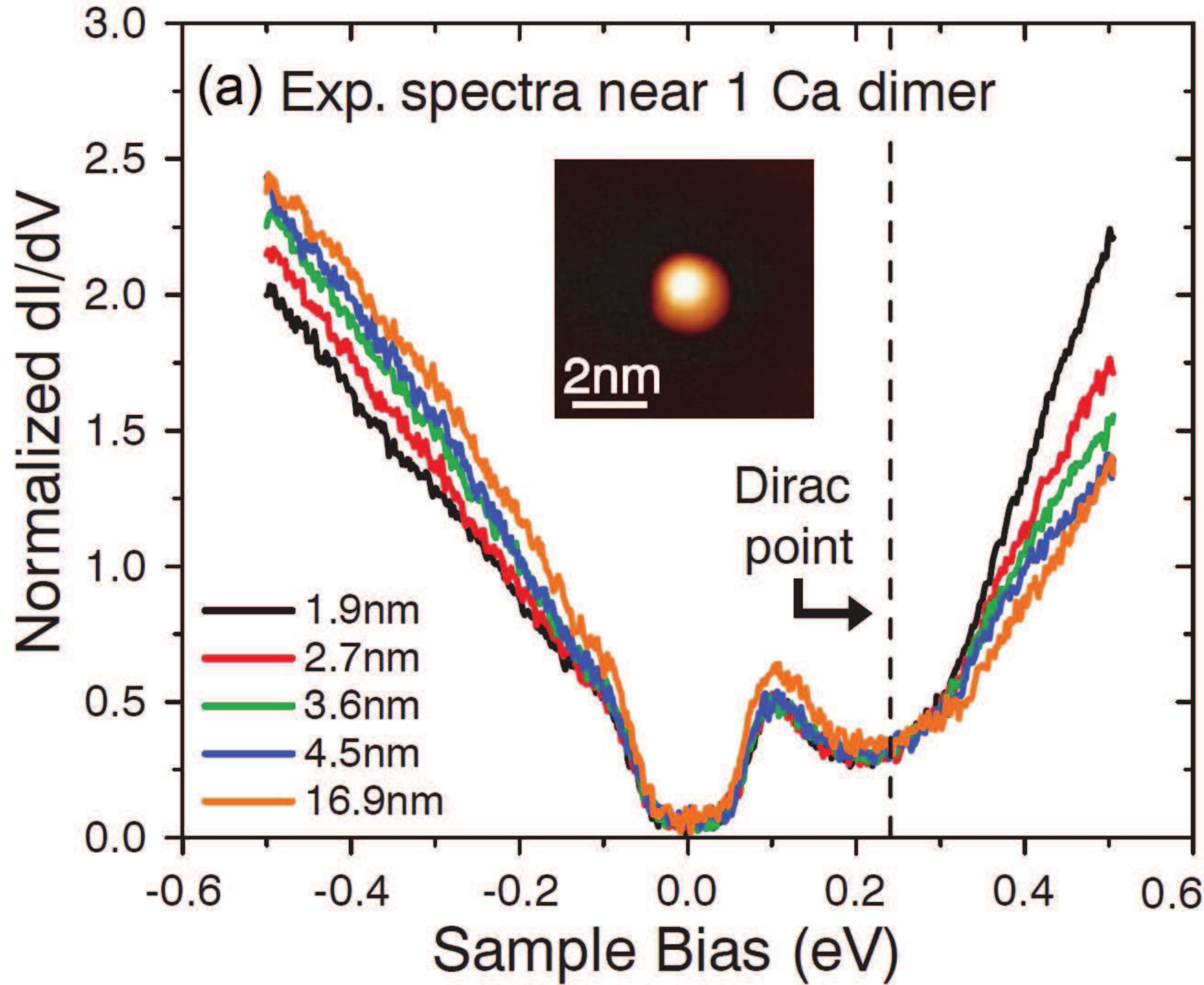}
\includegraphics[width=5.3cm]{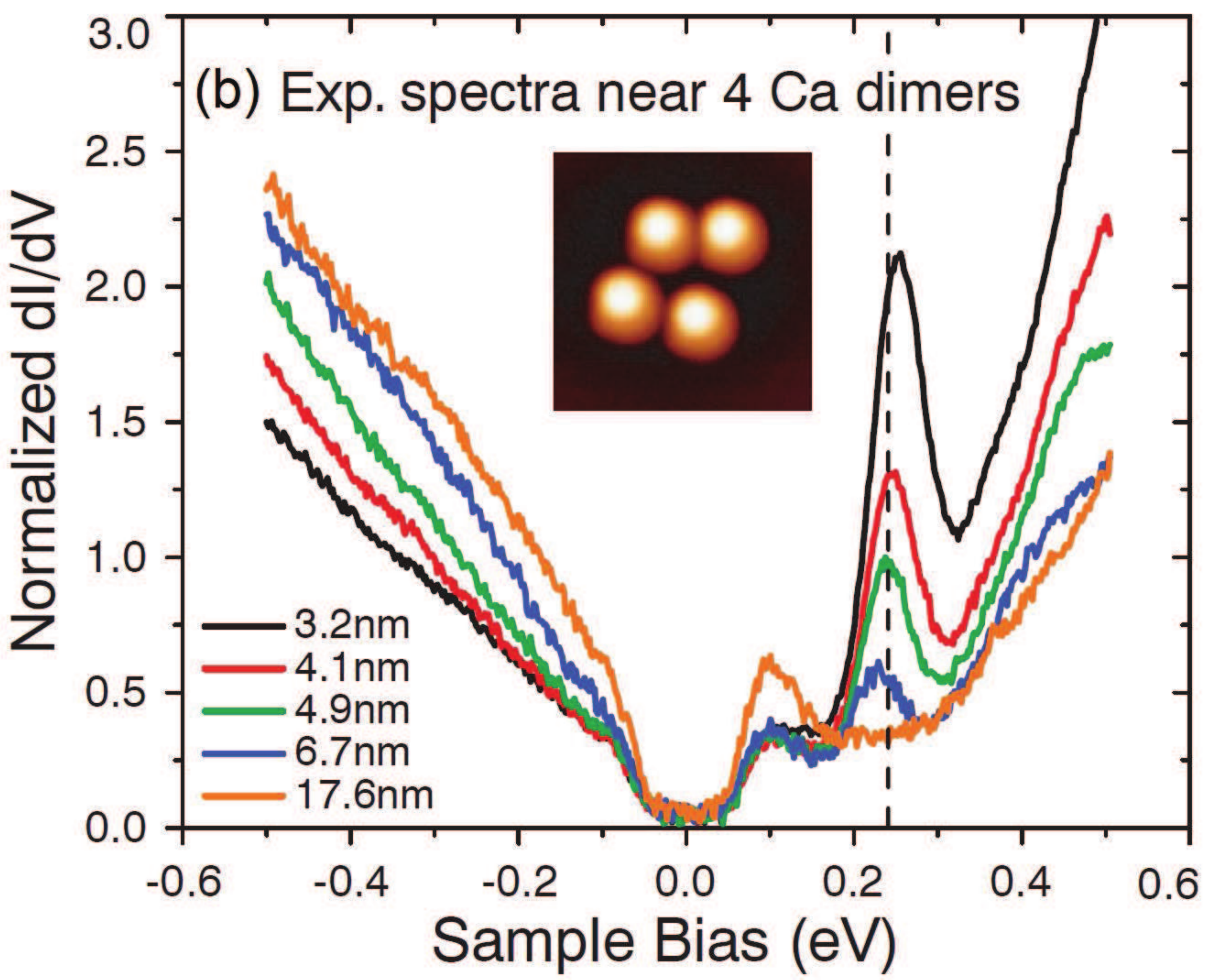}
\includegraphics[width=5.3cm]{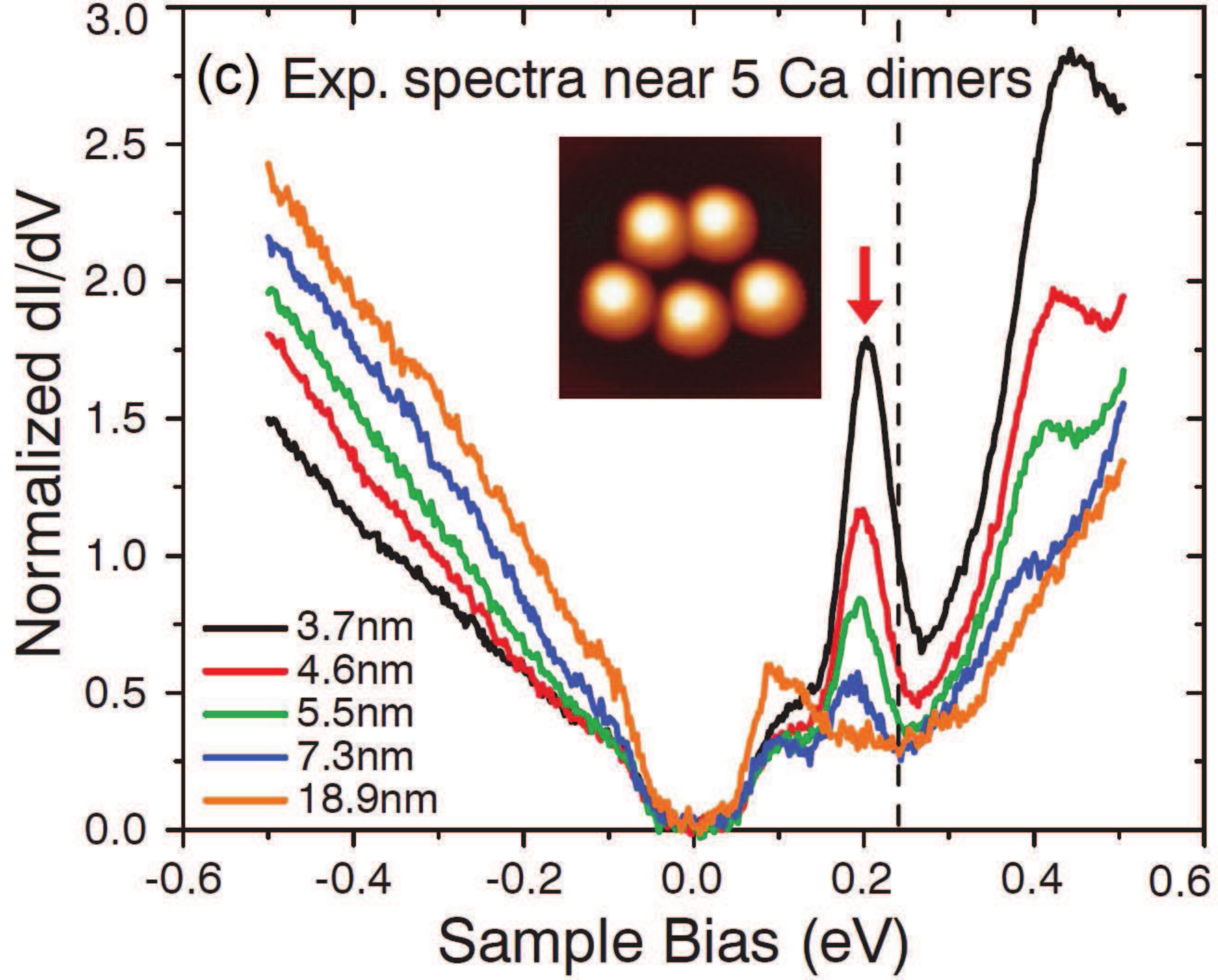}
	\caption{
		Evolution of charged impurity clusters from subcritical to supercritical regime. \textbf{(a-c)} $dI/dV$ spectra measured at
        different distances from the center of Ca-dimer clusters (i.e., artificial nuclei) composed of 1, 4, and 5 dimers.
        ``Center'' here is defined as
        the average coordinate of dimers within a cluster. All spectra were acquired at the same back-gate
		voltage ($V_g = -30\,{\rm V}$) and each was normalized by a different constant factor to account for
		exponential changes in conductivity due to location-dependent tip-height changes \cite{Wang2012,Li1997,Wittneven1998}.
        Insets: STM topographs of atomically fabricated Ca-dimer clusters.  The nuclear charges $Z/Z_{\rm c}$ of \textbf{(a)} 0.5 (1-dimer cluster),
		\textbf{(b)} 1.8 (4-dimer cluster), and \textbf{(c)} 2.2 (5-dimer cluster). Black dashed lines indicate Dirac point, red arrows
        indicate atomic collapse state observed in experiment. This figure is an adapted version of the corresponding figure from
        Ref.~\cite{Wang2013}.
	}
	\label{Crommie-fig1}
\end{figure}

The tunable charge centers were synthesized by pushing together Ca dimers using the tip of a scanning tunneling microscope (STM)
(see insets to Fig.~\ref{Crommie-fig1}, a to c), thus allowing creation of supercritical Coulomb potentials from subcritical charge elements.
The scanning tunneling spectroscopy was used to observe the emergence of atomic-collapse electronic states extending further than
10~nm from the center of artificial nuclei in the supercritical regime ($Z>Z_{\rm c}$). Here, the effective charge $Z$ is defined as the
screened cluster charge where the effects of intrinsic screening due to graphene band polarization and the substrate are taken into account,
and the critical value is $Z_c=\hbar v_F/2e^2\sim0.25$. By tuning the graphene Fermi level ($E_{\rm F}$) via electrostatic gating the atomic
collapse behavior was observed.

Experimentally, the local density of states (LDOS) is measured by means of the STM technique. A sharp STM tip scans
over a graphene piece and measures the electric current $I$ from the surface due to the tunneling effect. This current depends on the voltage
$V$ between tip and a sample and its derivative with respect to $V$ is proportional to the LDOS, $dI/dV\sim D(E,\mathbf{r})$ where $D(E,\mathbf{r})$
is given by Eq.(\ref{eq3}) below. The curves in Fig.\ref{Crommie-fig1} show the differential conductance $dI/dV$ (and thus the LDOS) as a
function of the bias voltage $V$, hence the energy $E=eV$. The various curves in panels (a-c) correspond to different distances from the charge center
in the range of about $2-20 \mbox{nm}$.

Spectra acquired near 1-dimer clusters (Fig.~\ref{Crommie-fig1}a) displayed electron-hole asymmetry  as well as an extra oscillation
in the LDOS at high energies above the Dirac point. For the 4-dimer cluster, the resonance is clearly observed close to
the Dirac point (Fig.~\ref{Crommie-fig1}b). For the 5-dimer cluster, the resonance shifts below the Dirac point (Fig.~\ref{Crommie-fig1}c).
The formation of this resonance (or quasi-bound state) as nuclear charge increases is the ``smoking gun'' for the atomic collapse.
The experimental data suggest that clusters with just one or two Ca dimers are in the subcritical regime. The clusters composed of
four or more dimers are either (for four dimers) transitioning into or (for five dimers) have fully entered the supercritical regime, as
evidenced from panels (b) and (c) in Fig.\ref{Crommie-fig1}. For these clusters, $Z/Z_{\rm c}$ is determined by matching the quasi-bound state
resonance energy between the simulation and experiment. The main features seen in the experimental
data are well reproduced by the Dirac equation simulations in Ref.\cite{Wang2013}.

In order to check that the magnitude of $Z/Z_{\rm c}$ extracted for Ca dimers from the Dirac equation fits is physically reasonable, a completely
separate density functional theory calculation of the charge state expected for a Ca dimer adsorbed to graphene was performed~[\onlinecite{Wang2013}]. This
calculation (which had no fitting parameters) yielded a single-dimer charge ratio of $Z/Z_{\rm c} = 0.6 \pm 0.3$. This is in agreement with the
value $Z/Z_{\rm c} = 0.5 \pm 0.1$ obtained via Dirac equation simulations, and thus lends further support to overall interpretation of the data.
The behavior of the quasi-bound state observed for high-$Z$ artificial nuclei depends on whether it is occupied by electrons or empty. For the
details of this doping dependence see the original paper \cite{Wang2013}.

\section{Supercritical instability in a magnetic field}
\label{section:magnetic}

As we discussed in the Introduction, the supercritical charge instability in a many-body system leads to much more dramatic consequences
compared to the single-particle problem of the Coulomb center. Like the Cooper instability in the theory of superconductivity, the QED
supercritical coupling instability is resolved only through the formation of a condensate of electron-positron pairs generating a mass gap
in the spectrum \cite{FominReview1983}. It was shown in \cite{Krive1992,Klimenko1991,Gusynin1994,Gusynin1996} that magnetic field catalyses
gap generation in relativistic-like systems and even the weakest attraction leads to the formation of a symmetry breaking condensate. Therefore,
the many-body system is always in the supercritical regime once there is an attractive interaction. The magnetic catalysis plays an important
role in quantum Hall effect  studies in graphene \cite{graphene-QHE}, where it is responsible for lifting the degeneracy of the Landau levels.

In QED in (3+1) dimensions, the Coulomb center problem in a magnetic field was studied for massive fermions in \cite{Oraevski1977,Schlutter1985}.
There it was found that the magnetic field confines the transverse electronic motion and the electron in a magnetic field is closer to the nucleus
than in the case where magnetic field is absent. Thus, it feels stronger Coulomb field. Therefore, $Z_c\alpha$ decreases with $B$.
The Dirac equation for (2+1)-dimensional quasiparticles in graphene in the Coulomb potential in a magnetic field was considered in
Ref.\cite{Khalilov-attempt} where exact solutions were found for certain values of magnetic field, i.e., this problem furnishes an example of
the so-called quasi-exactly solvable models. However, no instability or resonance was found.

We would like to stress that the presence of a constant magnetic field changes qualitatively the supercritical Coulomb center
problem. Indeed, if magnetic field is absent, then the supercritical Coulomb center instability leads to a resonance which describes an outgoing
positron propagating freely to infinity. However, since charged particles in a plane perpendicular to a magnetic field do not propagate freely to infinity, such
a behavior is impossible for the in-plane Coulomb center problem in graphene in an out-of-plane magnetic field. Therefore, a priori it is not clear how the supercritical instability
manifests itself in the Coulomb center problem in a magnetic field. This question was studied in Ref.~\cite{Gamayun2011}. We would like to note that the role of a magnetic
field for the atomic collapse in graphene is rather subtle and different conclusions on this issue were drawn in the literature \cite{Siedentop2012,Yang2014,Moldovan2017}.

In the presence of a charged impurity, degenerate Landau levels convert into bandlike structures due to lifting the orbital
degeneracy. For zero chemical potential, as the charge of impurity increases, the energy level with the quantum numbers $n=0$,
$j=-1/2$ comes close to the highest energy state of the level $n=-1$. In the absence of magnetic field, the corresponding bound
state would dive into the lower continuum and further increase of the charge of impurity would produce a resonance. The situation
is qualitatively different in the presence of a magnetic field as the energy curves with the same momenta $j$ never cross.
The results clearly demonstrate this phenomenon of the level repulsion between the sublevels with the same $j$ and the formation of a quasiresonance
state when the impurity charge exceeds a critical value. In such a case we observe a redistribution of profiles of radial distribution functions
with the same orbital momentum among lower Landau levels $n\le-1$.

\subsection{The Coulomb center in a magnetic field}
\label{Coulomb-center-magnetic}

Let us consider the electron states in gapped graphene with a single charged impurity in a magnetic field. The corresponding Hamiltonian could be obtained
from Eq.(\ref{Master-Hamiltonian}) by the standard substitution $\mathbf{p}\to \boldsymbol{\pi}=-i\hbar\boldsymbol{\nabla}+\frac{e}{c}\mathbf{A}$,
where $-e<0$ is the electron charge and the vector potential $\mathbf{A}=B/2(-y,\ x)$ in the symmetric gauge describes magnetic field perpendicular
to the plane of graphene. We regularize the Coulomb potential of an impurity by introducing a parameter $r_{0}$ of the order of the graphene lattice
spacing. Then the regularized interaction potential of the impurity with charge $Ze$ is given by
\begin{equation}
V\left(\mathbf{r}\right)=-\frac{Ze^2}{\kappa\sqrt{r^{2}+r_{0}^{2}}}.
\label{impurity-potential}
\end{equation}

It is convenient to introduce the magnetic length $l_{B}=\sqrt{{\hbar c}/{|eB|}}$ and the dimensionless quantity
$\zeta=Z e^{2}/(\kappa\hbar v_{\rm F})$ which characterizes the strength of the bare impurity. Since the total angular momentum is conserved, we use
the polar coordinates $(r,\ \phi)$ and seek eigenfunctions in the form (\ref{spinor}). Then the Dirac equation takes the form
\begin{equation}
\label{system-eq}
\left\{
\begin{array}{c}
a'-\frac{j+1/2}{r}a-\frac{r}{2l_{B}^{2}}a+\frac{E+\xi\Delta-V(r)}{\hbar v_{\rm F}}b=0,\\
b'+\frac{j-1/2}{r}b+\frac{r}{2l_{B}^{2}}b-\frac{E-\xi\Delta-V(r)}{\hbar v_{\rm F}}a=0.
\end{array}
\right.
\end{equation}

\begin{figure}[h]
\includegraphics[width=7.5cm]{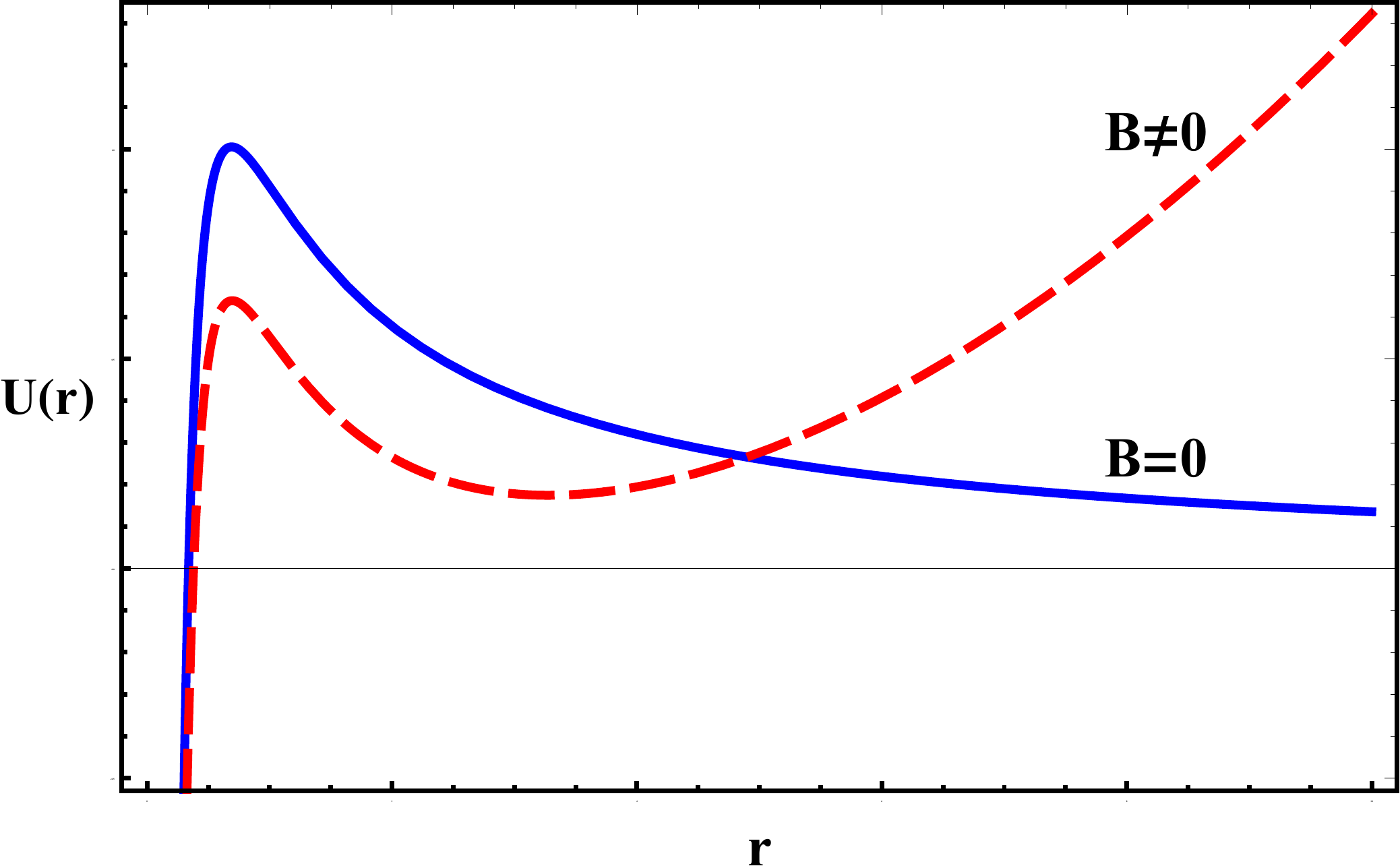}
\caption{The  potential $U(r)$ as a function of a distance from the Coulomb center
at zero and nonzero magnetic field, near the $K_{-}$ point for $E=-\Delta$  and $j=-1/2$. \label{pot}}
\end{figure}

Eliminating, for example the function $a(r)$, one can get a second order differential equation for $\chi(r)$ defined by the relation
\begin{equation}
[E-\xi\Delta-V(r)]^{1/2}\chi(r)=\frac{b(r)}{\sqrt{r}}.
\end{equation}
We obtain the following Schr\"{o}dinger-like equation:
\begin{equation}
-\chi''(r)+U(r)\chi(r)=\mathcal{E}\chi(r),
\label{effective-equation}
\end{equation}
where
\begin{equation}
\mathcal{E}=\frac{E^{2}-\Delta^{2}}{(\hbar v_{\rm F})^{2}},
\end{equation}
and the effective potential, $U=U_{1}+U_{2}$, reads
\begin{eqnarray}
U_{1}&=&\frac{V(2E-V)}{(\hbar v_{\rm F})^{2}}+\frac{j(j+1)}{r^{2}}+\frac{r^{2}}{4l_{B}^{4}}
+\frac{j-1/2}{l_{B}^2},\label{U-1}\\
U_{2}&=&\frac{1}{2}\left[\frac{V''}{E-\xi\Delta-V}+\frac{3}{2}\left(\frac{V'}
{E-\xi\Delta-V}\right)^{2}-\left(\frac{j}{r}+\frac{r}{2l_{B}^{2}}\right)\frac{2V'}
{(E-\xi\Delta-V)}\right]. \label{U-2}
\end{eqnarray}
We plot the effective potential $U(r)$ near the $K_{-}$ point ($\xi=-1$) for $E=-\Delta$  and $j=-1/2$ in
Fig.~\ref{pot}, where the energy barrier in the absence of magnetic field is clearly seen, which leads to the appearance of resonances
for sufficiently large charge. We note that the equations for spinor components $a(r)$ and $b(r)$ at the $K_{-}$ point can be
obtained from the equations in Sec.IIB1 at the $K_{+}$ point by interchanging $a \leftrightarrow b$ and changing $j \rightarrow -j$ since
two points are related by means of the time reversal transformation, $\Psi_{K_{-}}=\Theta\Psi_{K_{+}}$, introduced in Sec.\ref{section:monolayer-graphene}.
The presence of a magnetic field changes the asymptotic of the effective potential at infinity and, thus, forbids the occurrence of resonance
states. This feature distinguishes {\it qualitatively} the Coulomb center problem in a magnetic field from that at $B=0$.

\begin{figure}[ht]
	\centering
	\includegraphics[width=9.4cm]{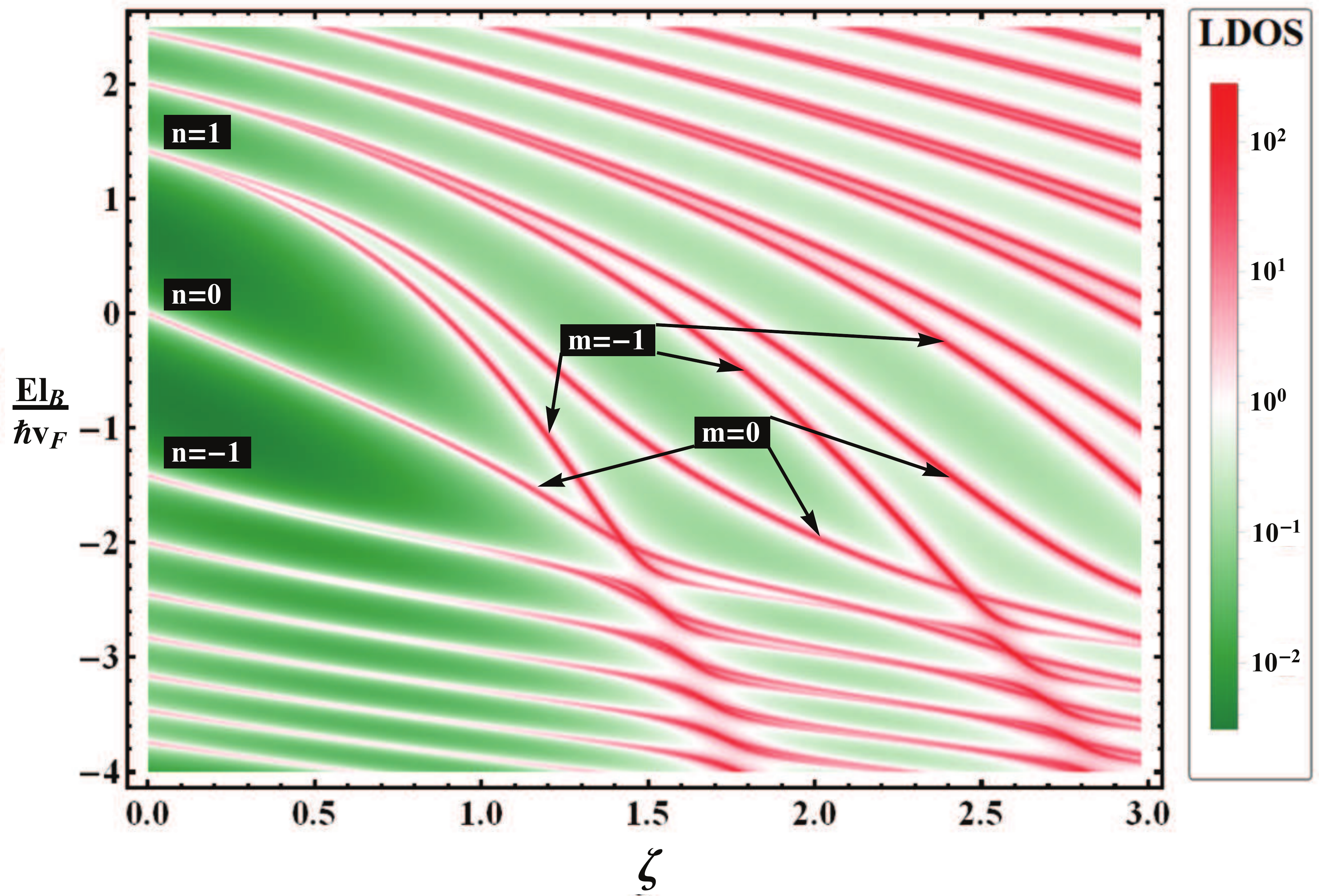}\hfill
    \includegraphics[width=8cm]{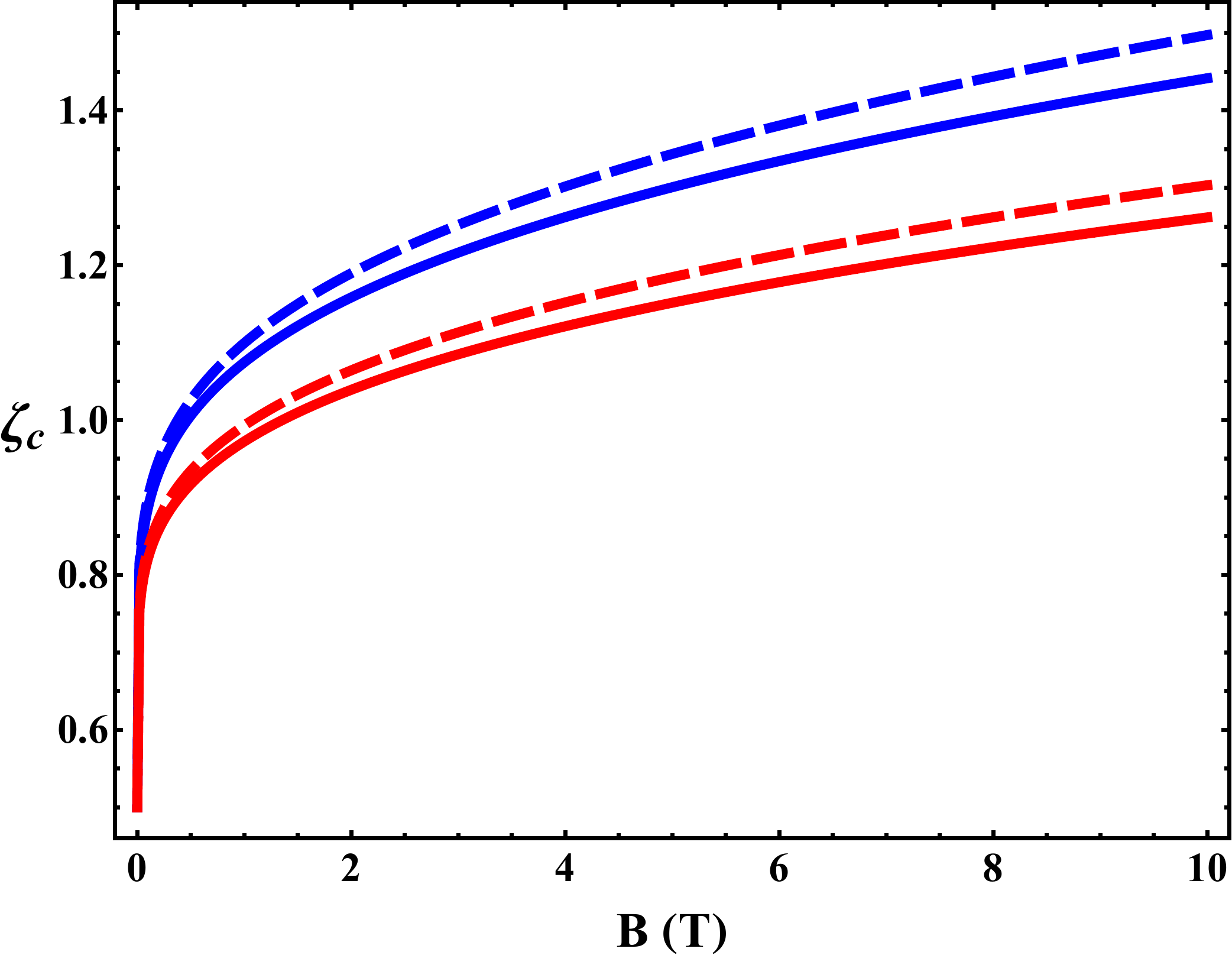}
	\caption{Left panel: The colormap of the LDOS at the impurity position ($r=0$) as a function of coupling $\zeta$ and energy $E$
in the magnetic field $B=10$~T. Black labels indicate the Landau level numbers $n$ and orbital quantum numbers $m$.		
Right panel: Critical coupling constant as a function of magnetic field for two types of regularized impurity potential:
	displaced impurity, Eq.~(\ref{impurity-potential}) -- blue lines; cut-off potential (\ref{potential-single-regularized}) -- red lines.
	Solid lines correspond to anticrossing of the $n=0$, $m=0$ and $n=-1$, $m=0$ levels, dashed lines correspond to anticrossing of the $n=1$,
	$m=-1$ and $n=-1$, $m=-1$ levels. In both panels the regularization parameter is chosen as $r_{0}=0.5$~nm.}
	\label{compare}
\end{figure}

Unfortunately, Eq.(\ref{effective-equation}) belongs to the class of equations with two regular and one irregular singular (at $r=\infty$) points,
and cannot be solved in terms of known special functions. In the regime $Z\alpha \to 0$, we can find it using perturbation theory. For $Z\alpha=0$,
the corresponding solutions are the well known Landau states degenerate in the total angular momentum $j$. The Coulomb potential of impurity removes
degeneracy in $j$ and the eigenenergies split into series of sublevels resulting in an $j$ dependent energy $E_{nj}$. The energy downshift is largest
for $E_{0j}$ and diminishes with increasing $-j$. For the $n=0$ level with $E_{0j}^{(0)}=E_{0}^{(0)}=\Delta$ the normalized wave function has the form
(at the $K_{-}$ point)
\begin{equation}
\Psi_M(r,\phi)=\frac{(-1)^{M}}{l_{B}\sqrt{2\pi M!}}e^{-r^2/4l_{B}^{2}} \left( \begin{array}{c} 0 \\
\left(\frac{r^{2}}{2l_{B}^{2}}\right)^{M/2}e^{-i M\phi}
\end{array} \right),
\label{unperturbed-solution}
\end{equation}
where $M=-(j+1/2)=0,\,1,\,2,...$ is the orbital quantum number.  Energy corrections $\varepsilon_{M}$ of perturbed states of the Landau level
$E_{0}^{(0)}=\Delta$ are found from the secular equation
\begin{equation}
|\varepsilon-V_{M_1M_2}|=0,
\end{equation}
where $V_{M_1M_2}$ is a matrix element of the potential on states (\ref{unperturbed-solution}). Since $V_{M_1M_2}$ is a diagonal matrix, we easily obtain
\begin{equation}
\varepsilon_M=V_{MM}=-\frac{Ze^{2}}{
M!2^{M}\kappa l_{B}}\int\limits_0^{\infty}\frac{d\rho\,\rho^{2M+1}\,e^{-\rho^2/2}}{\sqrt{\rho^2+\rho_0^2}}=-\frac{Ze^2}{2^{M+1}\kappa l_B}\rho_0^{2M+1}
\Psi\left(1+M,3/2+M;\rho_0^2/2\right)\,,
\label{energy-correction-regpotential}
\end{equation}
where $\Psi(a,c;z)$ is the confluent hypergeometric function, $\rho_0=r_0/l_B$. For small $Z\alpha\ll 1$ we can use the unregularized
Coulomb potential, then setting $\rho_0=0$ in Eq.(\ref{energy-correction-regpotential}) we get
\begin{equation}
\varepsilon_M=-\frac{Ze^{2}\Gamma(M+\frac{1}{2})}{\kappa l_{B}\sqrt{2}\Gamma(M+1)}.
\label{energy-correction}
\end{equation}
 Thus at large $M$ the energy levels accumulate near the value $E=\Delta$,
\begin{equation}
E_{0M}=E^{(0)}_{0}+\varepsilon_{M}\simeq \Delta-\frac{Ze^{2}}{\kappa l_{B}\sqrt{2M}}.
\end{equation}
The largest correction by modulus $\varepsilon_{0}=-Z\alpha\hbar v_{\rm F}\sqrt{\pi}/l_B\sqrt{2}$ is for the state with $M=0$.
Naturally, in the lowest order of perturbation theory, the energy linearly decreases with the increase of the impurity charge.
The numerical solution of Eq.~(\ref{effective-equation}) shows that this behavior changes when the charge exceeds a certain
critical value and after that the level repulsion occurs (see Fig.~\ref{compare}).

For finite $\Delta$ one can define the critical charge by the condition $E=E_{0}^{(0)}+\varepsilon_{0}=-\Delta$ when the lowest energy
empty level descending from the upper continuum crosses the energy level of a filled state. In the regime of small coupling,
$Z\alpha\ll1$ and $\Delta l_B\ll1$, this gives
\begin{equation}
Z_c\alpha=\frac{2\sqrt{2}\Delta l_B}{\sqrt{\pi}\hbar v_{\rm F}}\,.
\label{critical-charge}
\end{equation}
Clearly, this critical charge tends to zero as $\Delta \to 0$, while the state with $M=0$ of the zero Landau level
moves below zero energy for any small impurity charge (its energy is $\varepsilon_{0}=-Z\alpha\hbar v_{\rm F}\sqrt{\pi}/l_B\sqrt{2}$).
The states connected with the zero Landau level play an important role in the many-body problem, e.g., in the formation
of the excitonic condensate and gap generation for quasiparticles \cite{Gorbar2002,Semenoff1999} due to the magnetic catalysis.
In the case of a charged impurity in a magnetic field, the negative energy states are filled and it is physically more sensible to
connect the critical charge with the anticrossing of Landau levels in the negative energy region (see the discussion below).

Although, in view of the magnetic catalysis \cite{Gusynin1994}, a nonzero gap is always generated in graphene in a perpendicular magnetic field \cite{Khveshchenko2001,Gorbar2002,Gorbar2003,Khveshchenko2004}, this gap is rather small for realistic magnetic fields. Therefore, it makes sense to neglect it and see how levels with the same $j$ evolve. Let us solve Eqs.~(\ref{system-eq}) numerically by using the shooting method.
In order to utilize this method, one should determine the appropriate asymptote of the solution at
$r \to 0$ for $|V(r)|\approx |V(0)|=\frac{Z e^{2}}{\kappa r_{0}}\gg |E|$. At the $K_{+}$ point it is convenient to introduce the orbital quantum
number $m=j-1/2=-M-1$. Then, for $m \ge 0$ ($j\ge 1/2$) the upper component of (\ref{spinor}) dominates and the leading behavior
is $a(r)\sim r^{m+1}$, for $m < 0$ ($j<1/2$) the lower component dominates with $b(r)\sim r^{-m}$ (see Ref.~[\onlinecite{Sobol2016}]).

The numerical integration of Eq.~(\ref{system-eq}) proceeds as follows. We take a ``shot'' from $r=0$
at a fixed value of energy solving the differential equations with the correct initial conditions and check the behavior of the wave functions
at $r\to\infty$. The latter may tend to $+\infty$ for some values of energy or to $-\infty$ for other values. A physical solution is the solution
for which the exponentially growing behavior of the absolute value is absent. We find the corresponding value of the energy of this solution by
using the method of bisections. In all numerical calculations, we use $r_{0}=0.05 l_{B}$.

The magnetic field modifies the energy spectrum of electrons in the Coulomb field of the charged impurity making all continuum
states discrete and provides an effective scale given by the magnetic length. On the other hand, the charged impurity removes
the orbital degeneracy of Landau levels transforming the latter into bandlike structures. Figure~\ref{compare} (left panel) shows the colormap of
the LDOS at the impurity position as a function of coupling $\zeta$ and dimensionless energy ${E l_{B}}/{\hbar v_{\rm F}}$ in the magnetic field
$B=10$~T. Red lines correspond to Landau levels split into sublevels with different orbital numbers.
At the beginning, the curves decrease linearly in accordance with Eq.~(\ref{energy-correction}). As the charge of impurity increases, the
curves, which correspond to the $n=0,\,1$ Landau levels come close to lower curves, which form a ``quasicontinuum''. In the absence of
magnetic field, with further increase of the charge of impurity the corresponding bound state would dive into the lower continuum producing a
resonance.

According to Fig.~\ref{compare} (left panel), the situation is {\it qualitatively} different in the case of zero gap when a
magnetic field is present as the  curves with the same orbital number $m$ never cross each other. Instead, typical level repulsions are
realized (the well-known avoided crossing theorem \cite{Wigner1929} forbids a level crossing for two states with the same symmetry).
We clearly see the repulsion between the levels
$n=1,\ m=-1$ and $n=-1,\ m=-1$, as well as between the levels $n=0,\ m=0$ and $n=-1,\ m=0$. States with different quantum numbers
$m$ simply cross each other without repulsion. The situation is similar to that of a quantum electrodynamical system of finite size
\cite{Muller1972b,Greiner1985}. In Fig.~\ref{compare} (right panel) we plot the dependence of the critical charge on a magnetic field $B$
defined as the anticrossing points of the Landau levels $n=0$ and $n=-1$ with $m=0$, as well as of the levels $n=1$ and $n=-1$ with
$m=-1$, in the negative energy region. The corresponding dependence at zero gap $\Delta=0$ can be very accurately fit by
the following function:
\begin{equation}
\zeta_{cr}=\frac{1}{2}+\frac{A}{{\rm ln}^{2}(c r_{0}/l_{B})},
\end{equation}
where  $A$ and $c$ are fitting parameters, which can be determined numerically. For the displacement regularization, these parameters are
$A_{1}=15.75$, $c_{1}=0.305$ and, for the cut-off regularization, they are $A_{2}=13.94$, $c_{2}=0.25$. The increasing magnetic field
strength causes the anticrossings to appear at higher charge $\zeta$ in accordance with the observation in Ref.\cite{Moldovan2017}.
The dependence of the critical charge $\zeta_{\rm cr}$ on a magnetic field is similar to its dependence on a gap in the absence of the field
(see Eq.~(\ref{crit-charge-on-r0})).

\begin{figure}[ht]
	\centering
	\includegraphics[scale=0.25]{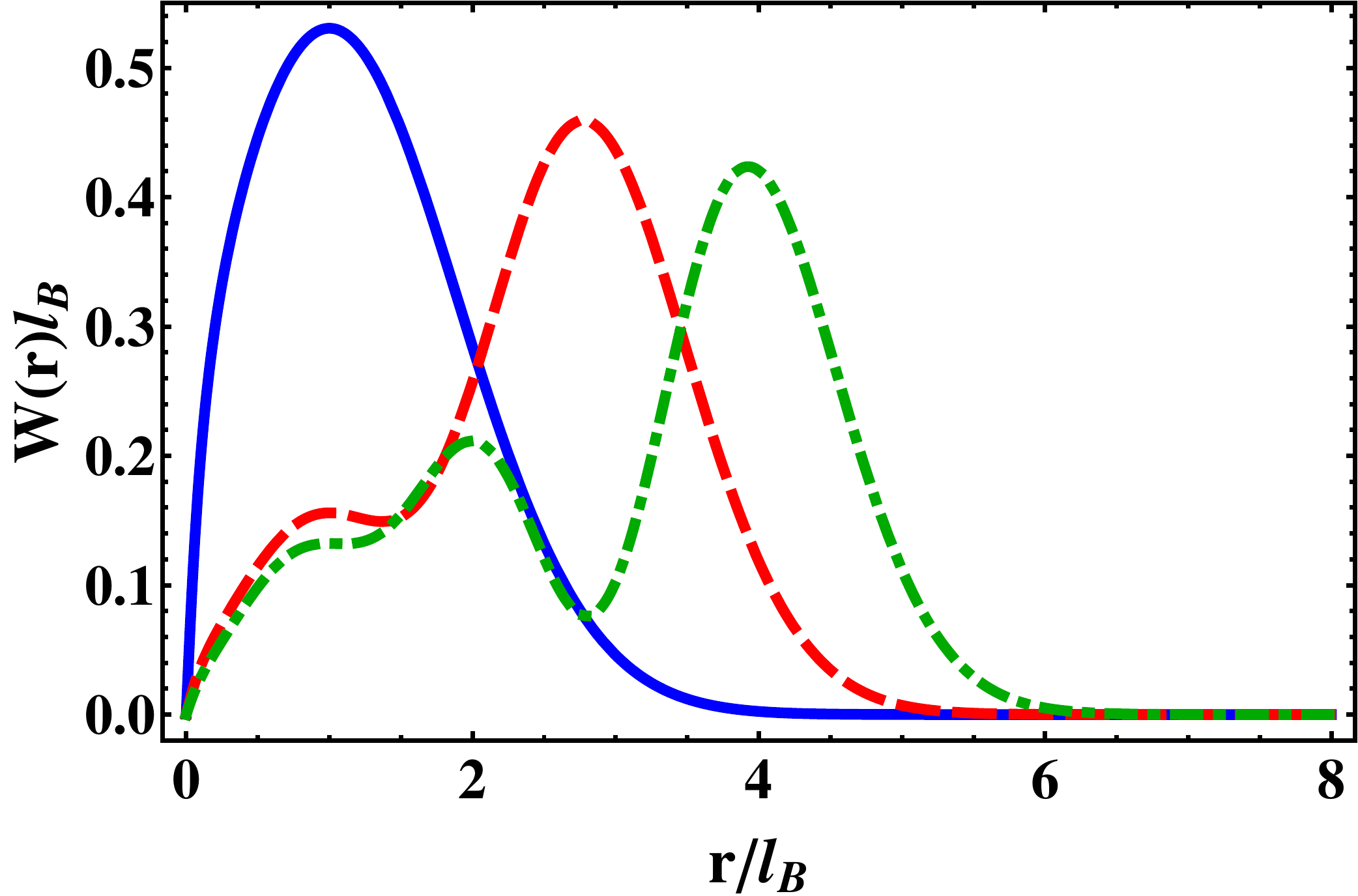}
	\includegraphics[scale=0.25]{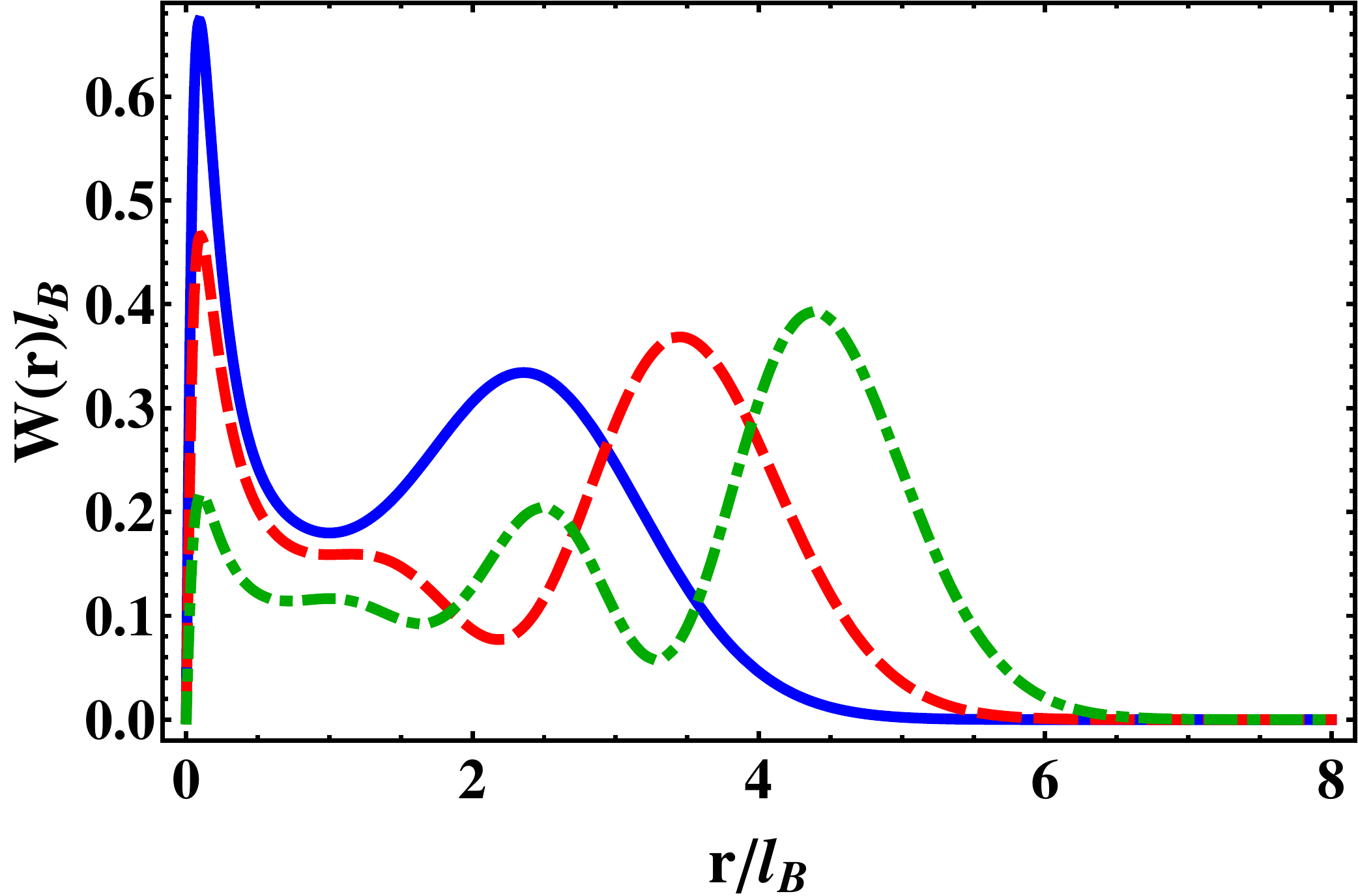}
	\includegraphics[scale=0.25]{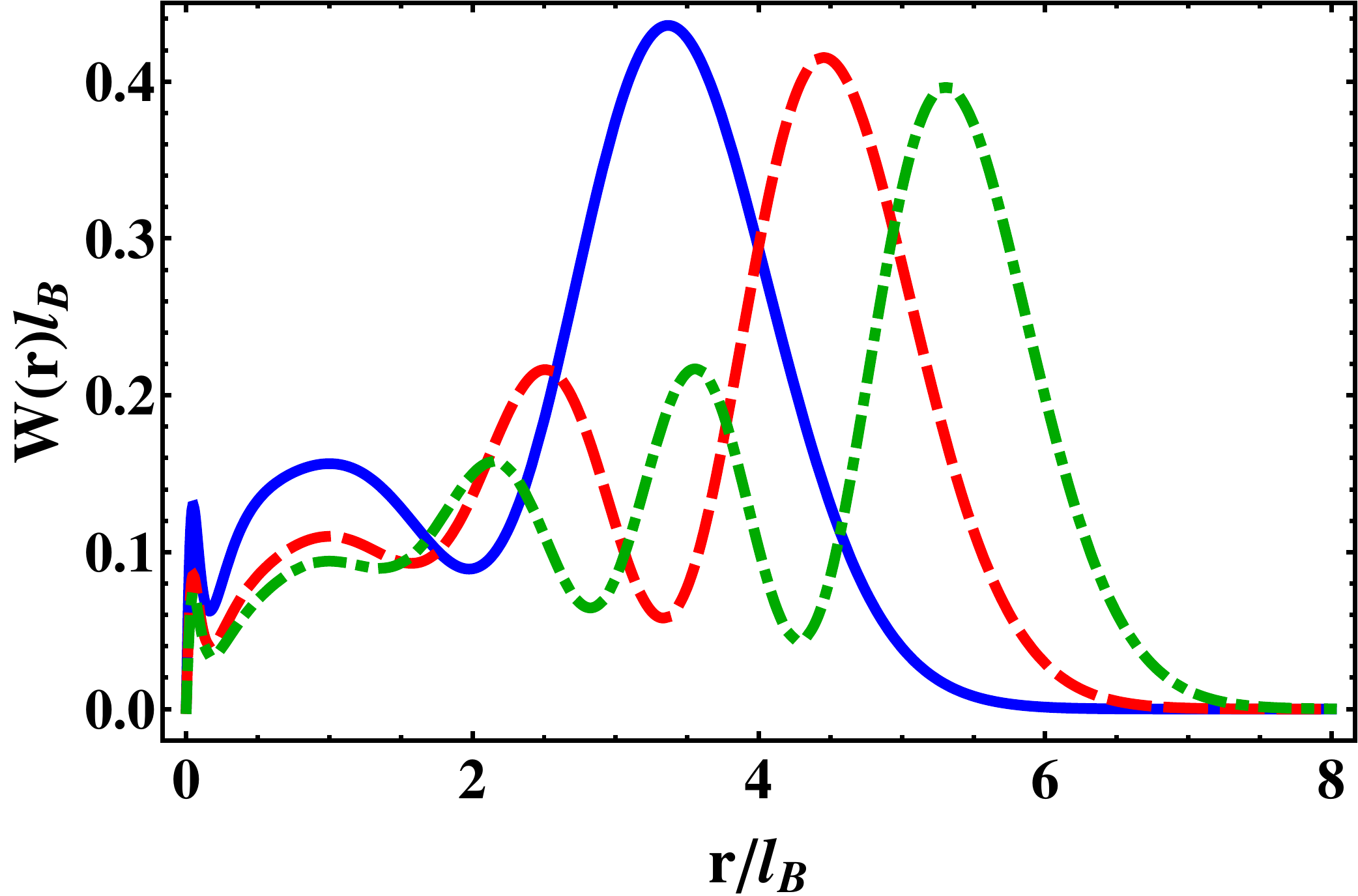}
	\caption{The radial functions of the electron density of the $m=0$ state for the Landau
		levels $n=0$ (blue solid lines), $n=-1$ (red dashed lines), and $n=-2$ (green dot-dashed lines) and
		three different values of the impurity charge: $\zeta=0.7$ (left panel), $\zeta=1.3$
		(middle panel), $\zeta=1.9$ (right panel).}
	\label{WF_mag}
\end{figure}

Figure~\ref{WF_mag} shows the radial distribution function $W(r)=2\pi r|\Psi_{nm}|^2$ for the $m=0$ and $n=0,\, -1,\, -2$ states for the three
values of the impurity charge $\zeta=0.7$, $1.3$, and $1.9$. The second value corresponds to the states in the vicinity of the avoided crossing.
For a small charge of the impurity (the leftmost panel), the electron density is weakly affected by the impurity and the radial
distribution functions of the above mentioned states have one, two, and three maxima, respectively. As the impurity charge increases, all leftmost
maxima in $W(r)$ move to the impurity position $r=0$ and attain their maximal values at $\zeta\approx 1.3$ (the middle panel). In addition, a new
maximum appears on the blue solid curve (as well as additional maxima on the other two curves), and the radial distribution function of the $n=0$
level begins to look qualitatively like the radial distribution function of the $n=-1$ level with two maxima.

Further, the middle panel implies that the peak in the radial distribution function of the $n=0$ level near the impurity is redistributed among
the $m=0$ states of the $n=-1,-2, ...$ Landau levels. Obviously, this is an analog of the phenomenon of the diving into continuum for a supercritical
charge in the absence of a magnetic field. In the latter case, the lowest bound state dives into the lower continuum producing a resonance
whose  wave function can be considered as redistributed over the lower continuum states with energies of the order of the resonance width $\gamma$.
All wave functions from this region have an additional sharp peak near the origin. As we see, when magnetic field is present, there is a similar
redistribution of the profiles of radial distribution functions near the impurity (note that as the impurity charge increases, the ``redistribution''
region shifts down to the lower Landau levels). According to the rightmost panel in Fig.~\ref{WF_mag}, the blue curve representing the
electron density is now similar to the red dashed curve in the leftmost panel and the red dashed curve is similar to the green dot-dashed
curve in the leftmost panel.

So far we did not take into account the screening of a charged impurity due the polarization effects in  graphene to which we turn
our attention in the next subsection.

\subsection{Tuning the screening of charged impurity with chemical potential}
\label{magnetic-screening}

Experimentally, as shown in Sec.~\ref{section-center-magnetic}, the strength of a charged impurity and splitting of Landau sublevels with
different orbital momenta in a magnetic field can be very effectively tuned by a gate voltage \cite{Luican-Mayer2014}. In this subsection, we
theoretically study this phenomenon by taking into account the polarization in a magnetic field which is controlled by the chemical
potential due to gate voltage.

It is natural to attribute the variation in the strength of the impurity potential to the screening properties of the 2D electron
system. To describe this effect theoretically, we follow Ref.~[\onlinecite{Sobol2016}] and use the polarization function
calculated in the absence of a charged impurity. Then the corresponding Poisson equation, which defines the screened impurity
potential, reads
\begin{equation}
\label{Poisson_eq}
\sqrt{-\Delta_{2D}}V_{tot}(\mathbf{x})=-\frac{2\pi Z e^{2}}{\kappa}\delta^{(2)}(\mathbf{x})
-\frac{2\pi e^{2}}{\kappa}\int\!\!d^{2}\mathbf{y}
\Pi(\mathbf{x}-\mathbf{y};\mu)V_{tot}(\mathbf{y}),
\end{equation}
where $\Pi(\mathbf{x}-\mathbf{y};\mu)$ is the static polarization function calculated by using the wave functions of free electrons
in a magnetic field. Notice the presence of the pseudodifferential operator $\sqrt{-\Delta_{2D}}$ in the equation above, which is necessary
to correctly describe the Coulomb interaction in a dimensionally reduced electrodynamic
system \cite{Kovner1990,Dorey1992,Marino1993,Gorbar2001}.

\begin{figure}[ht]
	\centering
	\includegraphics[scale=0.77]{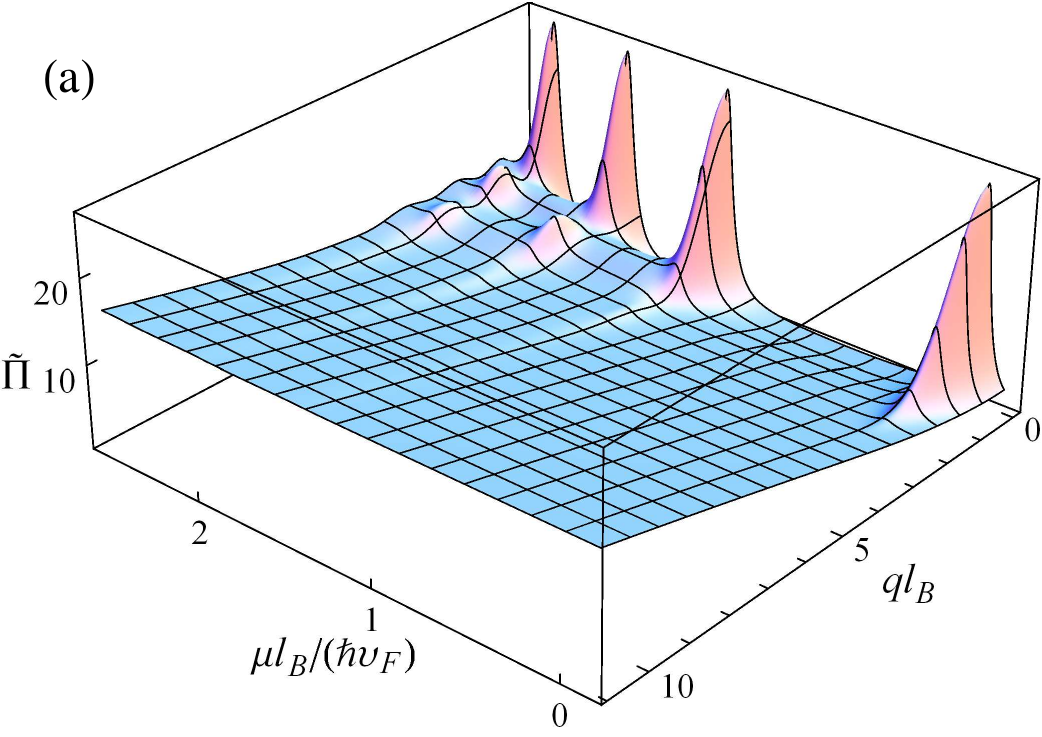}\qquad
	\includegraphics[scale=0.65]{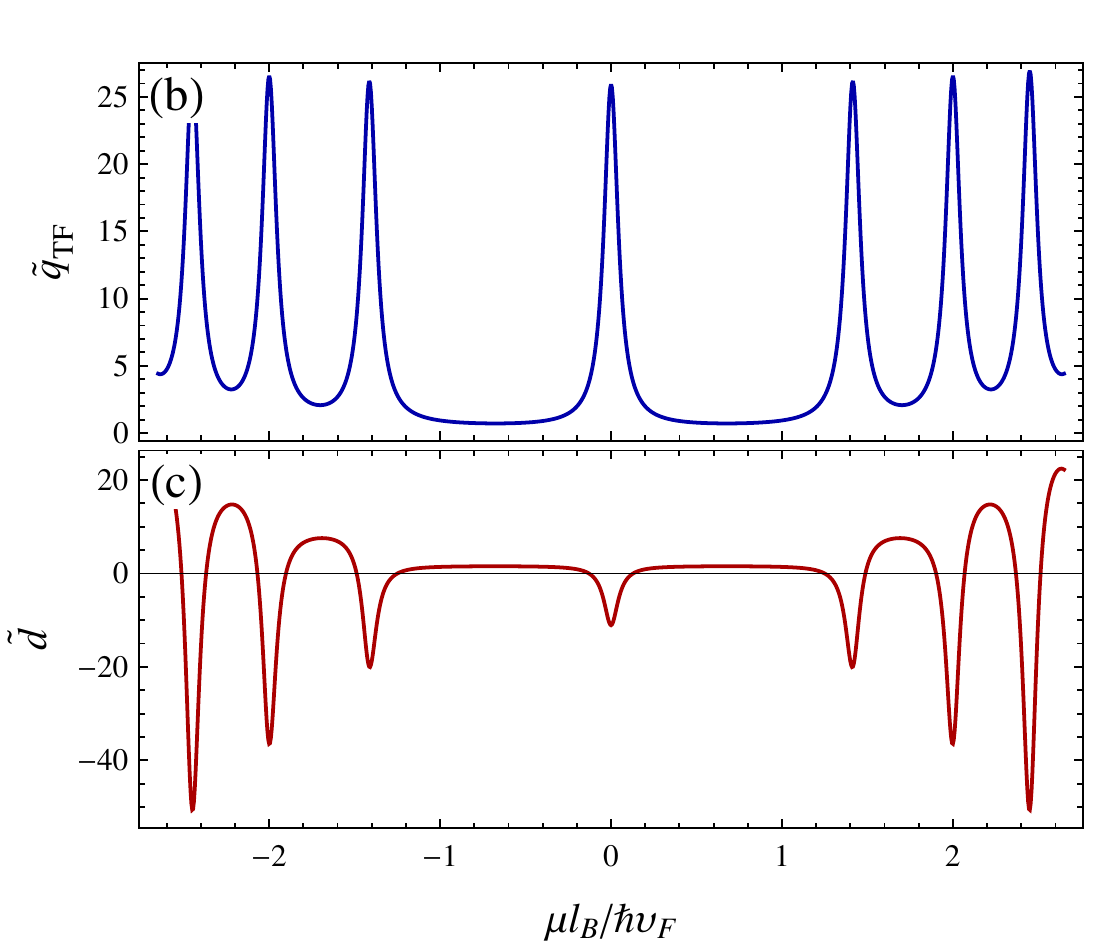}
	\caption{The dimensionless polarization function  $\tilde\Pi=(4\pi\hbar v_{\rm F}l_{B}/N_{f})\Pi(q;\mu)$
		as a function of the chemical potential and the wave vector (left panel) and the two coefficients
		$\tilde q_{\rm TF}=(2\kappa\hbar v_{\rm F}l_B/e^2N_f)q_{\rm TF}$ and $\tilde d=(2\kappa\hbar v_{\rm F}/e^2N_fl_B)d$ of its
		expansion~(\ref{pol_long_wavelength}) at small wave vectors (right panel).}
	\label{pol-function}
\end{figure}

\begin{figure}[ht]
	\centering
	\includegraphics[scale=0.42]{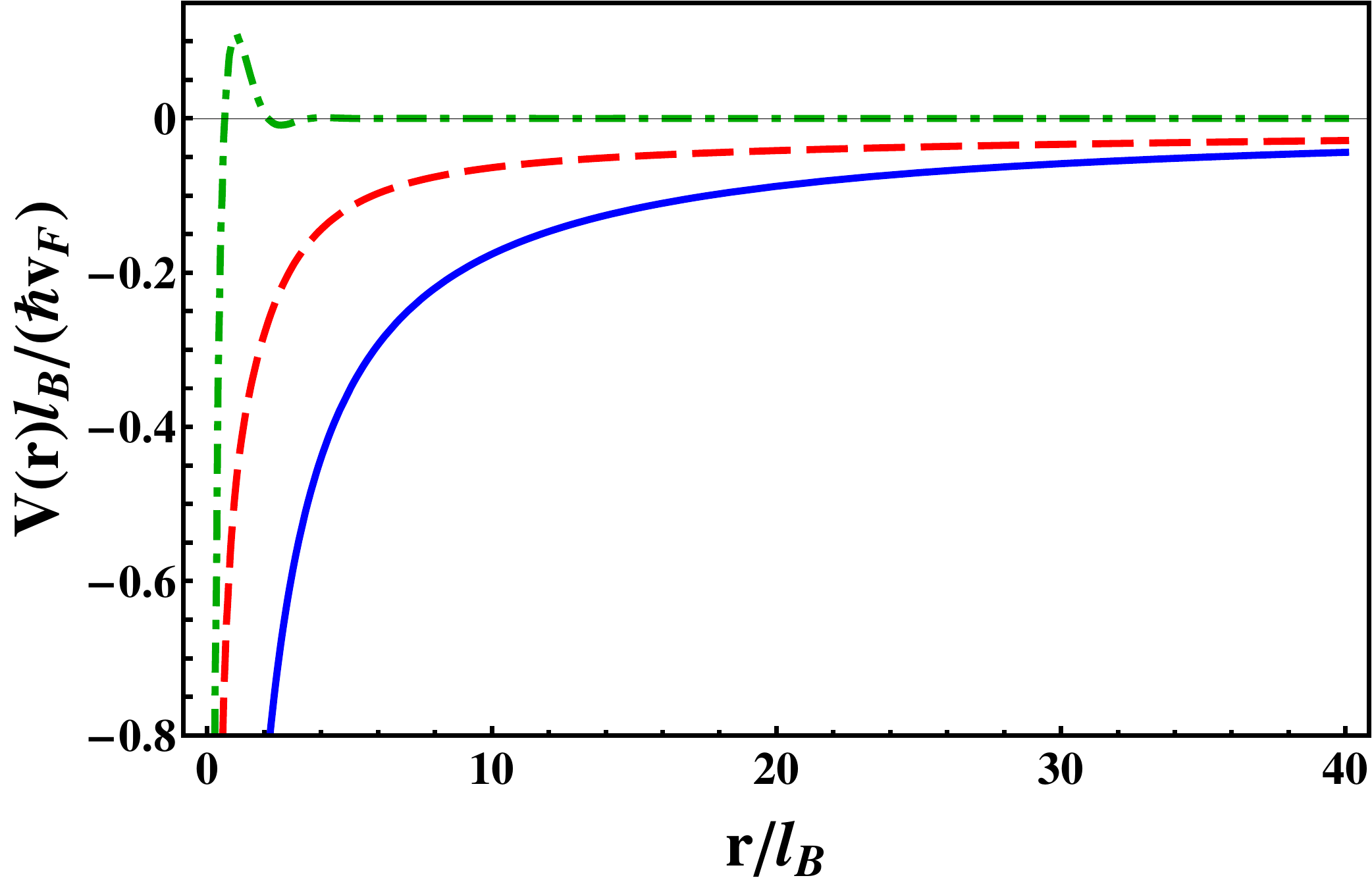}
	\caption{The unscreened regularized Coulomb potential (blue solid line) and the screened potential of the impurity as a function of $r/l_{B}$ in the cases where the chemical potential is situated between Landau levels (red dashed line) and lies inside the zeroth Landau level (green dash-dotted line).}
	\label{pot_scr}
\end{figure}

Since Eq.(\ref{Poisson_eq}) is algebraic in momentum space
\begin{equation}
\label{alg_eq}
\left(|\mathbf{q}|+\frac{2\pi e^{2}}{\kappa}\Pi(\mathbf{q};\mu)\right)V_{tot}(\mathbf{q})=-\frac{2\pi Z e^{2}}{\kappa},
\end{equation}
we easily find the screened impurity potential in coordinate space
\begin{equation}
\label{potential_screened}
V_{tot}(\mathbf{x})= -\frac{Z e^{2}}{\kappa}\int\frac{d^2q}{2\pi}\frac{\exp(i\mathbf{q}\mathbf{r})}
{|\mathbf{q}|+\frac{2\pi e^{2}}{\kappa}\Pi(\mathbf{q};\mu)}=
-\frac{Z e^{2}}{\kappa}\int\limits_{0}^{+\infty}\!\!dq
\frac{q\ J_{0}(q|\mathbf{x}|)}{q+\frac{2\pi e^{2}}{\kappa}\Pi(q;\mu)},
\end{equation}
where $q=|\mathbf{q}|$. The static polarization function at zero temperature has the form \cite{Pyatkovskiy2011},
\begin{equation}
\label{polarization}
\Pi(q;\mu)=\frac{N_{f}}{4\pi l_{B}^{2}}\left\{\sum\limits_{n=0}^{n_{c}}\sum\limits_{\lambda=\pm}Q^{\lambda\lambda}_{nn}
\left(\frac{q^2l^2_B}2\right)\delta_{\Gamma}(\mu-\lambda M_{n})-\underset{\lambda n\neq\lambda' n'}{\sum\limits_{n,n'=0}^{n_{c}}\sum
	\limits_{\lambda,\lambda'=\pm}}Q^{\lambda\lambda'}_{nn'}
\left(\frac{q^2l^2_B}2\right)\frac{\theta_{\Gamma}(\mu-\lambda M_{n})
	-\theta_{\Gamma}(\mu-\lambda' M_{n'})}{\lambda M_{n}-\lambda' M_{n'}}\right\},
\end{equation}
where $M_{n}=\frac{\hbar v_{\rm F}}{l_{B}}\sqrt{2n}$ are the Landau level energies, and we introduced the ultraviolet cutoff $n_c$ because
of the divergence of the sum over the Landau levels. Since the bandwidth is finite in graphene, $n_c$ is estimated as $n_c=10^4/B[T]$
\cite{Roldan2009,Roldan2010}. As in experiment \cite{Luican-Mayer2014}, we consider the system of two superposed graphene layers twisted
away from Bernal stacking by a large angle. This does not affect the spectrum of single-layer graphene but results in an additional twofold
layer degeneracy: the factor $N_{f}=2_{s}2_{l}=4$ takes into account spin degeneracy and the presence of the second graphene layer.
In experiment, this setup makes possible to diminish the random potential fluctuations due to substrate imperfections. The smeared delta function $\delta_{\Gamma}(x)=\frac{\Gamma}{\pi}\frac{1}{x^{2}+\Gamma^{2}}$ and the step function
$\theta_{\Gamma}(x)=\frac{1}{2}+\frac{1}{\pi}\arctan\left(\frac{x}{\Gamma}\right)$ account for the finite width of Landau levels, and the functions $Q^{\lambda\lambda'}_{nn'}(y)$ are defined as
\begin{equation}
Q^{\lambda\lambda'}_{nn'}(y)=e^{-y}y^{|n-n'|}\left(\sqrt{\frac{(1+\lambda\lambda'\delta_{0,n_{>}})
		n_{<}!}{n_{>}!}}L^{|n-n'|}_{n_{<}}(y)+\lambda\lambda'(1-\delta_{0,n_{<}})\sqrt{\frac{(n_{<}-1)!}
	{(n_{>}-1)!}}L^{|n-n'|}_{n_{<}-1}(y)\right)^{2},
\label{Qlambda-function}
\end{equation}
where $n_<=\min(n,n')$, $n_>=\max(n,n')$ and $L_n^m(y)$ are the generalized Laguerre polynomials. The first term in Eq.~(\ref{polarization})
describes the contribution from the intralevel transitions while  the second term represents the contribution from the interlevel transitions.
For a small width of Landau levels the first term looks like a sequence of delta functions  and  contributes only when the chemical potential
lies inside Landau levels. At small wave vectors ($q\ll l_B^{-1}$) the polarization function~(\ref{polarization}) behaves as \cite{Pyatkovskiy2011}
\begin{equation}
\Pi(q;\mu)\simeq\frac{\kappa}{2\pi e^2}(q_\text{TF}+d q^2),
\label{pol_long_wavelength}
\end{equation}
where
\begin{equation}
q_\text{TF}=\frac{e^2N_f}{\kappa l^2}\sum_{n=0}^{n_c}\sum_{\lambda=\pm}(2-\delta_{0n})\delta_\Gamma(\mu-\lambda M_n)
\label{q_TF}
\end{equation}
is the Thomas-Fermi wave vector which determines the strength of the long-wavelength screening, and parameter $d$ is given by
\begin{equation}
d=-\frac{e^2N_f}{2\kappa}\sum_{n=0}^{n_c}\sum_{\lambda=\pm}(4n+\delta_{0n})
\delta_\Gamma(\mu-\lambda M_n)-\frac{e^2N_fl}{2\sqrt2\kappa\hbar v_{\rm F}}\sum_{n=0}^{n_c-1}
\sum_{\lambda,\lambda'=\pm}\frac{\theta_\Gamma(\mu-\lambda M_{n+1})-\theta_\Gamma(\mu-\lambda' M_n)}
{(\lambda\sqrt{n+1}-\lambda'\sqrt n)^3}.
\label{a_coeff}
\end{equation}
In fact, the static polarization function $\Pi(0,\mu)$ is proportional to the density of states which at finite scattering rate has
the form of a series of broaden Landau levels \cite{Pyatkovskiy2011}. Figure~\ref{pol-function} illustrates the dependence of the static
polarization function~(\ref{polarization}) and its two leading long-wavelength terms~(\ref{q_TF}) and~(\ref{a_coeff}) on the chemical potential.
We plot for comparison the unscreened potential and the screened potential (\ref{potential_screened}) of the impurity in Fig.~\ref{pot_scr}.
Let us consider the case where the chemical potential is situated between Landau levels. Then the Thomas-Fermi wave vector~(\ref{q_TF}) is close
to zero [Figs.~\ref{pol-function}(a) and~\ref{pol-function}(b)] and the Coulomb potential of the impurity is weakly screened, although even in
this case graphene contributes to the total dielectric function at large and intermediate momenta, which effectively diminishes the charge of
the impurity and the screened potential. Indeed, while the screened potential tends to its bare value at $r\to\infty$, it is weakened for small
and intermediate distances (see the red dashed line in Fig.~\ref{pot_scr}). On the other hand, when the chemical potential lies inside any given
Landau level, the screening is much more effective due to large $q_\text{TF}$ (see the green dash-dotted line in Fig.~\ref{pot_scr}) providing
an excellent means of controlling the effective charge of impurity by the gate voltage which is directly related to the chemical potential
$\mu$. Moreover, the coefficient $d$ in Eq.~(\ref{a_coeff}) in this case is negative [see Fig.~\ref{pol-function}(c)] which means that
$\Pi(q;\mu)$ has a nonmonotonic momentum dependence with a peak at $q=0$. This behavior of the polarization function leads to the oscillations
of the screened potential (green dash-dotted line in Fig.~\ref{pot_scr}) with the sign change (i.e., the overscreening of the Coulomb potential)
at intermediate distances of the order of several magnetic lengths.

In Ref.~[\onlinecite{Sobol2016}] the backreaction of the charged impurity on the polarization properties was also taken into
account. Although the qualitative picture of screening is the same, it was shown that due to the downshift of the energy levels, the
polarization function no longer remains symmetric with respect to the exchange $\mu\leftrightarrow-\mu$. These features of a charged
impurity in graphene in the magnetic field are clearly observed in the recent experiments \cite{Luican-Mayer2014,Mao2016}.

It should be noted that the approximation of noninteracting electrons may become invalid when the chemical potential lies inside the Landau
level. Indeed the electron-electron interactions could lead in sufficiently clean graphene specimen to such interesting phenomena as the
fractional quantum Hall effect. Then, the chemical potential cannot be tuned continuously and instead jumps from one plateau to
another. Although our analysis becomes inapplicable in the fractional quantum Hall regime, the experimental results in \cite{Luican-Mayer2014}
show that the conclusion about the maximal screening remains unchanged.

\subsection{Screening charged impurities and lifting the orbital degeneracy in graphene by populating Landau levels}
\label{section-center-magnetic}

Charged impurities in undoped gapless graphene produce a spatially localized signature in the density of states (DOS) which is readily observed
with scanning tunneling microscope and spectroscopy (STM+STS)\cite{Mizes1989}. This effect is especially important in the presence of a magnetic
field when the quantization of the 2D electronic spectrum into highly degenerate Landau levels (LL) gives rise to the quantum Hall effect.
In this regime charged impurities are expected to lift the orbital degeneracy causing each LL in their immediate vicinity to split into discrete
sublevels \cite{Gamayun2011,Zhang2012}.

By making use of high quality gated graphene devices in a magnetic field, it was shown in Ref.\cite{Luican-Mayer2014} that the strength of a
charge impurity, as measured by its effect on the electron spectrum, can be effectively controlled by tuning the LL occupation with a back gate
voltage. The LL spectra were obtained by measuring the bias voltage dependence of the differential tunneling conductance, $dI/dV$, which is
proportional to the DOS, $D(E,\mathbf{r})$, at the tip position $\mathbf{r}$. Here $V=(E-E_F)/e$ is the bias voltage and $E$ is the energy
measured relative to the Fermi level, $E_F$. For almost empty LLs, the impurity is screened and essentially invisible whereas at full LL
occupancy screening is very weak and the potential due to the impurity attains maximum strength. The underlying discrete
quantum-mechanical spectrum arising from lifting the orbital degeneracy was experimentally resolved in the unscreened regime.

To explore the influence of the impurity on the LLs the spatial evolution of spectra along a trajectory traversing it for a series of gate
voltages was studied. For certain gate voltages  the spectra become significantly distorted close to the
impurity, with the $n=0$ level (and to a lesser extent higher order levels) shifting downwards toward negative energies. The downshift indicates
an attractive potential produced by a positively charged impurity. Its strength, as measured by the distortion of the $n=0$ LL, reveals a
surprisingly strong dependence on LL filling. In the range of gate voltages $-15V < V_g < 9 V$ corresponding to filling the $n=0$ LL
the distortion grows monotonically with filling.  At small filling the distortion is almost absent indicating that the impurity is effectively
screened attaining its maximum value close to full occupancy. At full occupancy the $n=0$ level shifts by as much as $\approx 0.1 eV$
indicating that the effect would survive at room temperature. The spectral distortion is only present in the immediate vicinity of the impurity.
Farther away no distortion is observed for all studied carrier densities.

The variation of the impurity strength with filling is related to the screening properties of the electron system. They were studied
theoretically in Sec.~\ref{magnetic-screening}. For a positively (negatively) charged impurity and almost empty (full) LLs, unoccupied states
necessary for virtual electron transitions are readily available in the vicinity of the impurity, resulting in substantial screening. By
contrast for almost filled (empty) LLs, unoccupied states are scarce, which renders local screening inefficient.

The downshift is the largest for the $n=0$ and $j=-1/2$ state and diminishes with increasing $|j|$ and/or $n$.
The local tunneling DOS was calculated in Ref.\cite{Luican-Mayer2014} assuming a finite linewidth $\Gamma$
\begin{equation}
\label{eq3}
D(E,\mathbf{r})= 4\sum_{nj} \delta_\Gamma(E-E_{nj}) \psi^\dagger_{nj}(\mathbf{r}) \psi_{nj}(\mathbf{r}),
\end{equation}
where $\delta_\Gamma(E-E_{nj})=\Gamma/[\pi((E-E_{nj})^2+\Gamma^2)]$ represents a broadened LL. The peak intensity is determined by the
probability density $ \psi^\dagger_{nj}(\mathbf{r}) \psi_{nj}(\mathbf{r})$ and is position dependent.
If $\Gamma < \Delta{E_{nj}}$ ($\Delta E_{nj}$ defines spacing between adjacent levels), the discreteness of the spectrum
is resolved, but for $\Gamma \ge \Delta{E_{nj}}$ the peaks of adjacent states overlap and merge into a continuous band.

Thus, even if the spectrum is discrete, but the resolution insufficient or if impurities are too close to each other, the measured $D(E,\mathbf{r})$
will still display ``bent'' LLs, whose energies seemingly adjust to the local potential. In particular, upon approaching the impurity the $n=0$
LL splits into well resolved discrete peaks connected with specific orbital states. As it could be seen from the experimental plots
in Ref.~[\onlinecite{Luican-Mayer2014}], the states  $\psi_{0j}(\mathbf{r})$ with $j=-1/2,\,-3/2,\,-5/2$ are well resolved close to the impurity but
higher order states are less affected and their contributions to $D(E,\mathbf{r})$ merge into a continuous line. Similarly, the discreteness of the
spectrum is not resolved for $n\neq 0$ LLs that is consistent with the weaker impurity effect at larger distances. For partial filling
($ V_{g} = -5\,{\rm V}, 0\,{\rm V} $) as screening becomes more efficient and orbital splitting is no longer observed, the unresolved sublevels merge into continuous
lines of ``bent'' Landau levels. The ability to tune the strength of the impurity in-situ demonstrated in Ref.\cite{Luican-Mayer2014} opens the
door to exploring Coulomb criticality and to investigate a hitherto inaccessible regime of criticality in the presence of a magnetic field \cite{Gamayun2011,Zhang2012}.

\begin{figure}[hpt]
	\centering
	\includegraphics[scale=0.35]{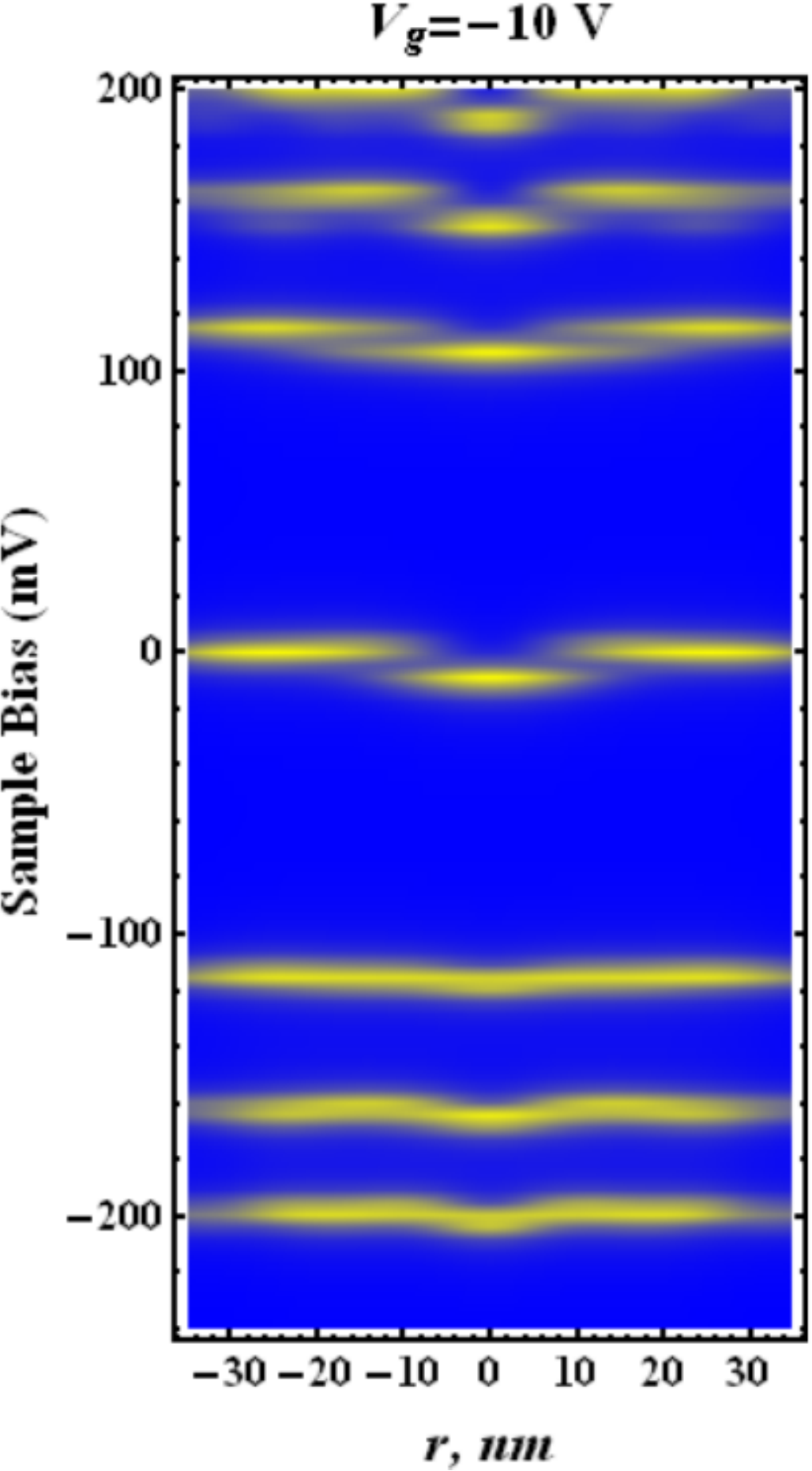}
	\includegraphics[scale=0.35]{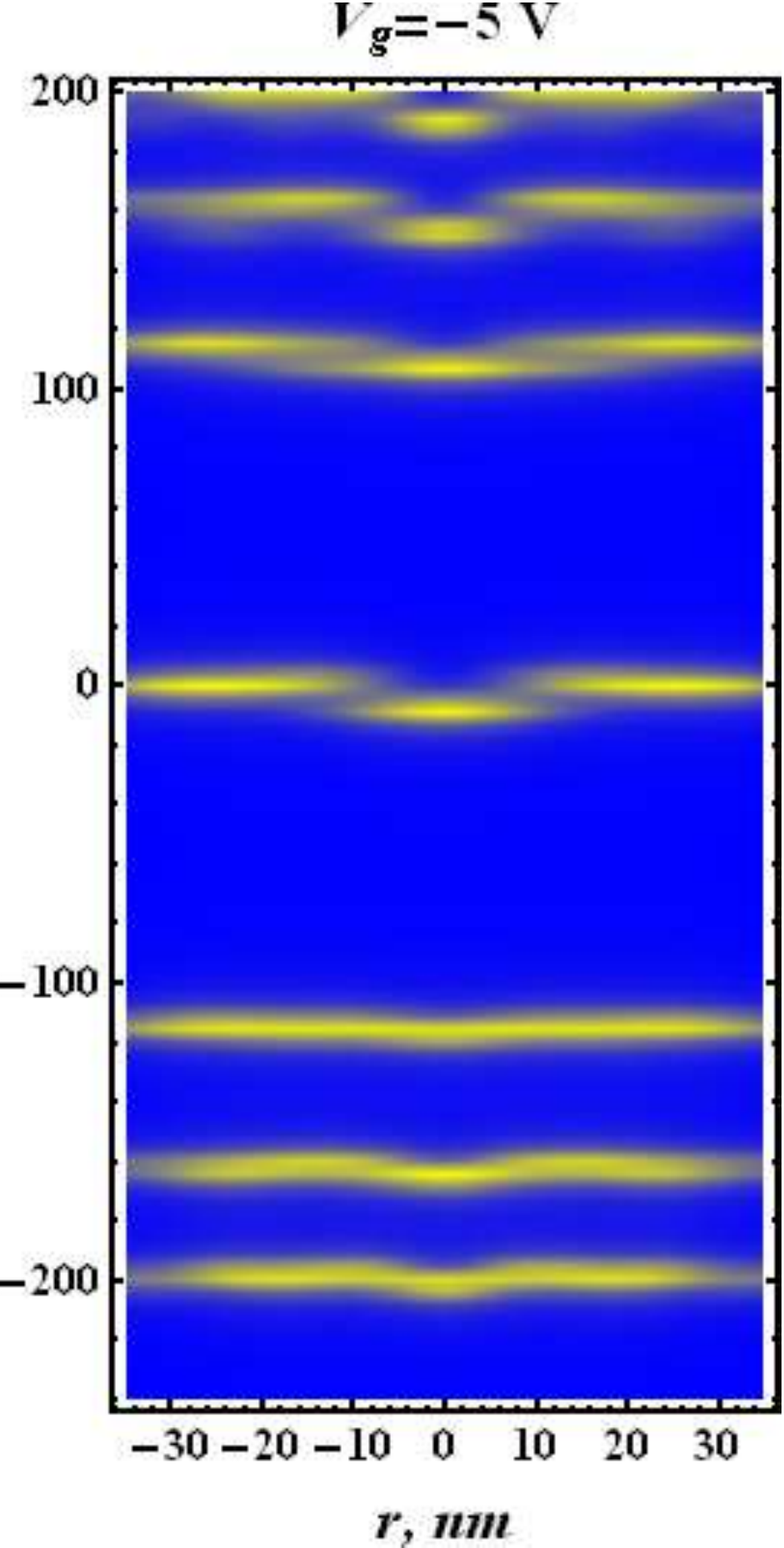}
	\includegraphics[scale=0.35]{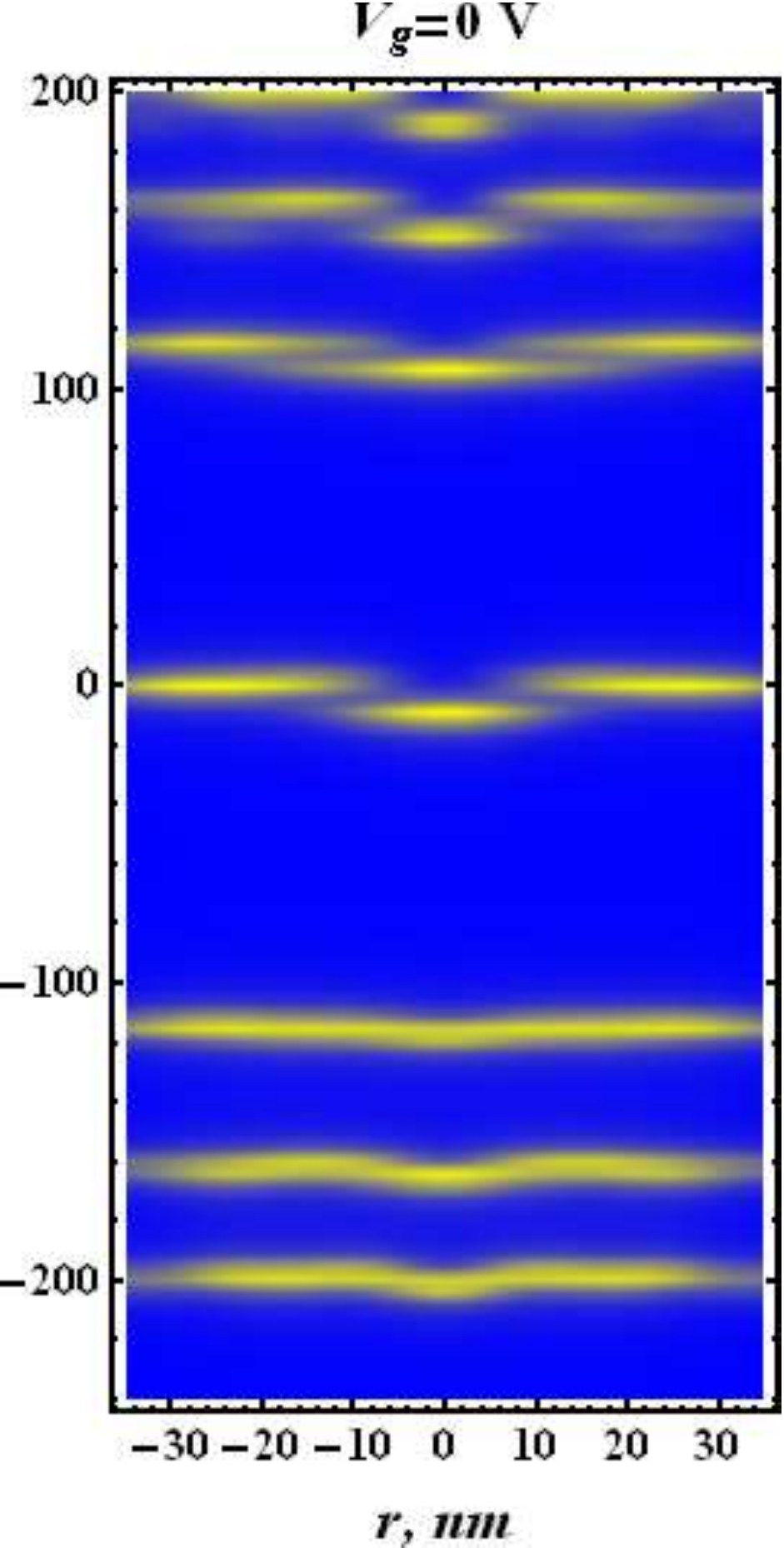}
	\includegraphics[scale=0.35]{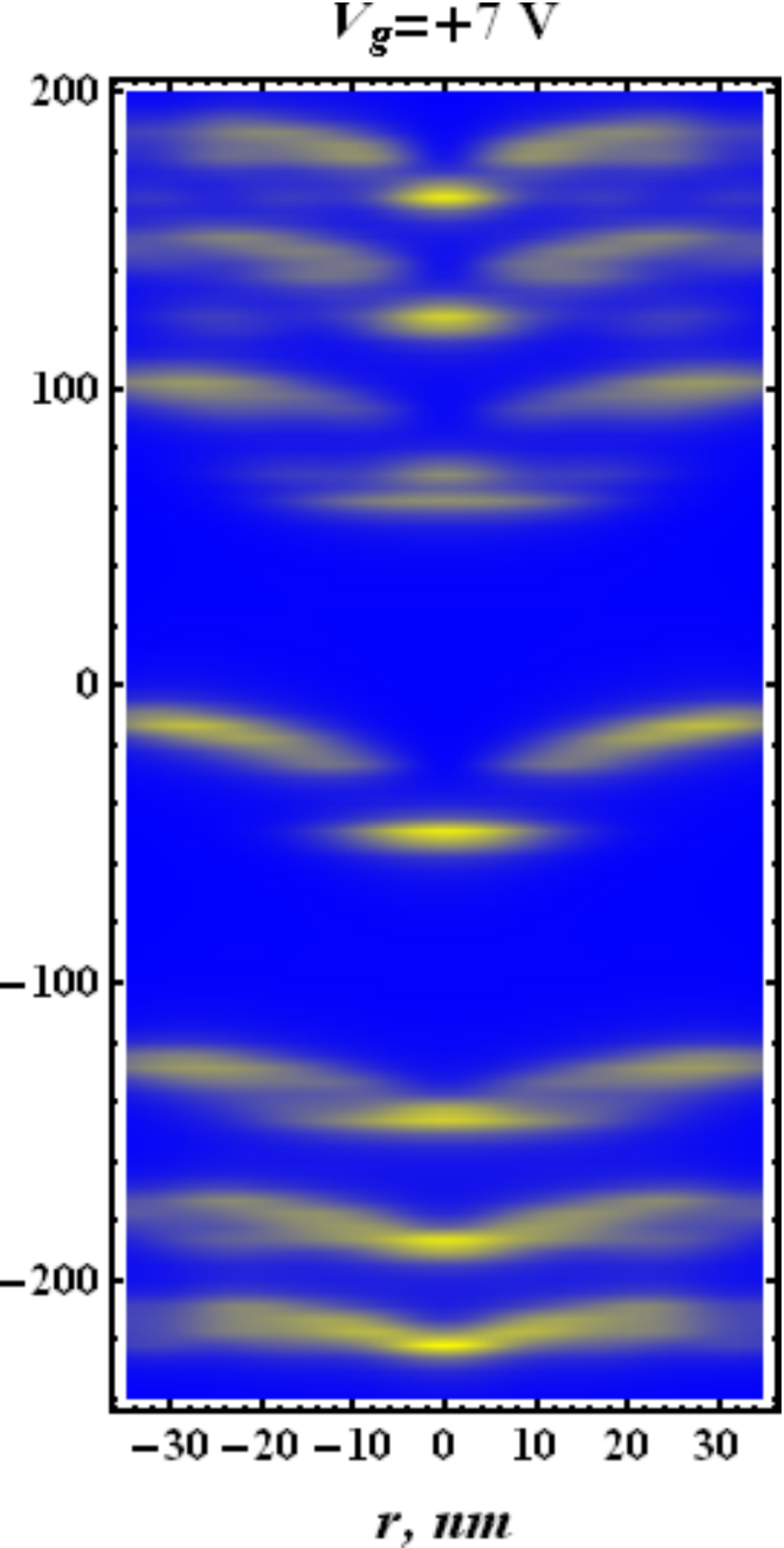}
	\caption{The local density of states calculated numerically in Ref.~[\onlinecite{Sobol2016}] is plotted at four values of gate voltage along
	the line cuts across the impurity. 
		\label{LDOS}}
\end{figure}

The tuning of the effective charge due to the polarization effects was studied theoretically by the three of us in
Ref.~[\onlinecite{Sobol2016}]. Numerically integrating the Dirac equation and determining the energies of several first Landau levels in the
screened impurity potential (\ref{potential_screened}), the local density of states along the line cuts across the impurity was determined and
is plotted in Fig.~\ref{LDOS}. The corresponding LDOS is in good agreement with the experimental results of Ref.~[\onlinecite{Luican-Mayer2014}]
(see Fig.3 therein).

\section{Two-electron bound states near a Coulomb impurity}
\label{section:bound-states}

In the weak interaction regime $\alpha\alt 0.4$, it was found in Ref.~\cite{DeMartino2017} that a pair of repulsively interacting Dirac fermions
in graphene in the attractive potential of a Coulomb impurity with charge $Ze$ forms a two-body bound state localized near the impurity. It could 
be observed by means of STM experiments similar to those previously reporting supercritical behavior in graphene \cite{Wang2012,Luican-Mayer2014,Wang2013} 
and trapped electron states in electrostatically defined graphene dots \cite{Lee2016,Gutierrez2016}.

The negatively charged two-electron hydrogen ion $H^-$ represents a classic problem of nonrelativistic quantum mechanics
\cite{Chandrasekhar1944,Bethe1957,Hill1977,Bransden1983,Andersen2004,Hogaasen2010}. As it was shown in Ref.~\cite{Hill1977},
there exists a single bound state in three spatial dimensions. Chandrasekhar proposed to construct a trial wave function for the ground state of
$H^-$ as follows \cite{Chandrasekhar1944}:
\begin{equation}
\label{chandr}
\Psi({\bf r}_1,{\bf r}_2)= e^{-a r_1 - b r_2} - \epsilon e^{ -b r_1-a r_2 },
\end{equation}
where $r_{l}=|{\bf r}_l|$, $l=1,2$ denote the distance of the corresponding electron to the nucleus, $a$ and $b$ are the variational
parameters, and $\epsilon=\mp 1$ corresponds to a spin singlet/triplet state, respectively. The variational calculation shows that the minimal
energy for a two-body bound state is obtained for $a\neq b$ in the spin singlet configuration ($\epsilon=-1$).

The nonrelativistic 2D counterpart of the above system, which is the $D^-$ problem, describes a donor impurity ion with two
electrons in a 2D semiconductor quantum well \cite{Phelps1983,Pang1990,Larsen1992a,Larsen1992b,Sandler1992,Ivanov2002}.
The effects of quantum confinement on two-body bound-state energies have been studied experimentally in Ref.~\cite{Huant1990}. In the absence of
a magnetic field, there exists only a single bound state in the spin singlet sector. The $D^-$ problem is also similar to the negatively
charged exciton ($X^-$) problem, which was experimentally studied in quantum wells \cite{Shields1995}.

The corresponding 2D relativistic problem could be realized in gapped graphene monolayers (or topological insulator surfaces) with a Coulomb
impurity. However, it was found long ago that the relativistic $H^-$ problem is subtle because the single-particle Dirac Hamiltonian is unbounded from below
\cite{Brown1951,Kolakowska1996,Nakatsuji2005}. In order to set a physically and mathematically well-posed problem,
it is necessary to project the interaction Hamiltonian onto the states with positive energy. It can be devised for
interacting Dirac fermions in graphene if (i) a single-particle gap exists ($\Delta>0$), and (ii) electron-electron interactions are weak, see
Refs.~\cite{Sucher1980,Sucher1984,Haeusler2009,Egger2010}.

We follow the derivation in Ref.~[\onlinecite{DeMartino2017}] and consider the interacting two-particle problem for a gapped graphene monolayer
in the presence of a charged impurity. The corresponding Dirac-Coulomb Hamiltonian at a given valley reads
\begin{equation} \label{2bdiraccoulomb}
H  = \sum_{l=1,2} H(l)  +V_{\rm 2b}.
\end{equation}
Here $H(l)$ is the single-particle Dirac Hamiltonian (\ref{Master-Hamiltonian}) for particle $l=1,2$ with the single-particle potential of
charge $Z$ impurity at the origin
\begin{equation}
V = -\frac{Ze^2}{\kappa r} = -\frac{Z\alpha_g \hbar v_{\rm F}}{\kappa r}
\end{equation}
and $V_{\rm 2b}$ is the standard two-body Coulomb interaction which equals
\begin{equation}
V_{\rm 2b} =\frac{e^{2}}{\kappa|\mathbf{r}_{1}-\mathbf{r}_2|}.
\end{equation}
However, as was mentioned above, the bound state problem is not well posed for the Dirac-Coulomb Hamiltonian of two particles \eqref{2bdiraccoulomb}
because the spectrum of the single-particle Dirac Hamiltonian is unbounded from below. A similar problem occurs in the study of relativistic
effects in the helium atom as was found long ago by Brown and Ravenhall \cite{Brown1951}.
According to Sucher \cite{Sucher1980,Sucher1984}, the Hamiltonian $H$ in Eq.~\eqref{2bdiraccoulomb} has to be replaced 
by the projected Hamiltonian \cite{Haeusler2009,Egger2010}
\begin{equation}\label{Hproj}
H_+= H(1) + H(2) + \Lambda_+ V_{\rm 2b} \Lambda_+,
\end{equation}
with the projection operator $\Lambda_+=\Lambda_+(1)\Lambda_+(2)$. {Since $\mathcal{E}(l)=[H^2(l)]^{1/2}$ is positive definite, the
single-particle operator $\Lambda_+(l) = \left[\mathcal{E}(l)+H(l)\right]/2\mathcal{E}(l)$, obviously, projects onto the positive energy states. As
shown in Refs.~\cite{Sucher1980,Sucher1984,Haeusler2009,Egger2010}, the projected Hamiltonian $H_+$ takes into account the most important
effects of the electron-electron interaction. Further, due to the presence of a band gap, the replacement $H\to H_+$ is reliable for
the ground state of the system in the limit of weak Coulomb repulsion. Moreover, the projection guarantees that the
Hamiltonian $H_+$ can possess the genuine two-particle bound states.
	
Taking the two-particle wave function in a factorized form $\Phi_{\rm tot} =\Phi |\chi \rangle$, where $|\chi \rangle$ is the
normalized spin part (singlet or triplet), its spatial part equals \cite{DeMartino2017}
\begin{equation}\label{cdans}
\Phi({\bf r}_1, {\bf r}_2) = \Psi_{I}({\bf r}_1) \Psi_{O}({\bf r}_2) - \epsilon
\Psi_{O}({\bf r}_1) \Psi_{I}({\bf r}_2) ,
\end{equation}
where $\epsilon=\mp 1$ for the spin singlet/triplet sector and $\Psi_I$ and $\Psi_O$ are the normalized ground-state eigenspinors of
the 2D relativistic hydrogen problem with charges $Z$ replaced by variational parameters $Z_I$ and $Z_O$. These eigenspinors are
given by
\begin{equation}\label{eigenspin}
\Psi_\lambda({\bf r}) =\frac{p_{\lambda}}{\sqrt{\pi\Gamma(1+\gamma_{\lambda})}}
\left(2p_{\lambda}r\right)^{(\gamma_{\lambda}-1)/2} e^{-p_{\lambda} r}
\left(\begin{array}{l}
\sqrt{1+\gamma_{\lambda}}\\
i e^{i\phi} \sqrt{1-\gamma_{\lambda}}
\end{array}\right).
\end{equation}
Here $\lambda=I,O$ and
\begin{equation}
\gamma_{\lambda} = \sqrt{1-4Z^2_{\lambda}\alpha^2},\qquad p_\lambda = 2 \Delta Z_{\lambda}\alpha/(\hbar v_{\rm F}).
\end{equation}

For the energy functional
\begin{equation}
\label{Efunctional}
E_{\epsilon}(Z_I,Z_O)=\frac{\langle \Phi_{\rm  tot} | H_+ | \Phi_{\rm tot} \rangle }{\langle \Phi_{\rm tot} | \Phi_{\rm tot} \rangle}
=\frac{\langle \Phi | H_+ | \Phi \rangle }{\langle \Phi | \Phi \rangle},
\end{equation}
calculating the matrix elements explicitly, we obtain the following energy functional (for details see Ref.\cite{DeMartino2017}):
\begin{equation}
E_{\epsilon}(Z_I,Z_O)=
\sum_\lambda \left( \Delta \gamma_\lambda +
\frac{(Z_\lambda-Z)({\cal V}_\lambda-\epsilon S {\cal U})}{1-\epsilon S^2}\right)+\frac{{\cal V}^{\rm dir}_{\rm 2b} -
\epsilon {\cal V}^{\rm exc}_{\rm 2b}}{1-\epsilon S^2},
\label{energyfunct}
\end{equation}
where the overlap integral equals
\begin{equation}
S=\frac{\sqrt{(1+\gamma_I)(1+\gamma_O)}+\sqrt{(1-\gamma_I)(1-\gamma_O)}}{2}
\frac{\Gamma(1+(\gamma_I+\gamma_O)/2)}{\sqrt{\Gamma(1+\gamma_I)\Gamma(1+\gamma_O)}} \frac{(2p_I)^{(1+\gamma_I)/2}(2p_O)^{(1+\gamma_O)/2}}{(p_I+p_O)^{1+(\gamma_I+\gamma_O)/2}},
\label{2-electron-overlap}
\end{equation}
one-particle matrix elements are
\begin{eqnarray}
{\cal V}_\lambda &=& \langle \Psi_\lambda |(\alpha/r_1)|
\Psi_\lambda \rangle  =2\alpha\frac{p_\lambda}{\gamma_\lambda}, \\
{\cal U} &=& \langle \Psi_I |(\alpha/r_1)| \Psi_O \rangle  =
2\alpha \frac{p_I+p_O}{\gamma_I+\gamma_O}S,
\end{eqnarray}
and two-particle matrix elements
\begin{eqnarray}
{\cal V}_{\rm 2b}^{\rm dir}  &=&
\int d{\bf r}_1d{\bf r}_2 \, \left| \Psi_I({\bf r}_1)\right|^2 \frac{\alpha}{r_{12}}
\left| \Psi_O({\bf r}_2)\right|^2 , \\
{\cal V}^{\rm exc}_{\rm 2b} & = &
\int d{\bf r}_1 d{\bf r}_2 \left[   \Psi^\dagger_I({\bf r}_1)
\Psi_O({\bf r}_1)  \right]  \frac{\alpha}{r_{12}}  \left[   \Psi^\dagger_O({\bf r}_2)
\Psi_I({\bf r}_2)  \right]
\end{eqnarray}
could be represented in terms of elliptic functions (see Ref.~[\onlinecite{DeMartino2017}]).

It should be noted that the energy functional (\ref{energyfunct}) includes the matrix elements of the full interaction operator rather than
those of the projected operator, $\Lambda_+ (\alpha/r_{12}) \Lambda_+$, which are more difficult to obtain and would require a detailed
numerical analysis. Both matrix elements coincide if the trial wave function has vanishing projection onto the negative energy eigenfunctions of
$H_{\rm D}$. In fact, in Ref.~[\onlinecite{DeMartino2017}], it was verified that for $\alpha\alt 0.4$ the cumulative weight of negative energy
states in the trial wave function is very small ($\alt 1\%$). Indeed, negative energy states will only be important if typical interaction
matrix elements can overcome the band gap $2\Delta$. For small $\alpha$, one therefore expects at most small quantitative corrections in the
bound-state energy due to this approximation.

The energy functional (\ref{energyfunct}) possesses the following features. First of all, it is symmetric under an
exchange of its arguments. Second, for the spin-singlet case ($\epsilon=-1$), this energy is bounded from below for $Z\alpha< 1/2$. Third, for
small $\alpha$, $E_{\epsilon}(Z_I,Z_O)$ reduces to the corresponding nonrelativistic energy functional for the $D^-$ problem in 2D
semiconductors \cite{Sandler1992}. However, in contrast to the nonrelativistic case, $E_{\epsilon}(Z_I,Z_O)$ is not homogeneous in $\alpha$, and
hence the bound-state energy explicitly depends on $\alpha$. As in the nonrelativistic case, this energy minimum is realized for unequal values
of $Z_I$ and $Z_O$.

It was shown in Ref.~\cite{DeMartino2017} that the energy functional for the singlet state with $Z=1$ has a minimum located
below the threshold, i.e., the binding energy is positive. In addition, there exists a two-body bound state. The situation is different in the
spin triplet sector, where the variational approach predicts that the energy functional has a minimum whose energy is above the threshold and,
thus, does not describe a bound state. For $Z=1$, the minimum is at $Z_O<1$ and $Z_I>1$ (or vice versa, due to the symmetry of
$E_\epsilon$). Physically, one quasiparticle partially screens the impurity charge seen by the other quasiparticle.

Also in Ref.~[\onlinecite{DeMartino2017}] the authors calculated the probability density and the pair distribution function for the bound
state, focusing on the two-body spin singlet state. They suggest that the bound state can be accessed experimentally, e.g., by means of STM techniques.

\section{Two-center problem}
\label{section:two-centers}

Although, naively, the supercritical instability should be easily realized for charged impurities with $Z > 1$, its experimental
observation remained elusive due to the difficulty of producing highly charged impurities. However, one can reach the supercritical regime by
collecting a large enough number of charged impurities in a certain region. As we saw in Sec.~\ref{section:experiment}, this approach was
successfully realized by creating artificial nuclei (clusters of charged calcium dimers) on graphene \cite{Wang2013} using the tip of a scanning
tunneling microscope. It is ironic that in spite of much larger value of coupling constant in graphene than in QED the first observation of the
supercritical instability in graphene still required the creation of supercritical potentials from subcritical charges like in the case of heavy
nuclei collisions in QED discussed in the Introduction. What crucially differs the graphene experiments \cite{Wang2013} from
those in QED is that the supercritical electric fields created by placing together ionized Ca impurities are static unlike the
fields created in heavy nuclei collisions in QED. This makes possible to observe and analyse reliably the supercritical regime.

The Hamiltonian of the two-center problem is the same as Hamiltonian (\ref{Master-Hamiltonian}) with the potential
\begin{equation}
V(\mathbf{r})=-\frac{e^{2}}{\kappa}\left(\frac{Z_1}{r_{1}}+\frac{Z_2}{r_{2}}\right),
\end{equation}
where $r_{1,2}=|\mathbf{r}\pm\mathbf{R}/2|$ are the distances from electron to impurities with charges $Z_{1,2}$. Since the experiments in
Ref.~[\onlinecite{Wang2013}] were performed for impurities of the same type, we will study in what follows the symmetric problem, i.e.,
$Z_1=Z_2=Z$. The alignment of the charges with respect to the origin $r=0$ is arbitrary due to translational and rotational invariance of the
free gapped Dirac Hamiltonian (\ref{Master-Hamiltonian}), and we choose them located at $(\mp R/2,0)$ with $R$ being a distance between two charges.

The main difficulty in solving the Dirac equation with two Coulomb centers in QED is that variables in this problem are not separable in any
known orthogonal coordinate system \cite{Marinov1975}. Unfortunately, this is true also for the Dirac equation for two Coulomb centers in the
$(2+1)$-dimensional problem in graphene. Therefore, we apply the approximate methods such as the linear combination of
atomic orbitals (LCAO) technique and variational method.

\subsection{LCAO approach for  symmetric two-center problem}
\label{2centers-LCAO}

The LCAO method is well known and widely used in molecular physics \cite{Cohen-Tannoudji}. Wave functions in this method are chosen as linear
combinations of basis functions, where the latter are usually the electron functions centered on the corresponding atoms of the molecule. By
minimizing the total energy of the system, the coefficients of the linear combinations are then determined. The LCAO approach for the symmetric
two-center Dirac problem in 2D was applied in Ref.\cite{Klopfer2014} (see also Ref.\cite{Klopfer_thesis}), where only the lowest
single-impurity bound state near each center is
retained.  This approximation is expected to yield accurate ground-state energies for large $R$ \cite{Matveev2000,Bondarchuk2007}, where the
molecular ground state is well approximated in terms of atomic orbitals. In addition, as we show below, the exact result for $R\to 0$ is also
captured by the LCAO solution.

For a single impurity of charge $Z$, the lowest bound state has the energy $\gamma\Delta$ with $\gamma = \sqrt{1- 4Z^2\alpha^{2}}$. In the
absence of short-distance regularization, the supercritical threshold is  reached at $Z_{c}\alpha=1/2$ \cite{Gamayun2009}, therefore, $Z<Z_c$ is assumed
henceforth. The corresponding normalized spinor, which is an eigenstate of the total angular momentum operator
$J_z=-i\hbar\partial_\phi +\hbar\frac{\sigma_z}{2}$ with eigenvalue $1/2$, has the same form as (\ref{eigenspin}) and reads
\begin{equation}
\label{singleimp}
\Psi_0 (r,\phi) = \frac{2Z\alpha}{\sqrt{\pi \Gamma(1+\gamma)} R_\Delta}
\left(\frac{4Z\alpha r}{R_\Delta}\right)^{(\gamma-1)/2}
e^{-2Z\alpha r/R_\Delta}
\left(
	\begin{array}{c}
	\sqrt{1+\gamma} \\
	ie^{i\phi}  \sqrt{1-\gamma}
	\end{array}
\right),
\end{equation}
where $R_\Delta=\hbar v_{\rm F}/\Delta$.

It is convenient to rewrite the interaction potential as follows:
\begin{equation}\label{qeffintro}
H= H_{0} - (Z_{\rm eff}+\delta Z)\frac{e^{2}}{\kappa}\left(\frac{1}{r_1}+\frac{1}{r_2}\right),
\end{equation}
where $\delta Z=Z-Z_{\rm eff}$ and the effective charge $Z_{\rm eff}$ is introduced. It is a variational parameter, which could be determined
by minimizing the total energy.

Following the standard LCAO approach \cite{Cohen-Tannoudji}, we seek the electron wave function $|\Phi\rangle$ in terms of atomic
orbitals, $|1\rangle$ and $|2\rangle$ centered near the Coulomb impurity at $(\mp R/2,0)$, which depend on $r_1$ and $r_2$, respectively, i.e.,
$|\Phi\rangle = v_1 |1\rangle+ v_2|2\rangle$. The atomic orbitals are chosen as eigenstates (\ref{singleimp}) in the field of a single impurity
of charge $Z_{\rm eff}$. The Dirac equation is thereby reduced to a linear system of equations for $v_1$ and $v_2$, and the energy
$E=E(Z_{\rm eff})$ follows from the condition
\begin{equation}\label{detc}
{\rm det} \left(\begin{array}{cc}  H_{11}-E & H_{12}-SE \\
H_{21}-SE & H_{22}-E \end{array}\right) = 0,
\end{equation}
where $H_{ij}=\langle i|H|j\rangle$ and the overlap integral $S= \langle 1|2\rangle=\langle 2|1\rangle$.

Defining the Coulomb integral,
\begin{equation}\label{coulint}
C = \langle 1|r_2^{-1} |1 \rangle = \langle 2 |r^{-1}_1|2\rangle,
\end{equation}
and the resonance integral,
\begin{equation}\label{resint}
A=\langle 1|r^{-1}_{1,2}|2\rangle=
\langle 2|r_{1,2}^{-1}|1\rangle,
\end{equation}
all matrix elements in Eq.~(\ref{detc}) can be written in compact form:
\begin{eqnarray}
H_{11}&=& H_{22}= \gamma\Delta - 4Z_{\rm eff}\delta Z\alpha^{2}\Delta/\gamma - Z\frac{e^{2}}{\kappa} C, \\  \nonumber
H_{12}&=& H_{21}= \gamma S\Delta -(Z+\delta Z)\frac{e^{2}}{\kappa} A.
\end{eqnarray}
While $S$, $C$, and $A$ can be directly evaluated \cite{Matveev2000,Bondarchuk2007} in the 3D Dirac problem, the 2D case is,
unfortunately, more involved. In order to compute them, it is useful to employ elliptic coordinates \cite{Gradshtein-book}
\begin{equation}
\label{elliptic-coordinates}
\xi=\frac{r_1+r_2}{R}\in [1,\infty),\quad \eta=\frac{r_1-r_2}{R}\in [-1,1].
\end{equation}

In these coordinates the integrals could be written as follows:
\begin{eqnarray}
S &=& \frac{u^{\gamma+1}}{\pi \Gamma(1+\gamma)} \int_1^\infty d\xi \int_{-1}^1 d\eta
\frac{(\xi^2-\eta^2)^{(\gamma-1)/2} e^{-u\xi}}{\sqrt{(\xi^2-1)(1-\eta^2)}} \left[\xi^2-1+\gamma(1-\eta^2)\right], \label{sinit} \\
A &=& \frac{2 u^{\gamma+1}}{\pi R \Gamma(1+\gamma)} \int_1^\infty d\xi \int_{-1}^1 d\eta
\frac{\xi (\xi^2-\eta^2)^{(\gamma-3)/2} e^{-u\xi}}{\sqrt{(\xi^2-1)(1-\eta^2)}} \left[\xi^2-1+\gamma(1-\eta^2)\right], \label{resonance-int} \\
C &=& \frac{2 u^{\gamma+1}}{\pi R \Gamma(1+\gamma)} \int_1^\infty d\xi
\int_{-1}^1 d\eta \frac{(\xi+\eta)^{\gamma} e^{-u(\xi+\eta)} } {
	\sqrt{(\xi^2-1)(1-\eta^2)} },
\end{eqnarray}
where $u=\frac{2RZ_{\rm eff}\alpha}{R_{\Delta}}$ and $\gamma=\sqrt{1-(2Z_{\rm eff}\alpha)^{2}}$.

Using the above integral representations for $S$, $C$, and $A$ (or the corresponding series from Ref.~[\onlinecite{Klopfer2014}]), it
is numerically straightforward to obtain an LCAO estimate for the ground-state energy $E(Z_{\rm eff})$ for given $Z_{\rm eff}$.
Then the minimal energy is determined, realized for $Z_{\rm eff}=Z^*$, where the numerical search is aided by noting
that $E(Z)$ depends quadratically on $Z_{\rm eff}-Z^*$.
Numerical results in Ref.~[\onlinecite{Klopfer2014}] show that the LCAO ground-state energy, $E(R)$, matches the
expected single-impurity values $\gamma\Delta$ in both limits, namely (i) for $R\to \infty$ with impurity charge $Z$, where we have
two decoupled copies of the single-impurity problem, and (ii) for $R\to 0$, where both centers conspire to form a single Coulomb impurity
of charge $2Z$. Furthermore, it was shown that the optimal effective charge $Z^*$ nicely matches
both limits as well.

Choosing larger $Z$ such that $\zeta=Z/Z_{c}=2Z\alpha$ is within the bounds $1/2<\zeta<1$, the supercritical regime can be realized by decreasing
$R$ through a transition value, $R=R_{\rm cr}$.  At the critical distance, the ground-state energy reaches the Dirac sea, $E(R_{\rm cr})=-\Delta,$
and for $R<R_{\rm cr}$, the two-center system with subcritical individual impurity charge becomes supercritical. The LCAO results in
Ref.~[\onlinecite{Klopfer2014}] show that, in practice, $Z$ has to be chosen quite close to $Z_{c}=1/(2\alpha)$, since otherwise $R_{\rm cr}$
becomes extremely small. This conclusion seems also in agreement with the reported experimental observations of supercriticality
\cite{Wang2013,Luican-Mayer2014}, where different ions first had to be pushed closely together, thereby forming charged clusters, before
supercriticality appears.

\subsection{Variational method}
\label{2centers-variational}

Another means to study the supercritical instability of two Coulomb centers in graphene is the variational method which was applied
to the corresponding two centers problem in QED in Ref.~\cite{Marinov1975}. As noted in Ref.~[\onlinecite{Popov1972b}], in order to obtain
a satisfactory accuracy it is necessary that trial functions correctly reproduce the asymptotics of the exact solution at infinity and near
the charged impurities. These asymptotics could be found from the direct analysis of the Dirac equation $\hat{H}\Psi=E\Psi$.
For two-component spinor $\Psi(\mathbf{r})=(\phi,\ \chi)^T$, expressing $\chi$ in terms of $\phi$ leads to the following second order equation
for the $\phi$ component of the Dirac spinor:
\begin{equation}
(\partial^2_x+\partial^2_y)\phi+\frac{\frac{\partial V}{\partial x}-i\frac{\partial V}{\partial y}}{E-V+\Delta}
\left(\frac{\partial\phi}{\partial x}+i\frac{\partial\phi}{\partial y}\right)+\frac{(E-V)^2-\Delta^2}{(\hbar v_{\rm F})^{2}}\phi=0.
\label{upper-component}
\end{equation}
According to Refs.~[\onlinecite{Zeldovich1972,Greiner1985}], the supercritical instability takes place when the bound state with the lowest
energy dives into the lower continuum. This occurs when $E=-\Delta$. For this solution, let us consider the asymptotic at large $r\gg R$, where
the potential equals
\begin{equation}
V\left(\mathbf{r}\right)=-\zeta \hbar v_{\rm F}\left(\frac{1}{r}+\frac{R^2}{4r^3} P_2(\cos\varphi)+O\left(\frac{1}{r^5}\right)\right),
\label{potential-asymptotic}
\end{equation}
$\zeta=2Z\alpha$ is a dimensionless charge, and $P_2(x)$ is the Legendre polynomial $P_n(x)$ with $n=2$. In what follows we consider the case
when charges of impurities  are subcritical whereas their total charge exceeds a critical one, $1/2<\zeta<1$. The case $\zeta<1/2$ corresponds
to the situation when the total charge is less than a critical one and is not relevant for the supercritical regime.

Neglecting the quadrupole and higher order multipole terms in the potential (this corresponds to the monopole approximation)
Eq. (\ref{upper-component}) reduces to the following equation for $\phi(r)$:
\begin{equation}
\label{e2}
\phi''+\frac{2}{r}\phi'+\left(\frac{\zeta^2}{r^2}-\frac{2m\zeta}{r}\right)\phi=0,
\end{equation}
where $m=\Delta/(\hbar v_{\rm F})$. The decreasing at infinity solution is expressed in terms of a Macdonald function,
\begin{equation}
\phi(r)=C_1r^{-1/2}K_{i\beta}(\sqrt{8m\zeta r}),\quad \beta=\sqrt{4\zeta^2-1}.
\label{phi-r-to-infinity}
\end{equation}
Its asymptotic
\begin{equation}
\phi_{\rm asym}(r)=C_1r^{-3/4}\exp(-\sqrt{8m\zeta r}),\quad r \to \infty
\label{asymptotic}
\end{equation}
agrees, of course, with the asymptotical behavior of a solution to the Dirac equation of one center with charge $2Ze$. This shows also
that the level which reached the boundary of the lower continuum remains localized.

In order to find the asymptotic of the solution in the vicinity of two Coulomb centers, it is conventional and convenient
to use the elliptic coordinate system (\ref{elliptic-coordinates}). We note that for charges of impurities such as
$Z\alpha< 1/2$ ($\zeta<1$) there is no ``collapse'' in the  Coulomb field of one impurity
\cite{Shytov2007a,Shytov2007b,Pereira2007,Gamayun2009,Khalilov2013,Chakraborty2013, Chakraborty2013a}, therefore, it is not necessary
to cut off the potential at small $r$ and the impurities may be considered as point-like. Therefore, for simplicity, in what follows,
we will consider the nonregularized Coulomb potential. In elliptic coordinates, it has the form
\begin{equation}
V\left(\mathbf{r}\right)=-\frac{2\zeta \hbar v_{\rm F} \xi}{R(\xi^2-\eta^2)}.
\end{equation}
To find the asymptotic of $\phi$ in the vicinity of impurities, i.e., for small $\xi^2-\eta^2$, we seek $\phi$ in the form
$\phi(\xi,\eta)=\phi(\mu)$, where $\mu=\xi^2-\eta^2=4r_1r_2/R^2$. Near the impurities $r_1\rightarrow0$ or $r_2\rightarrow0$, i.e.,
$\xi\rightarrow 1$ and $\eta\rightarrow \pm 1$ and, consequently, $\mu \to 0$, we obtain the following equation:
\begin{equation}
\frac{d^2\phi}{d\mu^2}+\frac{2}{\mu}\frac{d\phi}{d\mu}+\frac{\zeta^2}{4\mu^2}\phi=0,
\end{equation}
whose regular solution at $\mu \to 0$ is
\begin{equation}
\phi_{\rm imp}(\mu)=C_2\mu^{-\sigma/2},\quad \sigma=1-\sqrt{1-\zeta^2}.
\label{phi-mu-to-zero}
\end{equation}
This asymptotic describes the behavior of the wave function  at the impurities positions. Since at large distances $r \gg R$ the variable
$\mu$ equals $\mu\simeq 4r^2/R^2$, solution (\ref{phi-r-to-infinity}) can be rewritten as follows:
\begin{equation}
\phi(\mu)=C_1\mu^{-1/4}K_{i\beta}(2\sqrt{m\zeta R}\mu^{1/4}).
\label{phi-mu-to-infinity}
\end{equation}
Matching solutions (\ref{phi-mu-to-zero}) and (\ref{phi-mu-to-infinity}) at the point $\mu=1$ we can find an approximate estimate of the critical
distance $R_{\rm cr}(\zeta)$ as a function of $\zeta$. We obtain the following transcendental equation:
\begin{equation}
2\sqrt{1-\zeta^2}-1=2\sqrt{m\zeta R}\frac{K_{i\beta}^\prime(2\sqrt{m\zeta R})}{K_{i\beta}
	(2\sqrt{m\zeta R})}.
\label{trancend-eq}
\end{equation}
For $m R \ll 1$, i.e., when the distance between the impurities is much less than the Compton wavelength of quasiparticles, Eq.~(\ref{trancend-eq})
can be simplified using the asymptotic of $K_{i\beta}(z)$ for $z \to 0$. Then we obtain the following analytical solution:
\begin{equation}
mR_{\rm cr}=\frac{1}{\zeta}\exp\left[-\frac{2}{\beta}\left(\cot^{-1}\frac{1-2\sqrt{1-\zeta^2}}{\beta}
-{\rm arg}\Gamma(1+i\beta)\right)\right],
\label{appr-num-solution}
\end{equation}
where $\Gamma(z)$ is the Euler gamma function. It is amazing that Eq.~(\ref{appr-num-solution}) coincides with the corresponding
solution found in QED for scalar particles \cite{Popov1972}. Eq.~(\ref{appr-num-solution}) for $\sqrt{4\zeta^2-1}\ll1$  can be written in more
simple form
\begin{equation}
mR_{\rm cr}=\frac{1}{\zeta}\exp\left(-\frac{2\pi}{\sqrt{4\zeta^2-1}}\right).
\label{critical-distance-estimate}
\end{equation}
We find that the deviation of $R_{\rm cr}$ given by Eq.~(\ref{critical-distance-estimate}) from that determined by Eq.~(\ref{trancend-eq}) is
rather small up to $\zeta=0.8$. A numerical calculation of $R_{\rm cr}$ given by these equations is presented in Fig. \ref{appr-crit-line} in
comparison with $R_{\rm cr}$ determined in more refined calculations using a variational method.
\begin{figure}[ht]
	\centering
	\includegraphics[scale=0.37]{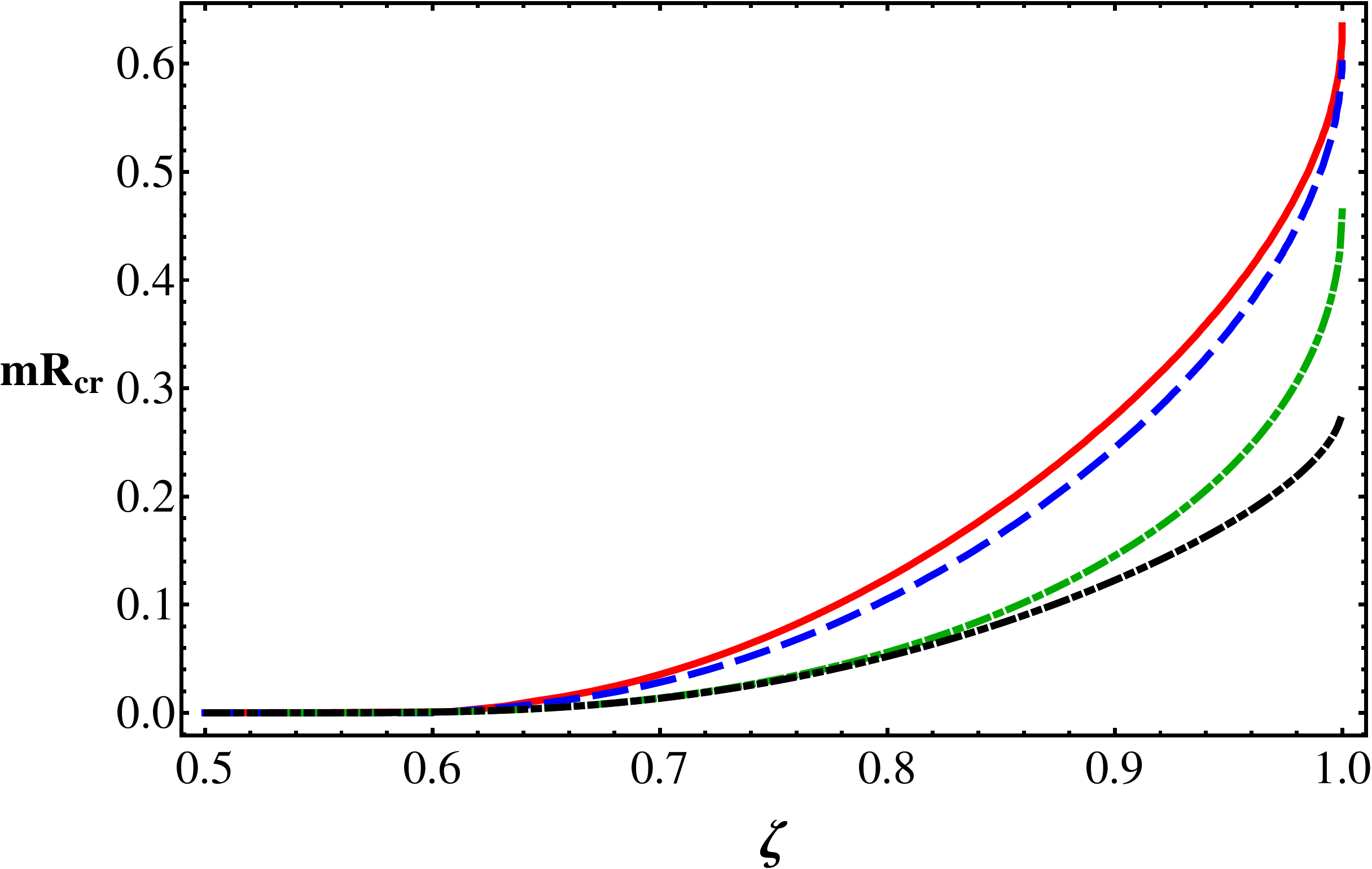}
	\caption{The dependence $mR_{\rm cr}(\zeta)$ given by Eqs.~(\ref{trancend-eq})
		(dash-dotted green line) and (\ref{appr-num-solution}) (dash-dotted with two dots black line), and calculated by variational method
with $N=1$ (blue dashed line) and $N=2$ (red solid line). \label{appr-crit-line}}
\end{figure}
Clearly, the approximation we used is rather crude because it matches only the asymptotics and, in particular, it does not take into account at
all the nonsphericity of the potential of two impurities described by $P_2(\cos\theta)$ and higher harmonics in potential (\ref{potential-asymptotic}).

To set up the variational problem, we note that the differential equation~(\ref{upper-component}) can be obtained as an extremum of the following
functional:
\begin{equation}
S[\phi]=\int\left((E-V+\Delta)^{-1}\left|\frac{\partial\phi}{\partial x}+i\frac{\partial\phi}
{\partial y}\right|^2-\frac{(E-V-\Delta)}{(\hbar v_{\rm F})^{2}}|\phi|^2\right) dxdy,
\label{functional}
\end{equation}
under the condition that the norm $N=\int\Psi^*\Psi dxdy$ is conserved (this condition is important for obtaining the correct
boundary conditions). Introducing a new field $\psi=W^{-1/2}\phi$, where $W=E-V+\Delta$, the functional $S[\phi]$ can be represented
in the form specific for nonrelativistic quantum mechanics
\begin{equation}
S[\psi]=\int\left[|\boldsymbol{\nabla}\psi|^2+i\left(\frac{\boldsymbol{\nabla} V}{2W}\times\boldsymbol{\nabla}\psi^*\right)
\psi-i\psi^*\left(\frac{\boldsymbol{\nabla} V}{2W}\times\boldsymbol{\nabla}\psi\right)+2(U-\epsilon)|\psi|^2\right]dxdy,
\label{functionalS-modified}
\end{equation}
where $\mathbf{a}\times\mathbf{b}=\epsilon_{ij}a_ib_j$, $\epsilon=(E^2-\Delta^2)/(2\hbar^{2} v_{\rm F}^2)$ is
the effective energy, and the effective potential $U$ is given by
\begin{equation}
U=\frac{2EV-V^{2}}{2(\hbar v_{\rm F})^2}+\frac{\triangle V}{4W}+\frac{3}{8}\frac{(\boldsymbol{\nabla} V)^2}{W^2}.
\end{equation}
The second and third terms in functional (\ref{functionalS-modified}) describe the pseudospin-orbit coupling with the field $\mathbf{F}=-{\boldsymbol{\nabla} V}/2W$,
they  do not contribute for the ground state wave function which is real. Functional (\ref{functionalS-modified}) is bounded from below, so one is
in position to apply to it the variational principle. In what follows we are interested in the case where the bound state with the lowest energy
crosses the boundary of the lower continuum, so we put $E=-\Delta$ i.e., $\epsilon=0$. Then $W=-V$ and the functional $S[\psi]$
is simplified.

In QED, the Ritz and Kantorovich methods were employed in order to solve the variational problem and find a critical distance $R_{\rm cr}$ (see
a discussion in Sec.~III in Ref.~[\onlinecite{Popov-review2001}]). In the Ritz method, the sought function $\psi$ is expanded over a fixed set of
basis functions $\psi(x,y)=\sum_nc_n\psi_n(x,y)$, where $c_n$ are variational constants. In the Kantorovich method, $\psi=\sum_nc_n(x)\psi_n(y)$,
where $\psi_n(y)$ are fixed functions, while $c_n(x)$ are variable functions. Obviously, the variational problem reduces to a system of linear
algebraic equations for $c_n$ in the Ritz method and to a system of linear ordinary differential equations for $c_n(x)$ in the Kantorovich method.

According to Eq.~(\ref{phi-mu-to-zero}), near impurities $\phi$ depends only on $\mu=\xi^2-\eta^2={4r_1 r_2}/{R^2}$. At the large distances,
$r\to\infty,$ the variable $\mu \to \infty$ and the asymptotic of $\phi$ is given by Eq.~(\ref{phi-mu-to-infinity}). Therefore, both asymptotics
of $\psi$ depend only on $\mu$. In order that variational ansatz for $\psi$ give appropriate results, it is essential to take into account
correctly the behavior of the exact solution near the Coulomb centers and at infinity. We choose the two variables $\mu,\nu$ so
that the function $\psi(x,y)$ has a singularity only in $\mu$. Then using the following ansatz in the Kantorovich method:
\begin{equation}
\psi=\sum\limits_{k=1}^{N}\psi_k(\mu)\nu^{k-1},
\label{Kantorovich-method}
\end{equation}
where $\psi_k(\mu)$ are variable functions of $\mu$ and $\nu(\xi,\eta)$ is a fixed function of $\xi$ and $\eta$, we can maximally correctly take
into account the behavior of the exact solution near the Coulomb centers. Since {\it a priori} we do not know what set of functions $\psi_n(x,y)$
is the best in the Ritz method, we use like in the QED studies \cite{Marinov1975} the Kantorovich method.

Two choices of function $\nu$ were considered in QED \cite{Marinov1975}: i) $\nu=\eta^2/(\xi^2-\eta^2)=(r_1-r_2)^2/4r_1r_2$ and ii) $\nu=\eta^2=
(r_1-r_2)^2/R^{2}$. The obtained results were close. Here, we will consider the case (i). Since the charges of impurities are identical, $Z_1=Z_2=Z$,
the wave function of the ground state is symmetric under the inversion $x\rightarrow -x, y\rightarrow -y$, therefore, the change of the variables
$x,y$ to $\mu,\nu$  is performed  by means of the formulas
\begin{equation}
x=\frac{R}{2}\mu\sqrt{\nu(\nu+1)},\quad y=\frac{R}{2}[(\mu(\nu+1)-1)(1-\mu\nu)]^{1/2}.
\end{equation}
Inserting ansatz (\ref{Kantorovich-method}) in Eq.~(\ref{functionalS-modified}) and integrating over $\nu$, we obtain
\begin{equation}
S_N(\psi)=4\sum\limits_{k,l=1}^{N}\int\limits_{0}^{\infty}d\mu \left(P_{kl}\psi_k'{\psi^*_l}'
+Q_{kl}\psi_k\psi^*_l+R_{kl}\psi_k'\psi^*_l+R_{kl}^{\dagger}{\psi_k}{\psi^*_l}'\right),
\label{functional2}
\end{equation}
where $P,Q$, and $R$ are $N\times N$ matrices which depend on $\mu$
\begin{equation}
\label{P} P_{kl}(\mu)=\int\limits_0^\infty
(\boldsymbol{\nabla}\mu)^2\nu^{k+l-2} |J|f(\mu,\nu) d\nu,
\end{equation}
\begin{equation}\label{Q}
Q_{kl}(\mu)=\int\limits_0^\infty\Biggl[(\boldsymbol{\nabla}\nu)^2(l-1)(k-1)\nu^{k+l-4}-i(l-k) \left(\!\frac{\boldsymbol{\nabla} V}{2V}\times\boldsymbol{\nabla}\nu\!\right)\nu^{k+l-3}+2U\nu^{k+l-2}\Biggr]|J|f(\mu,\nu) d\nu,
\end{equation}
\begin{equation}\label{R}
R_{kl}(\mu)=\int\limits_0^\infty\Biggl[\boldsymbol{\nabla}\mu\boldsymbol{\nabla}\nu(l-1)\nu^{k+l-3}+i\left(\!\frac{\boldsymbol{\nabla} V}{2V}
\times \boldsymbol{\nabla}\mu\!\right)\nu^{k+l-2}\Biggr]|J|f(\mu,\nu) d\nu.
\end{equation}
Here $f(\mu,\nu)=\theta(1-\mu\nu)[\theta(1-\mu)\theta(\mu(\nu+1)-1)+\theta(\mu-1)]$,
$|J|=\mu R^2/(16\sqrt{\nu(\nu+1)(\mu+\mu\nu-1)(1-\mu\nu)})$ is a Jacobian, $\boldsymbol{\nabla}$ is a gradient with respect to Cartesian coordinates, and $U$ is the effective potential. The explicit expressions for $\boldsymbol{\nabla}\mu$, $\boldsymbol{\nabla}\nu$, and $U$ are given in Appendix A in Ref.\,[\onlinecite{Sobol2013}].

Minima of functional~(\ref{functional2}) are given by solutions of the following set of
Euler-Lagrange equations:
\begin{equation}
\frac{d}{d\mu}\left(P_{kl}\frac{d\psi_k}{d\mu}+R_{kl}^{\dagger}\psi_k\right)-Q_{kl}\psi_k-R_{kl}
\frac{d\psi_k}{d\mu}=0.
\label{set-of-equations}
\end{equation}
The boundary conditions for functions $\psi_k$ follow from the requirement that the norm of the function $\psi$ be finite. The differential equation~(\ref{set-of-equations}) and these boundary conditions define our boundary value problem.

In the simplest case $N=1$, we have
\begin{equation}
\label{e4}
\frac{d}{d\mu}\left(P\frac{d\psi}{d\mu}\right)-Q\psi=0,
\end{equation}
where $P(\mu)=\pi\mu$ and $Q(\mu)$ is expressed through the complete elliptic integrals of the first and second kind:
\begin{eqnarray}
Q(\mu)&=&-\frac{\pi(\zeta^2-1)}{8\mu}\,+\frac{\zeta mR}{2}\left[\theta(1-\mu) {\rm K}(\sqrt{\mu})+
\frac{\theta(\mu-1)}{\sqrt{\mu}} {\rm K}\left(\!\frac{1}{\sqrt{\mu}}\!\right)\right]+\nonumber\\
&+&\theta(1-\mu)\biggl[\frac{3}{8\mu(1+\mu)} {\rm E}(\mu)-\frac{(2\zeta^2+1)(1+\mu)}{8\mu} {\rm K}(\mu)\biggr]+\nonumber\\
&+&\theta(\mu-1)\biggl[\frac{3}{8(1+\mu)} {\rm E}\left(\!\frac{1}{\mu}\!\right)-\frac{(\zeta^2-1)(1+\mu)+3\mu}{4\mu^2} {\rm K}\left(\!\frac{1}{\mu}\!\right)\!\biggr]. \label{Q-function}
\end{eqnarray}
We seek a wave function of the ground state which could be chosen real. Therefore, the function $R_{11}$, which is completely imaginary, does nor appear in Eq.\,(\ref{e4}).

The differential equation (\ref{e4}) determines the wave function of the critical bound state that just dives into the lower continuum.
Since the wave function of a bound state tends to zero at infinity, this translates in our case to the condition $\psi(\mu) \to 0$ as
$\mu \to\infty$. The asymptotic of the wave function near the impurities (where $\mu \to 0$) is given by Eq.~(\ref{phi-mu-to-zero}).
This equation completes the set-up of our boundary value problem which allows us to determine the critical distance $R_{\rm cr}$ between
the impurities as a function of $\zeta$. Since the function $Q(\mu)$ is given in terms of the complete elliptic integrals of the first
and second kind, the differential equation (\ref{e4}) cannot be solved analytically. We solve this equation numerically by using the
shooting method and proceed as follows. We fix the wave function and its first derivative at certain small $\mu$  using Eq.~(\ref{phi-mu-to-zero}).
Then, we fix $\zeta$ and solve Eq.~(\ref{e4}) numerically for different $mR$ [note that since the function $Q(\mu)$ depends only on the
product $mR$, parameters $m$ and $R$ cannot be separately varied]. The critical distance $R_{\rm cr}$ (for a given $m$) is then determined
as $R$ such that the wave function $\psi(\mu)$ tends to zero at infinity. Repeating this procedure for different $\zeta$, we find
how the critical distance between the impurities depends on $\zeta$. The corresponding dependence $mR_{\rm cr}$ on $\zeta$ is plotted
in Fig.~\ref{appr-crit-line} (blue dashed line).

The accuracy of computation can be improved taking $N>1$ in sum (\ref{Kantorovich-method}). In this case one should solve a set of
second-order differential equations. Since the shooting method is not well suited for this purpose, it is better then to follow the
corresponding calculations in QED in Ref.~[\onlinecite{Marinov1975b}] and reduce the set of Eqs.~(\ref{set-of-equations}) to a matrix
Riccati equation, which can be solved by the Runge-Kutta method. The case $N=2$ was considered in Ref.~[\onlinecite{Sobol2014UJP}] and
the obtained results are quite close to the case $N=1$ and are shown in Fig.~\ref{appr-crit-line} (red solid line).

\subsection{Quasistationary states}

When the distance between the impurities becomes smaller than the critical one the bound state dives into the lower continuum
transforming into a resonance.  The energy and width of this quasistationary state could be determined
using the Wentzel--Kramers--Brillouin method in the monopole approximation \cite{Sobol2013}. For distances $r > R$
(or more exactly $r \gg R$), the potential of the two-center problem is close to a spherically symmetrical one. Therefore, we
can consider one charged impurity with the charge $2Ze$ and restrict our consideration only to the region $r \ge \varkappa R$, where
$\varkappa\sim 1$ is a dimensionless constant. This is known as the monopole approximation.

For a spherically symmetric potential, the squared Dirac equation could be rewritten in the Schr\"{o}dinger-like form (\ref{Schrodinger-like:eq})
and the corresponding quasiclassical momentum $k(r,E)$ is given by Eqs.~(\ref{quasiclassical-momentum})--(\ref{U2-pot}). To find the energy of
quasibound states we use the Bohr--Sommerfeld quantization condition
\begin{equation}
\int\limits_{\varkappa R/2}^{r_{-}}k(r,E)dr=\int\limits_{\varkappa R_{\rm cr}/2}^{r_{-}^0}
k(r,-\Delta)dr=2\pi\hbar n,
\label{Bohr-Somm1}
\end{equation}
where $n=1,2...$ and $r_{-}$ is a boundary of the classically forbidden region determined by the equation $k(r_{\pm},E)=0$
and $r_{-}^0=r_{-}(E=-\Delta)$.

For energies close to the boundary of the lower continuum, $E\to-\Delta$, we find
\begin{equation}
{E(R,\zeta)}=-{\Delta}\cdot F(\zeta,R), \quad\quad F(\zeta,R)=\left(\frac{R_{\rm cr}}{R}+\frac{1+2j}{4\zeta^2}
-\frac{\zeta^2-j^2}{3\zeta^2}\right)/
\left(1+\frac{1+2j}{4\zeta^2}-\frac{\zeta^2-j^2}{3\zeta^2}\right).
\label{energy-resonance}
\end{equation}

The width of quasistationary states apart from a preexponential factor is determined by tunneling through the classically forbidden region.
For energies close to the boundary of the lower continuum, this gives the width
\begin{eqnarray}
\Gamma\propto \exp\hspace{-1mm}\left[-2\pi\hspace{-1mm}\left(\zeta\sqrt{\frac{E^2}{E^2-\Delta^2}}
-\sqrt{\zeta^2-j^2}\right)\hspace{-1mm}\right]\simeq\exp\hspace{-1mm}\left[-2\pi\hspace{-1mm}
\left(\tilde\beta\sqrt{\frac{R}{R_{\rm cr}-R}}-\sqrt{\zeta^2-j^2}\right)\hspace{-1mm}\right],
\tilde\beta=\sqrt{\frac{8\zeta^2+4j^2+6j+3}{24}},
\label{width-Gamma}
\end{eqnarray}
which tends to zero when $E\to-\Delta$ or $R\to R_{\rm cr}$.

\section{The dipole problem and migration of the wave function}
\label{section:dipole}

The electron states for Dirac fermions in the field of two oppositely charged ($\pm Q$, $Q=Ze$) nuclei at distance $R$ ($ZeR$ is
the corresponding electric dipole moment) in gapped graphene are described by Hamiltonian (\ref{Master-Hamiltonian}) with potential
\begin{equation}
\label{twocenter}
V(\mathbf{r})= \frac{Ze^{2}}{\kappa}\left(\frac{1}{\sqrt{(x+R/2)^2+y^2}}-
\frac{1}{\sqrt{(x-R/2)^2+y^2}}\right),
\end{equation}
where $\kappa$ is the dielectric constant.

The corresponding Hamiltonian has an intrinsic particle-hole symmetry $\Omega \hat{H}\Omega^+=-\hat{H}$, where the unitary operator $\Omega=\sigma_x {\cal R}_x$
satisfies $\Omega^2=1$ (${\cal R}_x$ is the operator of reflection $x\to -x$). It follows then that an eigenstate $\Psi_E(x,y)$ with energy $E$
has a partner $\Psi_{-E}(x,y)=\Omega\Psi_E(x,y)=\sigma_x\Psi_E(-x,y)$ with energy $-E$, hence, all solutions of the Dirac equation come in pairs
with $\pm E$.

This dipole problem was recently considered in Refs.~[\onlinecite{DeMartino2014,Klopfer2014}] (the 3D Dirac equation with the
electric dipole potential was also studied some time ago in Ref.~[\onlinecite{Matrasulov1999}]). It was shown that the
point electric dipole potential (this potential is defined as the limit $R \to 0$ of the finite-size electric dipole potential
(\ref{twocenter}) with fixed electric dipole moment) accommodates towers of infinitely many bound states exhibiting a universal Efimov-like
scaling hierarchy and at least one infinite tower of bound states exists for an arbitrary dipole strength. Notice that the Schr\"{o}dinger
equation in two dimensions for the electron in the field of an electric dipole also admits a bound state for any dipole strength
\cite{Connolly2007} unlike the three-dimensional case where a bound state exists only when the dipole moment exceeds a certain critical value
(see, e.g., a discussion including a historical one in Ref.~[\onlinecite{Turner1977}]). By combining  analytical and numerical methods,
the authors of Ref.~[\onlinecite{DeMartino2014}] found that the bound states do not dive into the lower continuum because the positive and negative energy
levels first approach each other and then go away. Actually this behavior is typical for an avoided crossing \cite{Wigner1929}, which forbids
level crossing for two states with the same quantum numbers. Since the bound states do not dive into the lower continuum, the authors of
Ref.~[\onlinecite{DeMartino2014}] concluded that supercriticality is unlikely to occur in the electric dipole problem in graphene.

We reconsidered the problem of supercriticality in our paper [\onlinecite{Gorbar2015}] for the case of two oppositely charged impurities
situated at finite distance (finite-size electric dipole). By using the LCAO technique and
variational Galerkin--Kantorovich method, we showed that for sufficiently large charges of impurities the wave function of the highest
energy occupied bound state changes its localization from the negatively charged impurity to the positive one as the distance between the
impurities changes (both methods gave similar results). The necessary condition for the instability to occur is the crossing of the electron
energy levels in the field of single positively and negatively charged impurities. This migration of the electron wave function of the
supercritical electric dipole is a generalization of the familiar phenomenon of the atomic collapse of a single charged impurity with holes
emitted to infinity to the case where both electrons and holes are spontaneously created from the vacuum in bound states with two oppositely
charged impurities thus partially screening them.

\subsection{Point electric dipole in graphene}
\label{dipole-point-like}

Let us start from the case of $1/r^2$ point electric dipole
potential, which was considered in Ref.\cite{DeMartino2014}.
Far away from the nuclei, $r\gg R$, Eq.~(\ref{twocenter})
is well approximated by the point-like dipole form
\begin{equation}
\label{pointdipole}
V_d(r,\theta)= -(\hbar v_{\rm F})^{2}\frac{p \cos\theta}{r^2},
\end{equation}
where $p=ZRe^{2}/(\kappa\hbar^{2}v_{\rm F}^{2})$ is the quantity proportional to the effective electric dipole moment. The $r\to 0$ singularity
requires regularization to avoid the usual fall-to-the-center problem, see below. For nonrelativistic Schr\"{o}dinger fermions, the dipole captures
bound states only above a finite critical dipole moment in three dimensions (3D) \cite{Abramov1972,Matrasulov1999,Camblong2001,Connolly2007,Schumayer2010}.
However, a dipole binds states for arbitrarily small $p$ in the 2D Schr\"{o}dinger case \cite{Connolly2007}.

The corresponding 3D Dirac electric dipole problem has not been discussed in (3+1)-dimensional QED presumably because of the lack of heavy anti-nuclei
preventing its experimental realization in atomic physics. However, it could be directly studied using STM spectroscopy in graphene
\cite{Wang2012,Luican-Mayer2014,Wang2013}. Bound states inside the gap, $E=\pm(\Delta-\varepsilon_b)$ with binding energy $\varepsilon_b \ll \Delta$,
come in ($j,\varkappa$) towers of definite ``angular'' quantum number, $j=0,1,2,\ldots$, and parity $\varkappa=\pm$ (with $j+\varkappa\ge 0$).
The $(j,\varkappa)$ tower is only present if the dipole moment exceeds a critical value, $p>p_{j,\varkappa}$, but then contains infinitely many bound
states. Since $p_{0,+}=0$, there is at least one such tower. Bound states in the same tower obey the scaling hierarchy
\begin{equation}\label{efimovscaling}
\frac{\varepsilon_{b,{n+1}}}{\varepsilon_{b,{n}}}= e^{-2\pi/s_{j,\varkappa}}, \quad n=1,2,\ldots
\end{equation}
where for $p$ close to (but above) $p_{j,\varkappa}$,
\begin{equation}\label{ssdef}
s_{j,\varkappa}(p) \simeq
\left\{ \begin{array}{ll} \sqrt{2} p\Delta, & (j,\varkappa)=(0,+),\\
\beta \sqrt{(p-p_{j,\varkappa})\Delta},& j>0, \end{array}\right.
\end{equation}
with $\beta\approx 0.956$. Equation (\ref{efimovscaling})
agrees with the universal Efimov law for the binding energies of three identical bosons with short-ranged particle interactions \cite{Efimov1970,Braaten2006,Gogolin2008}. Numerical diagonalization of the Dirac equation in a finite disc geometry indicates that as $p$ increases,
the bound states approach $E=0$ without ever reaching it. The absence of zero modes was shown analytically in Ref.~\cite{DeMartino2014}.

For energies close to the band edge $E=-\Delta+\varepsilon_b$  with $|\varepsilon_b|\ll \Delta$, where $\varepsilon_b>0$ corresponds to
bound states inside the gap and for $p \ll R^2 \Delta/(\hbar v_{\rm F})^{2}$, the upper spinor component stays always ``small'',
$\phi \simeq \frac{\hbar v_{\rm F}}{2\Delta} e^{-i\theta}\left(i\partial_r + \frac{1}{r}\partial_\theta\right)\chi$.
The equation, which determines the lower spinor component, has the form
\begin{equation}
\label{schr}
\left( -\frac{(\hbar v_{\rm F})^{2}}{2\Delta} \nabla^2- V+\varepsilon_b \right) \chi = 0,
\end{equation}
with the 2D Laplacian $\nabla^2$.  We proceed with the potential $V=V_d$ in Eq.~(\ref{pointdipole}), where Eq.~(\ref{schr}) is solved by the ansatz
$\chi(r,\theta) = F(r) Y(\theta)$. With separation constant $\gamma$, the angular function satisfies an $\varepsilon_b$-independent Mathieu equation,
\begin{equation}
\label{angular}
\left(\frac{d^2}{d\theta^2} + \gamma-2p\Delta\cos\theta \right)Y(\theta)=0,
\end{equation}
which admits $2\pi$-periodic solutions only for characteristic values $\gamma=\gamma_{j,\varkappa}(p)$, where $\varkappa=\pm$ is the parity, i.e.,
$Y_{j,\varkappa}(-\theta)= \varkappa Y_{j,\varkappa}(\theta)$, and due to the anisotropy, $j=0,1,2,\ldots$ differs from
conventional angular momentum, with $j+\varkappa\ge 0$. Using standard notation \cite{Gradshtein-book,AbramowitzStegun},
the solutions to Eq.~(\ref{angular}) are expressed in terms of Mathieu functions ${\rm ce}_{2j}$ and ${\rm se}_{2j}$, with eigenvalues $a_{2j}$ and
$b_{2j}$, respectively,
\begin{eqnarray}\label{mathieusol}
Y_{j,+}(\theta) &=& {\rm ce}_{2j}\left(\frac{\theta}{2}, 4p\Delta
\right) ,\quad \gamma_{j,+}= \frac{1}{4}
a_{2j}(4p\Delta), \\  \nonumber
Y_{j,-}(\theta) &=& {\rm se}_{2j}\left(\frac{\theta}{2}, 4p\Delta
\right),\quad \gamma_{j,-}=\frac{1}{4} b_{2j}(4p\Delta).
\end{eqnarray}
The characteristic values are ordered as $\gamma_{0,+}<\gamma_{1,-}< \gamma_{1,+}<\gamma_{2,-}<\ldots$ for given $p$.
With $\gamma=\gamma_{j,\varkappa}(p)$, the radial equation reads
\begin{equation}\label{radial}
\left(\frac{d^2}{dr^2} + \frac{1}{r} \frac{d}{dr} - \frac{\gamma}{r^2}-
2\frac{\Delta \varepsilon_b}{(\hbar v_{\rm F})^{2}} \right) F(r) = 0.
\end{equation}
To regularize the fall-to-the-center singularity, the Dirichlet condition $F(r_0)=0$ is imposed at
a short-distance scale $r_0\approx R$ (on the level of the Dirac equation, this corresponds to vanishing radial current at $r=r_0$).
It is found that this regularization does not affect universal spectral properties such as the Efimov law (\ref{efimovscaling}).

Let us now look for bound states, $\varepsilon_b>0$. The solution of Eq.~(\ref{radial}) decaying for $r\to \infty$
is the Macdonald function $K_{\sqrt{\gamma}}\left(\sqrt{2\Delta \varepsilon_b}\ r/(\hbar v_{\rm F})\right)$ \cite{Gradshtein-book},
and the condition $F(r_0)=0$ then yields an energy quantization condition within each ($j,\varkappa$) tower. Thereby the binding energies,
$\varepsilon_{b,n,j,\varkappa}=z_{n}^2 (\hbar v_{\rm F})^{2}/(2\Delta r_0^2)$, are expressed in terms of the positive zeroes,
$z_1>z_2>\ldots>0$, of $K_{\sqrt{\gamma_{j,\varkappa}}}(z)$. Since only $K_{is}(z)$ (with imaginary order) has zeroes \cite{Gradshtein-book},
bound states require $\gamma_{j,\varkappa}(p)<0$. This condition is satisfied for $p>p_{j,\varkappa}$ with
\begin{equation}
\label{pjdef}
\gamma_{j,\varkappa}\left (p_{j,\varkappa}\right)=0.
\end{equation}
The lowest few $p_{j>0,\varkappa}$ resulting from Eq.~(\ref{pjdef}) could be found in Ref.~[\onlinecite{DeMartino2014}].
With increasing dipole moment, each time that $p$ hits a critical value $p_{j,\varkappa}$, a new infinite tower of bound
states emerges from the continuum.  Since $\gamma_{0,+}(p)<0$ for all $p$ \cite{AbramowitzStegun}, we find $p_{0,+}=0$: at least
one tower is always present. Explicit binding energies follow from the small-$z$ expansion of $K_{is}(z)$ \cite{Gradshtein-book}. With the
positive numbers $s_{j,\varkappa}(p)= \sqrt{-\gamma_{j,\varkappa}(p)}$ for $p>p_{j,\varkappa}$,  see Eq.~(\ref{ssdef}), we obtain
\begin{equation}\label{spectrum}
\varepsilon_{b,n,j,\varkappa} = \frac{2\hbar^{2}v_{\rm F}^{2}}{\Delta r_0^2}
e^{\varphi(s_{j,\varkappa})}
e^{-2\pi n/s_{j,\varkappa}},
\end{equation}
where $\varphi(s) = (2/s)\ {\rm arg} \Gamma(1+is)$. This becomes more and more accurate as $n$ increases.  For $n\to \infty$, in view of
the particle-hole symmetry, the energies accumulate near both edges, $\varepsilon_{b,n}\to 0$.  Importantly, Eq.~(\ref{spectrum}) implies
the Efimov scaling law announced in Eq.~(\ref{efimovscaling}).  This relation has its origin in the large-distance behavior of the dipole
potential, and is thus expected to be independent of short-distance regularization issues.  A similar behavior has been predicted for the
quasi-stationary resonances of a supercritical Coulomb impurity in graphene \cite{Shytov2007a,Gamayun2009}, and for 3D Schr\"{o}dinger
fermions \cite{Abramov1972,Schumayer2010}.

\subsection{Migration of the wave function in the finite dipole potential}
\label{dipole-LCAO}

In this subsection we consider the case of finite electric dipole with potential (\ref{twocenter}). Since the variables in Dirac equation with
this potential are not separable in any orthogonal coordinate system, we will utilize the LCAO method. It is convenient to work with dimensionless
quantities $h=H/\Delta$ and $\epsilon_\Delta=E/\Delta$ and use dimensionless coordinates and distances defined in units of $R_{\Delta}=\hbar
v_{\rm F}/\Delta$. We use also dimensionless coupling constant $\zeta=Ze^2/(\hbar v_{\rm F}\kappa)$.

The LCAO method was used in Sec.~\ref{2centers-LCAO} to solve the two Coulomb centers problem in graphene. Here we apply LCAO method to the dipole
problem in graphene. As to the atomic orbitals, we take the wave function of the lowest energy bound state in the field of positively charged impurity
and the wave function of the highest energy bound state for negatively charged impurity (these wave functions are related to each other by charge conjugation).

We begin our analysis with the Dirac equation for the electron in graphene with one positively charged impurity $h_{p}\Psi_p=\epsilon_{\Delta}\Psi_p$
with the Hamiltonian
\begin{equation}
h_p=-i(\sigma_{x}\partial_{x}+\sigma_{y}\partial_{y})+\sigma_{z}-\frac{\zeta}{\sqrt{r^{2}+r_{0}^{2}}}.
\end{equation}
[The Hamiltonian $h_n$ for the electron in the field of negatively charged impurity is obtained from the Hamiltonian $h_p$ by the change of the sign
of the last term in $h_p$.]

We determine numerically the energy levels by using the shooting method with regular boundary conditions at $r=0$ for the wave functions and
requiring that the wave functions decrease at infinity.
\begin{figure}[t]
	\centering
	\includegraphics[scale=0.42]{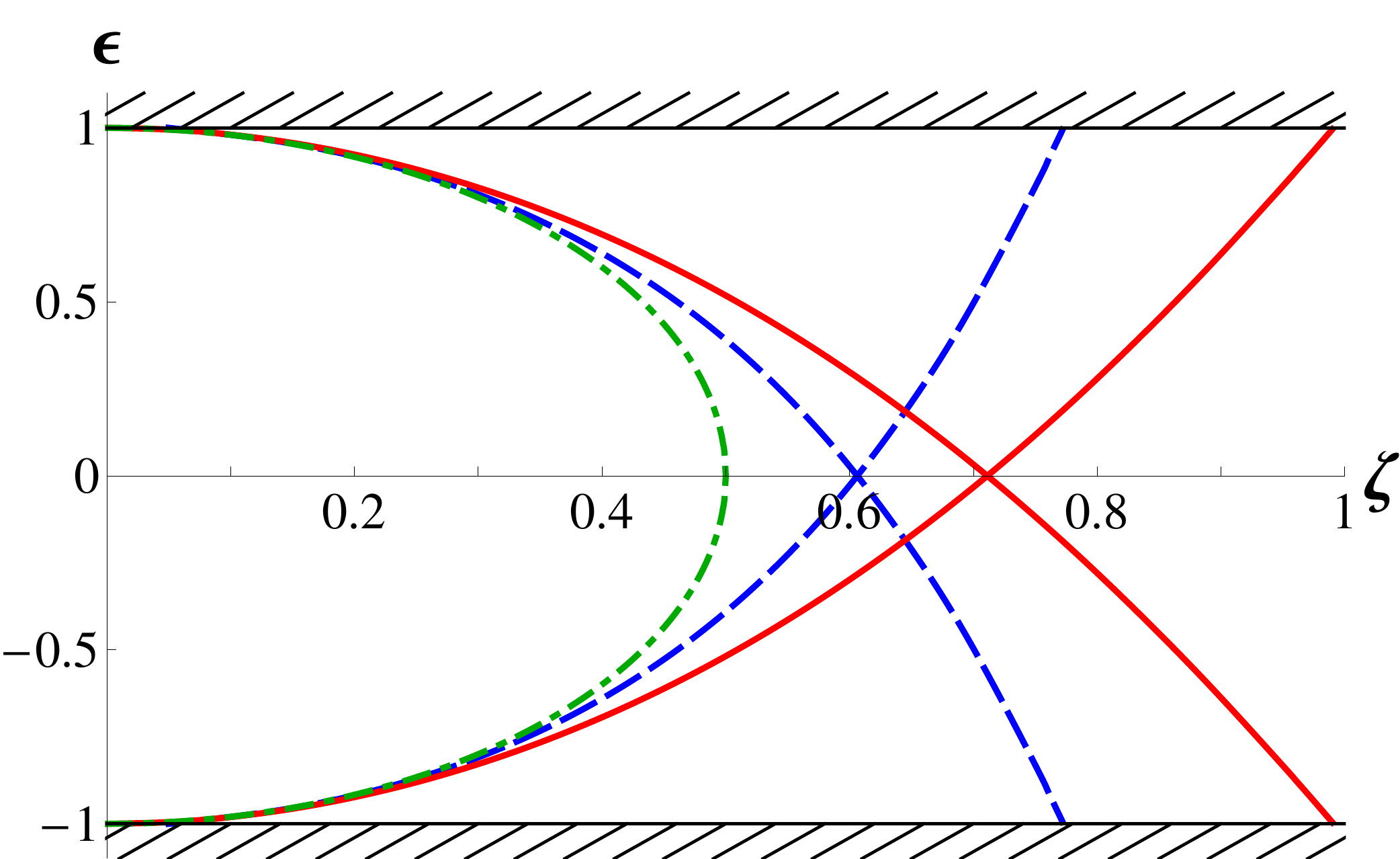}
	\caption{The energy of the electron bound state with $j=1/2$ in the regularized
		Coulomb potential with $\pm Ze$ charges  as a function of $\zeta$ for different values of the regularization parameter: $r_{0}=0.05 R_{\Delta}$
		(red solid lines), $r_{0}=0.01 R_{\Delta}$ (blue dashed lines), $r_{0}=0$ (green dash-dotted lines).}
	\label{fig1-LCAO}
\end{figure}
The energy of the lowest (highest) electron bound state with the total angular momentum $j=1/2$ in the regularized Coulomb potential with the
charge $+Ze$ ($-Ze$) is plotted in fig.\ref{fig1-LCAO} for different values of the regularization parameter $r_{0}$ as a function of $\zeta$.
The levels which descend from the upper continuum correspond to the positive charge $+Ze$ while those which are pushed from the lower continuum
and grow with $\zeta$ correspond to the negative charge $-Ze$.
These results are in accordance with calculations in Ref.\cite{Gamayun2009} (see fig.4 there) where slightly different regularization
for the one Coulomb center potential was used which admitted an analytical solution. They also reproduce qualitatively the behavior
seen directly at the tight-binding level on a honeycomb lattice \cite{Pereira2008}. For nonregularized
Coulomb potential with positive charge the lowest bound-state energy is always positive, it reaches the value $\epsilon=0$ for $\zeta=1/2$ and
becomes purely imaginary for $\zeta>1/2$ (the fall-to-center phenomenon \cite{Shytov2007a,Shytov2007b,Pereira2007,Fogler2007}). For
regularized Coulomb potential with charge $+Ze (-Ze)$, the lowest (highest) bound-state energy crosses $\epsilon_\Delta=0$ and
dives into the lower (upper) continuum at certain value of the charge. For example, for $r_{0}=0.05 R_{\Delta}$, this happens for
$\zeta\approx 1$.

Since the operator $U_{c}=\sigma_x K$ interchanges the $h_p$ and $h_n$ Hamiltonians, $U_{c}h_pU^{+}_{c}=-h_n$, the electron levels in the field
of negatively charged center described by the Hamiltonian $h_n$ are obtained by the reflection $\epsilon_\Delta\rightarrow-\epsilon_\Delta$
and intersect with the levels of the Hamiltonian $h_p$ at $\epsilon_\Delta=0$. The corresponding critical value $\zeta_c$ when this happens
will play a crucial role in the behavior of energy levels in the dipole potential because the behavior of these levels dramatically changes
depending on whether $\zeta<\zeta_c$ or $\zeta>\zeta_c$. For chosen values of the regularization parameter $r_0$ in fig.\ref{fig1-LCAO}, the
critical coupling $\zeta_c=0.6\, (r_0=0.01R_\Delta)$ and $\zeta_c=0.7\, (r_0=0.05R_\Delta)$. In general,  $\zeta_c$ increases with the increase
of $r_0$ (see Fig.\,\ref{cr1} and Eq.\,(\ref{crit-charge-on-r0})).

We are ready now to consider the Dirac equation for quasiparticles in graphene with two oppositely charged impurities. The corresponding
Hamiltonian has the form
\begin{equation}
h=-i(\sigma_{x}\partial_{x}+\sigma_{y}\partial_{y})+\sigma_{z}+
\frac{\zeta}{\sqrt{r_{n}^{2}+r_{0}^{2}}}-\frac{\zeta}{\sqrt{r_{p}^{2}+r_{0}^{2}}},
\end{equation}
where $r_{p,n}=\sqrt{(x\pm R/2)^{2}+y^2}$. We seek the wave function as a linear combination (hybridization),
\begin{equation}
\label{linear_combination}
|\Psi\rangle=v_{p}|\Psi_p\rangle+v_{n}|\Psi_n\rangle,
\end{equation}
of the wave functions $\Psi_p$ and $\Psi_n$ which are eigenstates of the Hamiltonians $h_p$ and $h_n$, respectively, with eigenvalues
$\pm\epsilon_{0}$, and $\epsilon_{0}$ is the energy of the lowest-energy electron bound state in the field of one Coulomb center with the charge
$+Ze$. Explicitly, the functions $|\Psi_p\rangle, |\Psi_n\rangle$ with the total angular momentum $j=1/2$ are given in polar coordinates by
\begin{equation}
\Psi_p=\left(\begin{array}{c}f(r_p)\\-i e^{i\theta_p}g(r_p)\end{array}\right),\quad\Psi_n=\left(\begin{array}{c}e^{-i\theta_n}g(r_n)\\
-i f(r_n)\end{array}\right),
\end{equation}
where $\exp[-i\theta_{p,n}]=(x\pm R/2-i y)/r_{p,n}$. The radial functions $f(r)$ and $g(r)$ are computed numerically in the regularized potential
of one positively charged impurity.

It is crucial for our analysis below to use the electron wave functions $|\Psi_p\rangle, |\Psi_n\rangle$ in the field of a single
regularized Coulomb center whose energies may cross zero. By making use of the wave function (\ref{linear_combination}), we project the Dirac
equation $h|\Psi\rangle=\epsilon_\Delta|\Psi\rangle$ on the states $|\Psi_p\rangle$ and $|\Psi_n\rangle$ and find the following
secular equation:
\begin{equation}
\det \left|
\begin{array}{cc}
h_{pp}-\epsilon_\Delta & h_{pn}-S\epsilon_\Delta\\
h_{np}-S\epsilon_\Delta & h_{nn}-\epsilon_\Delta
\end{array}
\right|=0,
\end{equation}
where $h_{ij}=\langle i|h|j\rangle$, $\langle i|j\rangle=\delta_{ij}$, $i,j=\Psi_p,\Psi_n$,
$S=\langle \Psi_p|\Psi_n\rangle=\langle \Psi_n|\Psi_p\rangle$, $\langle \Psi|\Psi\rangle=1$.
It is easy to see that the overlap integral
\begin{eqnarray}
S=\int dxdy \left(f(r_p)g(r_n)\frac{x-R/2}{r_p}+f(r_n)g(r_p)\frac{x+R/2}{r_n}\right)
\nonumber
\end{eqnarray}
vanishes after changing $x \to -x$ in the first term of the brackets.

\begin{figure}[ht]
	\centering
	\includegraphics[scale=0.32]{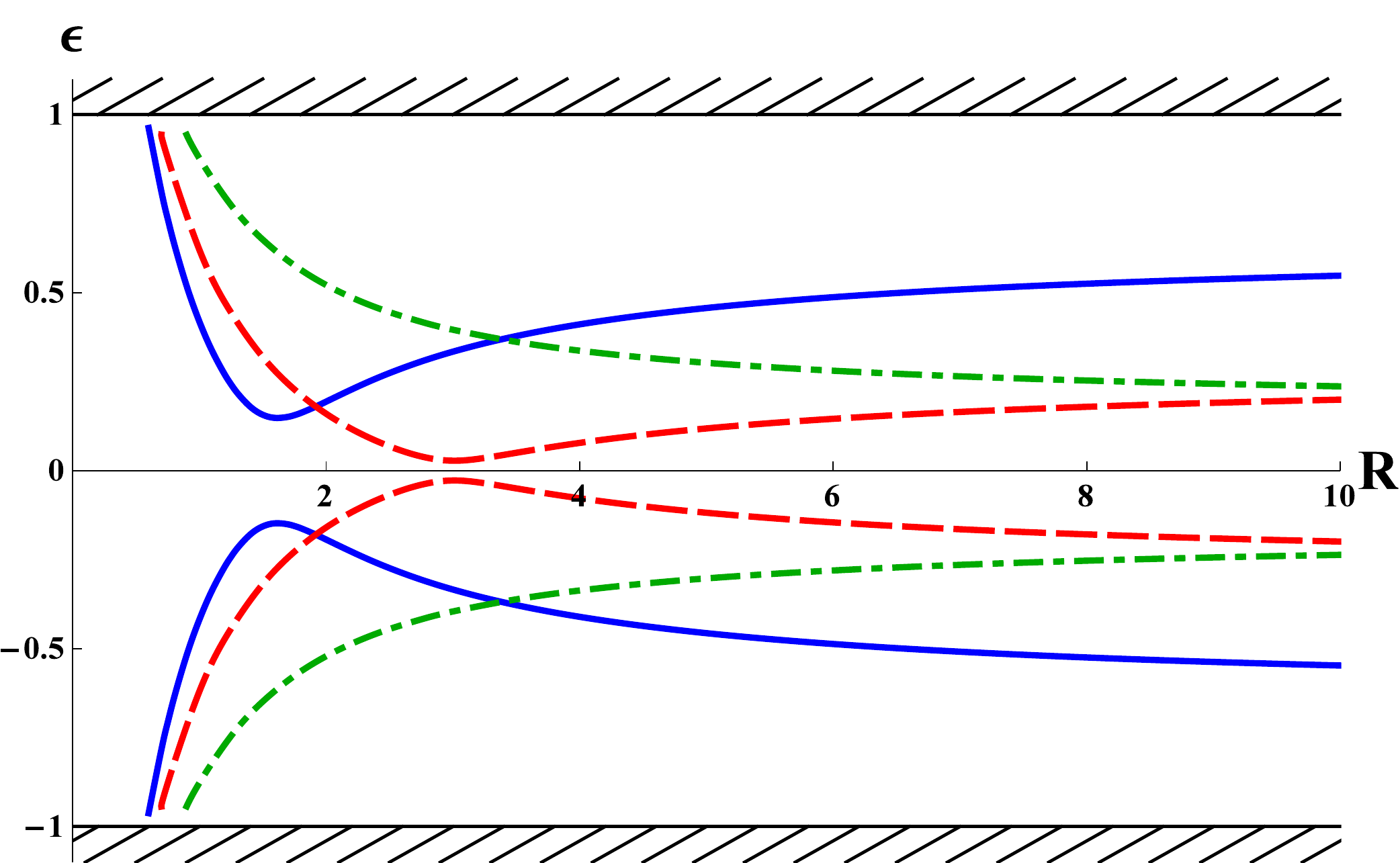}
	\caption{The energy of the bound state levels for $r_0=0.05R_\Delta$ as a function of
		distance in the LCAO method: $\zeta=0.65$ (green dash-dotted lines) and $\zeta=0.8$
		(red dashed lines), $\zeta=0.9$ (blue solid lines).}
	\label{fig2-LCAO}
\end{figure}

In order to calculate the coefficients $h_{ij}$, it is convenient to represent the Hamiltonian in the form
$h=h_p+{\zeta}/{\sqrt{r_{n}^{2}+r_{0}^{2}}}=h_n-{\zeta}/{\sqrt{r_{p}^{2}+r_{0}^{2}}}$. Then we obtain
\begin{equation}
h_{pp}=\epsilon_{0}+\langle \Psi_p|\frac{\zeta}{\sqrt{r_{n}^{2}+r_{0}^{2}}}|\Psi_p\rangle
=\epsilon_{0}+\zeta C=-h_{nn},\quad h_{pn}=-\langle \Psi_p|\frac{\zeta}{\sqrt{r_{p}^{2}+r_{0}^{2}}}|\Psi_n\rangle=-\zeta A=h_{np}.
\label{h-ij}
\end{equation}
We find that the Coulomb integral $C$ equals
\begin{equation}
C=\langle p|\frac{1}{\sqrt{r_{n}^{2}+r_{0}^{2}}}|p\rangle=
\int\limits_{0}^{\infty}\frac{4 r dr}{\sqrt{(r+R)^{2}+r_{0}^{2}}}{\rm K}\left(\sqrt{\frac{4r R}{(r+R)^{2}+r_{0}^{2}}}\right)
\left(f^{2}(r)+g^{2}(r)\right),
\end{equation}
where ${\rm K}(k)$ is the complete elliptic integral of the first kind. We note that the Coulomb integral $C$ is positive definite
and monotonously decreases with increasing $R$. Further, the resonance integral $A$ equals
\begin{equation}
A=\langle p|\frac{1}{\sqrt{r_{p}^{2}+r_{0}^{2}}}|n\rangle=
2\int\limits_{-\infty}^{\infty}dx\int\limits_{0}^{\infty}dy \,f(r_{p})g(r_{n})\frac{x- R/2}{r_{n}}\left(\frac{1}{\sqrt{r_{p}^{2}+r_{0}^{2}}}
-\frac{1}{\sqrt{r_{n}^{2}+r_{0}^{2}}}\right).
\end{equation}
The Coulomb integral $C$ and resonance integral $A$  can be computed numerically with the functions $f$ and $g$ found for an isolated
impurity problem. At small $R\sim0$ we have that $C=const$ while $A\rightarrow0$. Asymptotically at large $R$ they behave
as $C\simeq 1/R$ and $A\sim\exp(-\sqrt{1-\epsilon_0^2} R)$.

Finally, we obtain the energy levels
\begin{equation}
\label{spectrum-LCAO}
\epsilon_\Delta=\pm\sqrt{(h_{pp})^{2}+(h_{pn})^{2}}=\pm\sqrt{(\epsilon_{0}+\zeta C)^{2}+\zeta^{2}A^{2}},
\end{equation}
which are obviously symmetric with respect to the replacement $\epsilon_\Delta \to -\epsilon_\Delta$ in accord with the charge conjugation
symmetry of the problem under consideration. We note that the energy levels never cross in agreement with the avoided crossing theorem
\cite{Wigner1929}. Since the Coulomb and resonance integrals, $C$ and $A$, tend to zero as $R\rightarrow \infty$, the energy of the system
for large distances between the impurities tends to $\epsilon_\Delta\rightarrow\pm |\epsilon_{0}|$ as expected. The coefficients of the wave
function of the negative energy level are
given by
\begin{eqnarray}
v_p&=&-\frac{h_{pn}}{\sqrt{(h_{np})^{2}+(h_{pp}+\sqrt{(h_{pp})^{2}+(h_{pn})^{2}})^{2}}},\label{coef-p}\\
v_n&=&\frac{h_{pp}+\sqrt{(h_{pp})^{2}+(h_{pn})^{2}}}{\sqrt{(h_{np})^{2}+(h_{pp}+\sqrt{(h_{pp})^{2}
			+(h_{pn})^{2}})^{2}}}.\label{coef-n}
\end{eqnarray}

We plot the energy levels of the system for $r_{0}=0.05 R_{\Delta}$ in Fig.\ref{fig2-LCAO} as functions
of $R$ for $\zeta=0.65$, $\zeta=0.8$ and $\zeta=0.9$. In the first case, we have $\zeta<\zeta_c=0.7$ and the bound state levels
monotonously converge to each other as $R$ increases and never cross. For the  couplings $\zeta=0.8$ and $\zeta=0.9$ which are
larger than $\zeta_c=0.7$, their behavior is no longer monotonous. For small $R$, the levels converge like in the previous case.
However, after the maximal convergence of the levels they go away with the subsequent increase of $R$. This behavior is typical
for the avoided crossing \cite{Wigner1929}.

Eq.(\ref{spectrum-LCAO}) and the facts that $C$ and $A$ monotonously depend on $R$ and $C\geq 0$ imply that the level repulsion
can take place only for $\epsilon_{0}<0$ and the energy levels converge most closely for $\zeta C=|\epsilon_{0}|$ from which we
can determine the corresponding distance $R_{m}$. Exactly at this distance $R_{m}$ we have $h_{pp}=h_{nn}=0$ from
Eq.(\ref{h-ij}), hence $v_p=v_n=1/\sqrt{2}$ and, consequently, the probability to find the electron near the positively and
negatively charged impurities is the same. Furthermore, for $\zeta C \approx |\epsilon_{0}|$, the difference of the coefficients
$v_p$ and $v_n$ squared equals
\begin{equation}
v_n^{2}-v_p^{2}=\frac{2h_{pp}\left(h_{pp}+\sqrt{(h_{pp})^{2}+(h_{pn})^{2}}\right)}
{(h_{np})^{2}+(h_{pp}+\sqrt{(h_{pp})^{2}+(h_{pn})^{2}})^{2}}\sim{\rm sign}(h_{pp})={\rm sign}
(\epsilon_0 +\zeta C).
\label{difference}
\end{equation}

The following qualitative picture appears for $\epsilon_0 < 0$. For small $R$, $h_{pn}=h_{np}\sim0$, hence $|v_n| \gg |v_p|$ and,
therefore, the electron wave function of the negative energy level in the LCAO method is localized mainly on the negatively charged
impurity. Although this result seems to be counter-intuitive, it is quite natural. The point is that the energy spectrum of the system
for $R=0$ is composed of the upper and lower continua. Since the chemical potential in neutral graphene is zero, the electron states of
the lower continuum are occupied. For small $R$, the positively charged impurity produces electron bound states which descend from the upper
continuum. Obviously, there are also charged conjugated states localized near the negatively charged impurity, which rise from the lower continuum
as $R$ increases. These states are occupied for sufficiently small $R$. Since $|v_n|=|v_p|$ at the point of maximal convergence $R=R_{m}$,
the probability to find the electron near the negatively and positively charged impurities is then equal. As the distance between impurities
$R$ increases further, the difference of the square moduli (\ref{difference}) changes sign because the Coulomb integral $C$ decreases with
increasing $R$. This means that the electron wave function changes its localization to the positively charged impurity. This change of the
wave function localization (the relocalization or "migration" effect) is explicitly shown in Fig.\ref{coef-lcao} for $\zeta=0.85$ when
$\epsilon_{0}=-0.453<0$. The value of $R_{m}$, in general, depends on $r_0$ and on the value of the coupling constant $\zeta$. As the coupling
increases, $R_{m}$ decreases. A similar phenomenon of the change of localization of wave function takes place in the fission of quarkonium
resonances consisting of the heavy quark and antiquark \cite{Greiner1985}.
\begin{figure}[ht]
	\centering
	\includegraphics[scale=0.35]{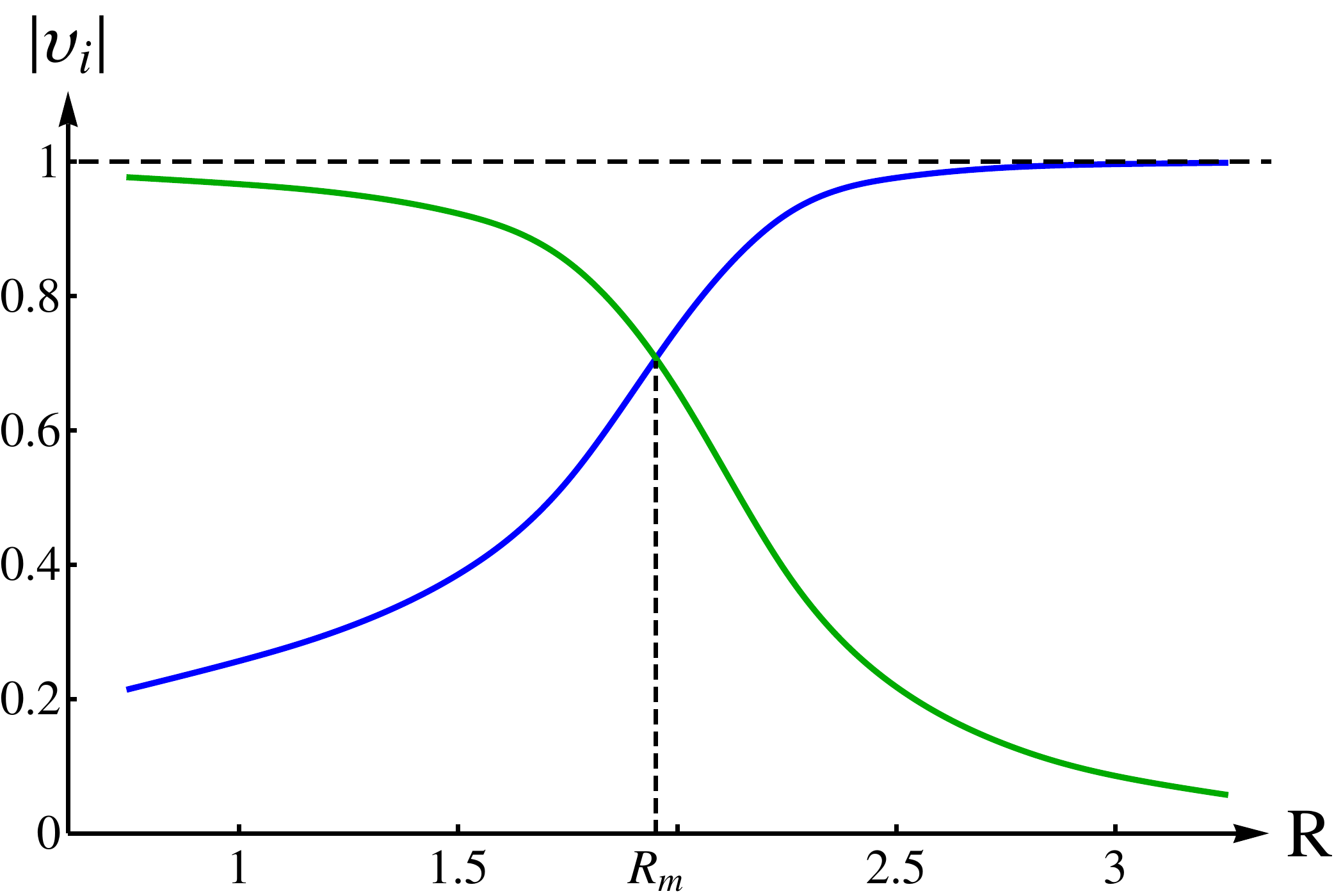}
	\caption{The dependence of coefficients (\ref{coef-p})-(\ref{coef-n}) on the distance $R$ between the impurities for
the negative-energy branch of the spectrum for $\zeta=0.85$: $|v_{p}|$ -- decreasing line, $|v_{n}|$ -- increasing line. The point of
the change of localization $R_{m}$ is marked by dashed line. }
	\label{coef-lcao}
\end{figure}
It is clear that if $\epsilon_{0}>0$, then nothing interesting happens. Indeed, since $h_{pp}>0$ for $\epsilon_{0}>0$, we find that
$|v_n|>|v_p|$ for any distance $R$ between the impurities. Therefore, the wave function is always localized on the negatively charged impurity
and, consequently, the wave function of the highest occupied state does not change its localization. Note that the behavior of the energy levels
in this case given by green dash-dotted lines in Fig.\ref{fig2-LCAO} as a function of $R$ is monotonous unlike the case where $\epsilon_0$ is
negative.

In Ref.~\cite{Gorbar2015a} we considered the numerical Galerkin--Kantorovich variational method and obtained qualitatively similar results.

\section{Coulomb center problem in bilayer graphene}
\label{section:bilayer}

The Bernal stacked bilayer graphene forms a very interesting two-dimensional physical system whose electron excitations are described by the
Hamiltonian with non-trivial chiral properties \cite{McCann2006,Novoselov2006}. These chiral fermions do not have  analogues in high energy physics
because their energy spectrum $E \sim \pm|\mathbf{p}|^2$ is parabolic in the two band model, like in non-relativistic systems described by the
Schr\"{o}dinger equation. Still since there are positive and negative energy bands related through the charge conjugation, this suggests
that the instability connected with diving of the lowest energy electron bound state into the lower band may take place in gapped bilayer graphene
too. In addition, it was shown in Ref.~[\onlinecite{Shytov-arXiv-2015}] that the chirality of quasiparticles in gapless bilayer graphene results
in cloaked bound states that do not hybridize with the hole continuum.

Actually, there is more to this. The parabolic spectrum in bilayer graphene is more soft than in monolayer graphene. Therefore, the electron-electron
interactions are effectively enhanced and may open a gap connected with the condensation of electron-hole pairs in bilayer graphene. This
conclusion agrees with the experimental data. While no gap is observed in monolayer graphene at the neutrality point, a small gap 2 $\mbox{meV}$ is
realized in bilayer graphene \cite{Martin2010,Weitz2010,Freitag2012} in the absence of external electromagnetic fields. A much larger gap
($\sim42$ meV) is observed in high-mobility ABC-stacked trilayer graphene \cite{Lee2014} where electron excitations have a softer dispersion
$E(\mathbf{p})\sim \pm|\mathbf{p}|^3$. As we mentioned above, the supercritical instability in the Coulomb center is a precursor of the excitonic
instability and gap opening in the electron spectrum of a many-body problem. Reversing this argument and taking into account the fact that bilayer
graphene is gapped due to the electron-electron interactions, one may expect that the supercritical instability for the electron in the field of
Coulomb center should take place in bilayer graphene.

The following qualitative consideration is important for the analysis of the supercritical instability in bilayer graphene. Since the quasiparticle
kinetic energy scales like $1/r^2$ with distance from the charged impurity in bilayer graphene and is negligible compared to the Coulomb interaction
$1/r$, this suggests the absence of the critical charge for the formation of a bound state in this material. On the other hand, since the kinetic
energy  at small distances is larger than the Coulomb interaction energy, this implies that the fall-to-center should not take place in bilayer graphene.
Therefore, {\it a priori}, one may expect that the supercritical instability in bilayer graphene should not be necessarily related to the phenomenon of
the fall-to-center. The situation is different in monolayer graphene where both the kinetic energy and Coulomb interaction scale equally with distance.
In fact, this equal scaling is the physical reason why the supercritical instability in monolayer graphene is related to the phenomenon
of the fall-to-center where the electron wave function shrinks towards the impurity as its charge increases.

Clearly, the above heuristic reasoning is based on the scaling in the effective low-energy theory in bilayer graphene. However, the parabolic spectrum
in bilayer graphene \cite{McCann2006} is valid only up to momenta $|\mathbf{p}| \approx \gamma_1/(2v_{\rm F})$, where $\gamma_1=0.39\,\mbox{eV}$ is the
interlayer hopping amplitude. For larger momenta, the energy dispersion in linear as in the monolayer graphene. Therefore, it is possible, in principle,
that as the charge of impurity increases the wave function of the electron bound state becomes more localized in the vicinity of the impurity such that
the parabolic energy spectrum does not apply. If this happens, then the supercritical instability in bilayer graphene would proceed like in monolayer
graphene and would be connected with the phenomenon of the fall-to-center. Actually, this scenario was advocated in Ref.[\onlinecite{Kolomeisky2016}],
where it was argued by using the semiclassical approach that the supercritical instability is an``ultra-relativistic'' effect in gapless bilayer graphene
and, therefore, the value of the critical charge is the same as in monolayer graphene.

In order to see whether the supercritical instability is indeed related to the atomic collapse in bilayer graphene, the three of us studied in
Ref.~[\onlinecite{Oriekhov2017}] the electron bound states in the field of a charged impurity in bilayer graphene by using the two continuum models:
the low-energy two-band model as well as the four-band model. The main principal advantage of the four-band model for the study of supercritical instability
in bilayer graphene is that it takes into account the evolution of
the electron energy dispersion from the low-energy quadratic to high energy linear energy dispersion. Since the trigonal
warping is small in bilayer graphene, it could be neglected. On the other hand, the screening effects play a very essential role in view of the
finite density of states at zero energy in bilayer graphene and should be taken into account. It will be shown that while the lowest energy
bound state in the Coulomb center in gapped bilayer graphene indeed dives into the lower continuum, the wave function of the electron bound
state does not shrink toward the impurity as its charge increases. This means that the supercritical instability in bilayer graphene is not
related to the phenomenon of the fall-to-center.

\subsection{Two-band model}
\label{bilayer-two-bands}

The two-band Hamiltonian, which describes low-energy  electron excitations in gapped bilayer graphene in the field of an impurity with charge $Z e$,
reads \cite{McCann2006}
\begin{equation}
H=\frac{v^2_F}{\gamma_1}\left(\begin{array}{cc}
0&(p_{-})^2\\
(p_{+})^2&0
\end{array}\right)+\Delta\left(\begin{array}{cc}
1&0\\
0&-1
\end{array}\right)+V(r),\quad V(r)=-\frac{Ze^{2}}{\kappa\,r},
\label{Hamiltonian-2band}
\end{equation}
where $p_\pm=p_x\pm i p_y$ and $\mathbf{p}=-i\hbar\boldsymbol{\nabla}$ is the two-dimensional momentum operator, $\gamma_1 \approx 0.39~\mbox{eV}$ is
the strongest interlayer coupling between pairs of orbitals that lie directly below and above each other, and $\kappa$ is the dielectric
constant which we choose equal to  4 in our analysis.

The energy spectrum of the free part of Hamiltonian (\ref{Hamiltonian-2band}) is given by $E({\bf p})=\pm\sqrt{({\bf p}^2/2m^{*})^2+\Delta^2}$,
so that we have a gap $2\Delta$ between the lower and upper continua (the valence and conductivity bands), where
$m^{*}=\gamma_1/2v_{\rm F}^2\approx0.054m_e$ is the quasiparticle mass and $m_e$ is the mass of the electron. Although the kinetic part of
Hamiltonian (\ref{Hamiltonian-2band}) has the same matrix structure as in monolayer graphene leading to chirality and the particle-hole symmetry
characteristic of relativistic systems, the dispersion relation for quasiparticles in bilayer graphene is quadratic like in non-relativistic systems.
Thus, the low-energy Hamiltonian (\ref{Hamiltonian-2band}) non-trivially combines relativistic and non-relativistic properties.
Clearly, while a regularized potential is crucially needed for the study of a charge instability in monolayer graphene, it suffices to
use the standard potential $\sim 1/r$ for the two-band model of bilayer graphene with the parabolic dispersion of quasiparticles where the kinetic
energy dominates over the potential one at small distance.

As we discussed in the Introduction and Sec.\ref{section:monolayer-graphene}, there is a critical charge of impurity in
monolayer graphene above which the atomic collapse occurs. It is interesting what happens in bilayer graphene in view of the quadratic dispersion
relation in this material. Although the two-band model is applicable only up to the energies of order
$\gamma_1/4$ when the next band becomes important, it still makes sense to study how the supercritical Coulomb center instability is realized in
the low-energy effective two-band model. We consider this problem in this section and discuss briefly the results obtained in the
four-band model in the next subsection.

It is convenient to define the coupling constant as $\xi=\frac{Z\alpha_g}{\kappa}\sqrt{\frac{\gamma_{1}}{\Delta}}$ and express distances in terms
of a characteristic length $\lambda_{\Delta}={\hbar v_{\rm F}}/{\sqrt{\gamma_{1}\Delta}}$ and wave vectors in its inverse, i.e.,
$\mathbf{\tilde{k}}=\mathbf{k}\,\lambda_{\Delta}$ (in what follows we will omit tilde over dimensionless momenta). The eigenstates of
Hamiltonian (\ref{Hamiltonian-2band}) for $\Psi^T=(\chi,\varphi)$ are determined in momentum space by the following system of equations:
\begin{equation}
\label{35}
\left\{
\begin{array}{c}
{(k_{x}-ik_{y})^2\varphi\left(\mathbf{k}\right)}+{(1-\epsilon_\Delta)\chi\left(\mathbf{k}\right)}+
{\int\!\!\frac{d^2q}{(2\pi)^{2}}\,{{\chi{\left(\mathbf{q}\right)}}V_{\rm eff}\left(\mathbf{k}-\mathbf{q}\right)}} = 0,\\
{(k_{x}+ik_{y})^2\chi\left(\mathbf{k}\right)}-{(1+\epsilon_\Delta)\varphi\left(\mathbf{k}\right)}
+{\int\!\!\frac{d^2q}{(2\pi)^{2}}\,{{\varphi{\left(\mathbf{q}\right)}}V_{\rm eff}
		\left(\mathbf{k}-\mathbf{q}\right)}} = 0,
\end{array}
\right.
\end{equation}
where $\epsilon_\Delta=E/\Delta$ and $V_{\rm eff}$ describes the screened potential of impurity with charge $Ze$.
It is defined by an analog of the Poisson equation of the form
\begin{equation}
\label{Poisson_eq-bilayer}
\sqrt{-\Delta_{2D}}V_{\rm eff}(\mathbf{x})=-2\pi\xi \delta^{(2)}(\mathbf{x})
-\frac{4\pi\alpha_g}{\kappa}\int\!\!d^{2}\mathbf{y}\,\Pi(\mathbf{x}-\mathbf{y})V_{\rm eff}(\mathbf{y}),
\end{equation}
where $\Pi(\mathbf{x})$ is the static polarization function in bilayer graphene. 
In momentum space, we easily find the following solution to Eq.~(\ref{Poisson_eq-bilayer}):
\begin{equation}
\label{Vtot1}
V_{\rm eff}(q)=-\frac{2\pi\xi}{q+\frac{4\pi\alpha_g}{\kappa}\Pi(q)} \equiv -2\pi\xi
\left(\frac{1}{q}-\delta V(q)\right),
\end{equation}
where
\begin{equation}
\delta V(q)=\frac{4\pi\alpha_g}{\kappa}\frac{\Pi(q)}{q\left[q+\frac{4\pi\alpha_g}{\kappa}\Pi(q)\right]}
\label{correction-interaction}
\end{equation}
is the correction to the Coulomb interaction due to the screening effects. The one-loop polarization function as an integral
over momentum was derived in Ref.[\onlinecite{Nandkishore2010}] and equals
\begin{equation}
\Pi(\mathbf{q})=\sqrt{\frac{\gamma_1}{\Delta}}\int \frac{d^2 k}{(2\pi)^2} \frac{2(\epsilon_{-}\epsilon_{+}-1)-(\mathbf{k}
	-\frac{\mathbf{q}}{2})_{+}^{2}(\mathbf{k}
	+\frac{\mathbf{q}}{2})_{-}^{2}-(\mathbf{k}+\frac{\mathbf{q}}{2})_{+}^{2}(\mathbf{k}
	-\frac{\mathbf{q}}{2})_{-}^{2}}{\epsilon_{-}\epsilon_{+}(\epsilon_{-}
	+\epsilon_{+})},\quad   \epsilon_{\pm}=\sqrt{1+\left|\mathbf{k}\pm\frac{\mathbf{q}}{2}\right|^{4}}.
\label{exact_polariz}
\end{equation}
For $q \ll 1$, the polarization function is proportional to $q^{2}$ and tends to zero for $q\to 0$, therefore, $\delta V(q)$ is
not singular at $q=0$. For $q \gg 1$, $\Pi(q)=\sqrt{\gamma_{1}/\Delta}\ln2/\pi$.

In polar coordinates, the angle and the absolute value of momentum variables can be separated. In order to do this, we note that
the total $z$ component of the pseudospin-orbital momentum commutes with the Hamiltonian (\ref{Hamiltonian-2band})
\begin{equation}
J_z=L_z+S_z=-i\hbar\frac{\partial}{\partial\theta}+\hbar\sigma_z,\quad\quad [J_z,\hat{H}]=0.
\end{equation}
Thus, we seek the spinor function in the form
\begin{equation}
\label{spinor-function}
\Psi=\left(\begin{array}{c}
\chi(\mathbf{k})\\
\varphi(\mathbf{k})
\end{array}\right)=\left(\begin{array}{c}
a_j(k)e^{i(j-1)\theta}\\
b_j(k)e^{i(j+1)\theta}
\end{array}\right),
\end{equation}
where $j$ is the total angular momentum which is integer in bilayer graphene. Substituting the spinor (\ref{spinor-function}) into the system (\ref{35}) and integrating over angle, we obtain
\begin{equation}
\label{system}
\left\{
\begin{array}{c}
k^{2}b_j(k) + a_j(k) - {\frac{\xi}{2\pi}}{\int\limits_{0}^{\Lambda}dq q\,a_j(q) [K_{j-1}(k,q)-
	\delta V_{j-1}(k,q)]} = \epsilon a_j(k),\\
k^{2}a_j(k)-b_j(k)-{\frac{\xi}{2\pi}}{\int\limits_{0}^{\Lambda}dq q\,b_j(q) [K_{j+1}(k,q)-
	\delta V_{j+1}(k,q)]} = \epsilon b_j(k),
\end{array}
\right.
\end{equation}
where
\begin{equation}
K_{j}(k, q)=\int\limits_{0}^{2\pi} \frac{d\theta\cos(j\theta)}{\sqrt{k^2+q^2-2kq \cos\theta}}
= \frac{2}{\sqrt{kq}}Q_{|j|-1/2}\left(\frac{k^2+q^2}{2kq}\right),
\label{kernel}
\end{equation}
\begin{equation}
\delta V_{j}(k, q)=\int\limits_{0}^{2\pi}d\theta\,\delta V\left(\sqrt{k^2+q^2-2kq \cos\theta}\right)\,
\cos(j\theta).
\end{equation}
Here $Q_\nu(z)$ is the Legendre function of the second kind. The kernels $K_{j}(k,q)$ can be expressed in terms of the full elliptic integrals
of the first and second kind. Unfortunately, it is not possible to find analytically a solution to the above system
of equations. Its numerical solutions will be given in the next subsection. However, in Ref.~[\onlinecite{Oriekhov2017}] the variational method
was applied in order to have an analytic insight into the problem. It is based on the squared Schr\"{o}dinger equation in coordinate space. The
main result is that the critical charge of impurity $Z_{\rm cr}$ for the unscreened Coulomb potential tends to zero as $\sqrt{\Delta}$ as
$\Delta\to 0$, while in the case of the screened interaction potential it tends to a finite value for $\Delta=0$.

\subsection{Numerical results}
\label{bilayer-numerical}

In order to find a numerical solution to the system of equations (\ref{system}), we split the momentum interval $(0,\Lambda)$ in $N$ equal
intervals and approximate the integrals by the sums of values of integrands at the ends of intervals multiplied by the weight function $w_i$ of
the Newton-Cotes formula of the fifth order
\begin{equation}
{\int\limits_{0}^{\Lambda}dq q [K_{j}(k,q)-\delta V_{j}(k,q)]b_j(q)}=\sum_{i=0}^{N}q_{i} w_{i} [K_{j}(k,q_{i})-\delta V_{j}(k,q_{i})]b_j(q_{i}).
\label{kernelsum}
\end{equation}
The kernel $K_{j}(k,q)$ defined in Eq.~(\ref{kernel}) has a logarithmic singularity at $q=k$. In order to deal with this singularity, we use
the following regularization for the first term in the square brackets on the left-hand side of Eq.(\ref{kernelsum}):
\begin{equation}
\label{reg}
{\int\limits_{0}^{\Lambda}dq q K_{j}(k,q)b_j(q)}=\sum_{i=0}^{N}q_{i} w_{i} K_{j}(k,q_{i})[b_j(q_{i})-b_j(k)]
+b_j(k){\int\limits_{0}^{\Lambda}dq q K_{j}(k,q)}.
\end{equation}
The last integral in Eq.(\ref{reg}) is not singular and could be expressed through the elliptic integrals and generalized hypergeometric
functions by using formulas from Sec.~5.11 in Ref.[\onlinecite{Gradshtein-book}]. For $j=0,1,2$, and $3$, the corresponding expressions are
listed in Ref.~[\onlinecite{Oriekhov2017}].

The quadrature step $h$ corresponds to the inverse of the size of a graphene disc, therefore, the energy spectrum of the system is discrete and
the upper and lower continua existing in an infinite system appear now as the sets of closely situated discrete levels with the distance between
them proportional to $h$.  The energy levels of the corresponding upper and lower quasicontinua drift very slowly as the impurity charge
increases. On the other hand, the bound levels inside the band gap shift towards the lower quasicontinuum much faster. Therefore, there exists a
critical value of the impurity charge when the lowest energy bound state approaches the highest energy state of the lower quasicontinuum. We
define this value as the critical charge. For different values of total angular momentum $j$, we find the different
values of the critical charge (see the lower panel in Fig.\ref{FH_bilayer_spectrum_m-2}). The minimal value of the critical charge is obtained
for $j=1$. Therefore, all numerical computations in this section
are performed for this value of the total momentum.

\begin{figure}[ht]
	\centering
	\includegraphics[scale = 0.45]{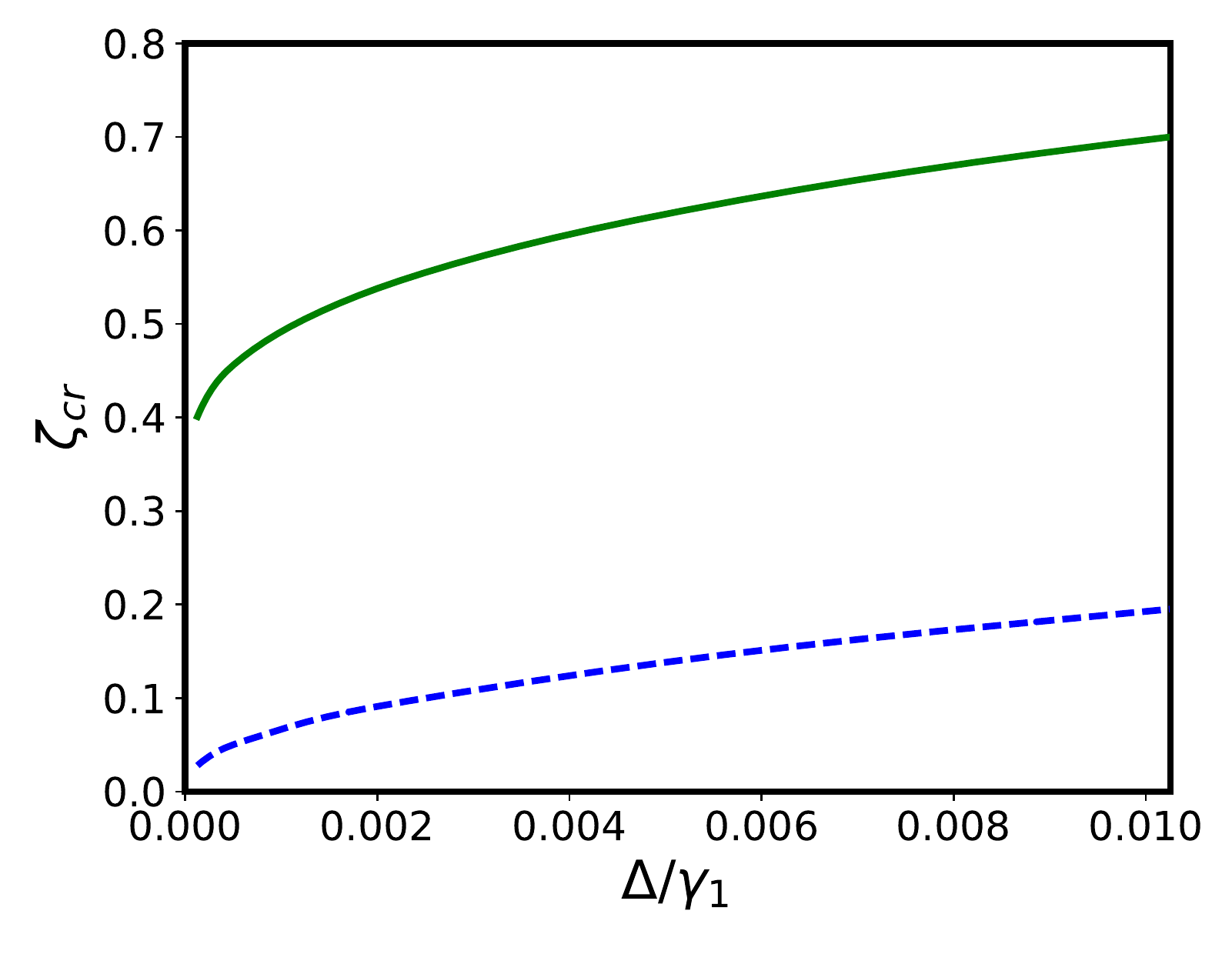}
	\caption{The critical coupling constant as a function of dimensionless gap $\Delta/\gamma_1$ with (green solid line) and
		without (blue dashed line) the screening effects taken into account.}
	\label{spectrum_fig}
\end{figure}

The dependence of the critical coupling constant on gap $\Delta$ is plotted in Fig.~\ref{spectrum_fig}
in the cases where the screening effects are absent and taken into account. Clearly, the critical charge decreases in both cases as gap
decreases. According to our numerical calculations, the critical charge in the absence of the screening effects is very well approximated by the
function $\zeta^{fit}_{cr}=1.88\sqrt{\Delta/\gamma_{1}}$. Such a gap dependence excellently agrees with
the conclusion obtained from the variational method in Ref.~[\onlinecite{Oriekhov2017}]. If the screening effects are taken into account,
then the critical coupling constant does not tend to zero as $\Delta \to 0$. The results of the numerical solution to system
(\ref{system}) are quite similar to those obtained in Ref.~[\onlinecite{Oriekhov2017}] by using the variational method and can be fit by the
function
\begin{equation}
\zeta^{fit}_{cr}=0.36+4.32\sqrt{\Delta/\gamma_{1}}.
\label{eq:zeta_screened_fitting}	
\end{equation}

In order to quantify the localization properties of the wave function of the electron bound state in the near-critical regime and
check the consistency of the use of the two-band model for the study of the supercritical instability in bilayer
graphene, we plot for $Z \approx Z_{cr}$ in the right panel of Fig.~\ref{FH_bilayer_spectrum_m-2} the square of the wave function in the
two-band model in momentum representation $W(k)=2\pi k[a_{j}^2(k)+b_{j}^2(k)]/N$ multiplied by the weight factor $2\pi k$, where
$N$ is the normalization constant $N=\int_{0}^{\Lambda}2\pi k [a_j^2(k)+b_j^2(k)]\,dk$. Obviously, if the wave function is localized in
momentum space in the region of momenta $k \ll \Lambda$, then the use of the low-energy model for the description of the supercritical
phenomena is consistent. According to the right panel in Fig.~\ref{FH_bilayer_spectrum_m-2}, the maximum of the wave function corresponds
to $k \approx 0.075$. This value is 10 times less than the cutoff of the two-band model. This result suggests that the low-energy two-band
model consistently describes the supercritical behavior in bilayer graphene.

Although our results show that the analysis of the supercritical instability is consistent in the two-band model, it is still necessary to study
the supercritical instability in the four-band model of bilayer graphene. The point is that the quadratic energy dispersion in the two-band
model is replaced by the linear energy dispersion as in monolayer graphene for momenta larger than $\gamma_1/(4v_{\rm F})$. Therefore, it
is possible, in principle, that the supercritical instability will be affected by the electron dynamics at short distances. The four-band model
whose energy dispersion smoothly interpolates between the low-energy quadratic and high-energy linear in momentum energy dispersion allows us to
investigate whether the conclusions made in the present section survive in the four-band model.

In Ref.~[\onlinecite{Oriekhov2017}] we studied the Coulomb center problem in bilayer graphene in the four-band model.  The right panel of
Fig.~\ref{FH_bilayer_spectrum_m-2} compares the localization properties of the wave function of the electron bound state in the two- and four-band
models. Clearly, the wave functions are localized in the same domain of momentum space in these models. This proves that the low-energy two-band
 model gives fairly accurate results for the Coulomb center problem in the case of small gaps.
\begin{figure}[ht]
	\centering
	\includegraphics[scale = 0.4]{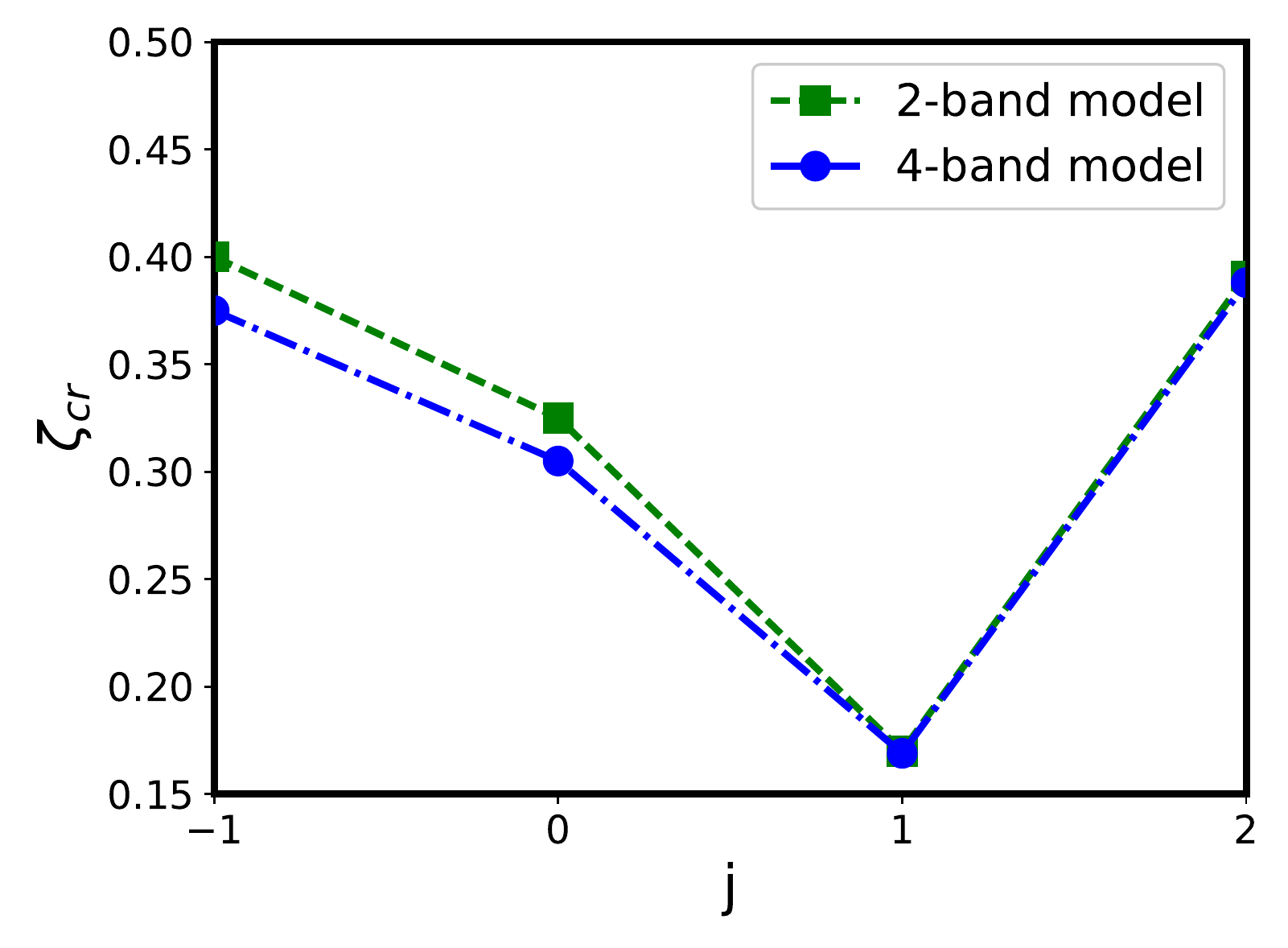}
	\includegraphics[scale = 0.4]{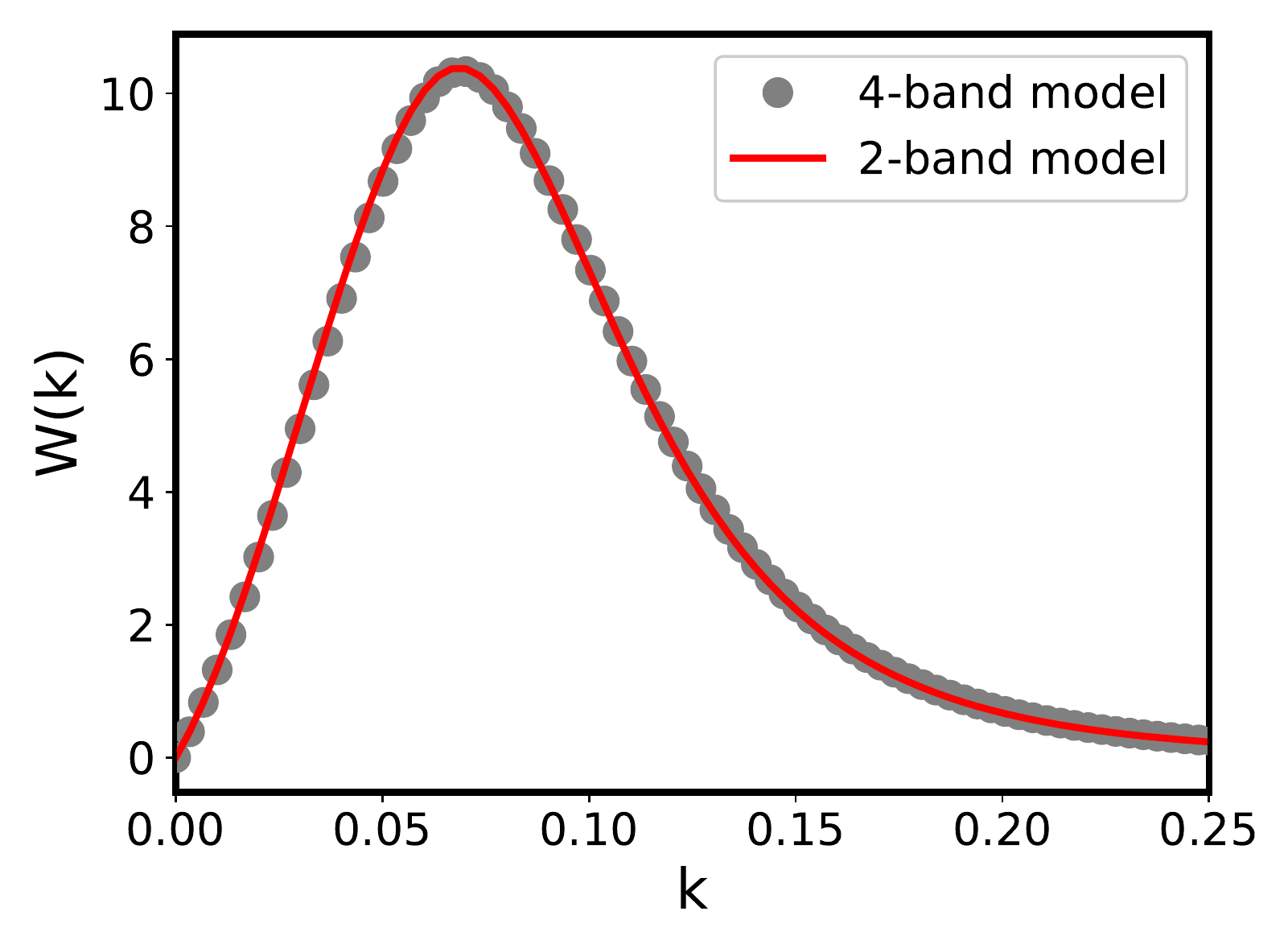}
	\caption{Left panel: The critical coupling constant as a function of total angular momentum $j$ for $\Delta=3$ meV. Right panel: The square
	of the wave function of the electron bound state for the near-critical charge $\zeta=\frac{Z\alpha_g}{\kappa}=0.17$ and the gap $\Delta=3$
	meV are plotted for the two-band and four-band models as a function of dimensionless momentum, which is expressed in terms of
	$1/\lambda_{\gamma_{1}}$. }
	\label{FH_bilayer_spectrum_m-2}
\end{figure}

\section{Summary}
\label{section:Summary}

In this review, we considered the electron states in the field of charged impurities in single layer graphene as well as
bilayer graphene paying the special attention to the phenomenon of supercritical instability and its realization in the presence of
several charged impurities and the external magnetic field. Such states are physically relevant for the actively developing area
of graphene quantum dots whose shape and boundary conditions at edges strongly affect the confined states \cite{Peeters2017}. In
addition, the electron states in the field of charged impurities are crucially important for the transport properties of graphene
\cite{Rossi2011}.

In the supercritical Coulomb center problem, we showed that the ``fall-to-center'' phenomenon arises if $Z\alpha$
exceeds the critical value $1/2$ leading to the appearance of quasistationary levels with complex energies. The energy of quasistationary states
in the case of gapless quasiparticles has a characteristic essential-singularity type dependence on the coupling constant reflecting the scale
invariance of the Coulomb potential. It is found that a quasiparticle gap stabilizes the system decreasing the imaginary part
$|{\Im m\,}E|$ of quasistationary states.

We reviewed also the experiments in which the supercritical behavior was observed. The STM measurements revealed the formation of
resonances around artificial nuclei (clusters of charged calcium dimers) fabricated on gated graphene devices via atomic manipulation
techniques. In a magnetic field, the STM and Landau level spectroscopy measurements demonstrate that the strength of the impurity
can be tuned by controlling the occupation of Landau-level states  with a gate-voltage. At low occupation the impurity is screened becoming
essentially invisible. The screening of the potential of the impurity diminishes as states are filled until, for fully occupied
Landau levels, the unscreened impurity significantly perturbs the spectrum in its vicinity. In this regime it is possible to observe the
Landau-level splitting into discrete states due to lifting the orbital degeneracy.

Further, we studied theoretically the electron states in the field of a charged impurity in graphene in a magnetic field. The charged
impurity removes the degeneracy of Landau levels converting them into band like structures. As the charge of impurity grows, the repulsion of
sublevels of different Landau levels with the same value of orbital momentum takes place leading to the redistribution of the wave function
profiles of these sublevels near the impurity \cite{Gamayun2011,Moldovan2017}. This qualitatively corresponds to the formation of
resonance states in the traditional version of supercritical instability. Taking into account the polarization effects in a magnetic field
allowed us to explain the tuning of the effective charge of the impurity by the gate voltage in agreement with the recent experiments.

In addition, the two-particle bound state problem for gapped graphene in the presence of a Coulomb impurity was studied. A variational approach,
using the projected Hamiltonian and Chandrasekhar-Dirac spinors as trial wave functions, predicts the existence of at least one bound state. In
contrast to the Schr\"{o}dinger case, the variational energy functional is not a homogeneous function of the coupling constant $\alpha$. As a
consequence, the optimal values of the variational parameters depend on $\alpha$ and the optimal binding energy has a more
complicated functional dependence on $\alpha$. In particular, the binding energy increases with respect to the nonrelativistic case.

We studied also the supercritical instability in gapped graphene with two charged impurities separated by distance $R$ in the case where
the charges of impurities are subcritical, whereas their total charge exceeds a critical one. The critical distance $R_{cr}$ in the system of
two charged centers is defined as that at which the electron bound state with the lowest energy reaches the boundary of the lower continuum.
Since the variables in the Dirac problem with two Coulomb centers are not separable in any known orthogonal coordinate system, this problem does
not admit an analytic solution. Therefore, the critical distance $R_{cr}$ could be calculated only with the help of approximate methods. The
LCAO technique and variational Kantorovich method give quite similar results. They show that the critical distance $R_{cr}$ increases as the
quasiparticle gap decreases. The transition to the supercritical regime is signaled by the appearance of
quasistationary states in the lower continuum.

A new type of supercritical behavior in gapped graphene with two oppositely charged impurities was revealed by studying the two-dimensional
Dirac equation for quasiparticles with the Coulomb potential regularized at small distances. By
utilizing the technique of linear combination of atomic orbitals and the variational Galerkin-Kantorovich method, it was shown that for
supercritical electric dipole the wave function of the electron bound state changes its localization from the negatively charged impurity to the
positively charged one as the distance between the impurities changes. Such a migration of the wave function corresponds to the electron and
hole spontaneously created from the vacuum in bound states screening the positively and negatively charged impurities of the supercritical
electric dipole, respectively. The obtained results were generalized in Ref.\cite{Gorbar2015a} to a particle-hole asymmetric
case, where the charges of impurities differ in signs and absolute values, and it was demonstrated that the necessary energetic condition for
the supercriticality of novel type to occur is that the energy levels of single positively and negatively charged impurities traverse together
the energy distance separating the upper and lower continua.

Finally, the supercritical instability in gapped bilayer graphene was studied in the low energy two-band as well as four-band continuum models.
The different scalings of the kinetic energy of quasiparticles and the Coulomb interaction with respect to the distance to the charged impurity
ensure that the wave function of the electron bound state does not shrink toward the impurity as its charge increases. This results in the absence
of the fall-to-center phenomenon in bilayer  graphene although the supercritical instability is realized. It was found that the screening effects
are crucially important in bilayer graphene. If they are neglected, then the critical value for the impurity charge as the lowest energy bound
state dives into the lower continuum tends to zero as the gap $\Delta$ vanishes. If the screened Coulomb interaction is considered, then the critical
charge tends to a finite value when gap goes to zero.

\begin{acknowledgments}
The work of E.V.G. and V.P.G. is partially supported by the National Academy of Sciences of Ukraine (project 0116U003191) and by its Program of
Fundamental Research of the Department of Physics and Astronomy (project No. 0117U000240). V.P.G. acknowledges the support of the RISE Project
CoExAN GA644076.

\end{acknowledgments}


\begin{thebibliography}{999}
	
	\bibitem{Landau:t3}
	L.~Landau and E.~Lifshitz,
	\newblock {\em {Quantum Mechanics. Non-relativistic Theory}},
	\newblock Butterworth-Heinemann, Oxford, 2004.
	
	\bibitem{Pomeranchuk1945}
	I.~Y. Pomeranchuk and Y.~A. Smorodinsky,
	\newblock {\em {On energy levels in systems with $Z>137$}},
	\newblock J. Phys. USSR {\bf 9}, 97 (1945).
	
	\bibitem{Zeldovich1972}
	Y.~B. Zeldovich and V.~S. Popov,
	\newblock {\em {Electronic structure of superheavy atoms}},
	\newblock Sov. Phys. Usp. {\bf 14}, 673 (1972).
	
	\bibitem{Greiner1985}
	W.~Greiner, B.~M{\"{u}}ller, and J.~Rafelski,
	\newblock {\em {Quantum Electrodynamics of Strong Fields. With an Introduction
			into Modern Relativistic Quantum Mechanics}},
	\newblock Texts and Monographs in Physics, Springer-Verlag, Berlin-Heidelberg,
	1985.
	
	\bibitem{Gershtein1970}
	S.~S. Gershtein and Y.~B. Zeldovich,
	\newblock {\em {Positron production during the mutual approach of heavy nuclei
			and the polarization of the vacuum}},
	\newblock Sov. Phys. JETP {\bf 30}, 358 (1970).
	
	\bibitem{Rafelski1971}
	J.~Rafelski, L.~P. Fulcher, and W.~Greiner,
	\newblock {\em {Superheavy Elements and an Upper Limit to the Electric Field
			Strength}},
	\newblock Phys. Rev. Lett. {\bf 27}, 958 (1971).
	
	\bibitem{Muller1972}
	B.~M{\"{u}}ller, H.~Peitz, J.~Rafelski, and W.~Greiner,
	\newblock {\em {Solution of the Dirac Equation for Strong External Fields}},
	\newblock Phys. Rev. Lett. {\bf 28}, 1235 (1972).
	
	\bibitem{Falkovsky1968}
	L.~A. Fal'kovsky,
	\newblock {\em {Physical properties of bismuth}},
	\newblock Sov. Phys. Usp. {\bf 11}, 1 (1968).
	
	\bibitem{Edelman1976}
	V.~S. Edel'man,
	\newblock {\em {Electrons in bismuth}},
	\newblock Adv. Phys. {\bf 25}, 555 (1976).
	
	\bibitem{Herring1937}
	C.~Herring,
	\newblock {\em {Accidental Degeneracy in the Energy Bands of Crystals}},
	\newblock Phys. Rev. {\bf 52}, 365 (1937).
	
	\bibitem{Armitage2017}
	N.~P. Armitage, E.~J. Mele, and A.~Vishwanath,
	\newblock {\em {Weyl and Dirac Semimetals in Three Dimensional Solids}},
	\newblock \textit{Rev. Mod. Phys.} \textbf{90}, 015001 (2018).
	
	\bibitem{Shytov2007a}
	A.~V. Shytov, M.~I. Katsnelson, and L.~S. Levitov,
	\newblock {\em {Atomic collapse and quasi-Rydberg states in graphene}},
	\newblock Phys. Rev. Lett. {\bf 99}, 246802 (2007).
	
	\bibitem{Shytov2007b}
	A.~V. Shytov, M.~I. Katsnelson, and L.~S. Levitov,
	\newblock {\em {Vacuum polarization and screening of supercritical impurities
			in graphene}},
	\newblock Phys. Rev. Lett. {\bf 99}, 236801 (2007).
	
	\bibitem{Pereira2007}
	V.~M. Pereira, J.~Nilsson, and A.~H. {Castro Neto},
	\newblock {\em {Coulomb Impurity Problem in Graphene}},
	\newblock Phys. Rev. Lett. {\bf 99}, 166802 (2007).
	
	\bibitem{Fogler2007}
	M.~M. Fogler, D.~S. Novikov, and B.~I. Shklovskii,
	\newblock {\em {Screening of a hypercritical charge in graphene}},
	\newblock Phys. Rev. B {\bf 76}, 233402 (2007).
	
	\bibitem{Wang2013}
	Y.~Wang, D.~Wong, A.~V.~Shytov, V.~W.~Brar, S.~Choi, Q.~Wu, H.-Z.~Tsai, W.~Regan, A.~Zettl, R.~K.~Kawakami, S.~G.~Louie, L.~S.~Levitov, and M.~F.~Crommie,
	\newblock {\em {Observing atomic collapse resonances in artificial nuclei on
			graphene}},
	\newblock Science {\bf 340}, 734 (2013).
	
	\bibitem{Feher2009}
	A.~Feher, I.~A.~Gospodarev, V.~I.~Grishaev, K.~V.~Kravchenko, E.~V.~Manzhelii, E.~S.~Syrkin, and S.~B.~Feodos'ev,
	\newblock {\em Effect of defects on the quasiparticle spectra of graphite and
		graphene},
	\newblock Low Temp. Phys. {\bf 35}, 679 (2009)
	\newblock [Fiz. Nizk. Temp. \textbf{35}, 862 (2009)].
	
	\bibitem{Loktev2012}
	V.~M. Loktev and Y.~G. Pogorelov,
	\newblock {\em Impurity and vacancy effects in graphene},
	\newblock Low Temp. Phys. {\bf 38}, 792 (2012)
	\newblock [Fiz. Nizk. Temp. \textbf{38}, 993 (2012)].
	
	\bibitem{Skrypnik2007}
	Y.~V. Skrypnyk and V.~M. Loktev,
	\newblock {\em Impurity induced Dirac point smearing in graphene},
	\newblock Low Temp. Phys. {\bf 33}, 762 (2007)
	\newblock [Fiz. Nizk. Temp. \textbf{33}, 1002 (2007)].
	
	\bibitem{Dora2008}
	B.~D\'{o}ra,
	\newblock {\em Disorder effect on the density of states in Landau quantized
		graphene},
	\newblock Low Temp. Phys. {\bf 34}, 801 (2008)
	\newblock [Fiz. Nizk. Temp. \textbf{34}, 1020 (2008)].
	
	\bibitem{Gamayun2009}
	O.~V. Gamayun, E.~V. Gorbar, and V.~P. Gusynin,
	\newblock {\em {Supercritical Coulomb center and excitonic instability in
			graphene}},
	\newblock Phys. Rev. B {\bf 80}, 165429 (2009).
	
	\bibitem{Wang2010}
	J.~Wang, H.~A. Fertig, and G.~Murthy,
	\newblock {\em {Critical behavior in graphene with Coulomb interactions}},
	\newblock Phys. Rev. Lett. {\bf 104}, 186401 (2010).
	
	\bibitem{Sabio2010}
	J.~Sabio, F.~Sols, and F.~Guinea,
	\newblock {\em {Two-body problem in graphene}},
	\newblock Phys. Rev. B {\bf 81}, 45428 (2010).
	
	\bibitem{Khveshchenko2001}
	D.~V. Khveshchenko,
	\newblock {\em Ghost excitonic insulator transition in layered graphite},
	\newblock Phys. Rev. Lett. {\bf 87}, 246802 (2001).
	
	\bibitem{Gorbar2002}
	E.~V. Gorbar, V.~P. Gusynin, V.~A. Miransky, and I.~A. Shovkovy,
	\newblock {\em Magnetic field driven metal-insulator phase transition in planar
		systems},
	\newblock Phys. Rev. B {\bf 66}, 045108 (2002).
	
	\bibitem{Gorbar2003}
	E.~V. Gorbar, V.~P. Gusynin, V.~A. Miransky, and I.~A. Shovkovy,
	\newblock {\em Fractal structure of the effective action in (quasi)planar
		models with long-range interactions},
	\newblock Phys. Lett. A {\bf 313}, 472 (2003).
	
	\bibitem{Khveshchenko2004}
	D.~V. Khveshchenko and H.~Leal,
	\newblock {\em Excitonic instability in layered degenerate semimetals},
	\newblock Nucl. Phys. B {\bf 687}, 323  (2004).
	
	\bibitem{Gamayun2010}
	O.~V. Gamayun, E.~V. Gorbar, and V.~P. Gusynin,
	\newblock {\em {Gap generation and semimetal-insulator phase transition in
			graphene}},
	\newblock Phys. Rev. B {\bf 81}, 75429 (2010).
	
	\bibitem{Gonzalez2012}
	J.~Gonz{\'{a}}lez,
	\newblock {\em {Electron self-energy effects on chiral symmetry breaking in
			graphene}},
	\newblock Phys. Rev. B {\bf 85}, 85420 (2012).
	
	\bibitem{Drut2009a}
	J.~E. Drut and T.~A. L{\"a}hde,
	\newblock {\em Is graphene in vacuum an insulator?},
	\newblock Phys. Rev. Lett. {\bf 102}, 026802 (2009).
	
	\bibitem{Drut2009b}
	J.~E. Drut and T.~A. L{\"a}hde,
	\newblock {\em Critical exponents of the semimetal-insulator transition in
		graphene: A Monte Carlo study},
	\newblock Phys. Rev. B {\bf 79}, 241405 (2009).
	
	\bibitem{Armour2010}
	W.~Armour, S.~Hands, and C.~Strouthos,
	\newblock {\em Monte Carlo simulation of the semimetal-insulator phase
		transition in monolayer graphene},
	\newblock Phys. Rev. B {\bf 81}, 125105 (2010).
	
	\bibitem{Buividovich2012}
	P.~V. Buividovich and M.~I. Polikarpov,
	\newblock {\em Monte Carlo study of the electron transport properties of
		monolayer graphene within the tight-binding model},
	\newblock Phys. Rev. B {\bf 86}, 245117 (2012).
	
	\bibitem{FominReview1983}
	P.~I. Fomin, V.~P. Gusynin, V.~A. Miransky, and Y.~A. Sitenko,
	\newblock {\em {Dynamical symmetry breaking and particle mass generation in
			gauge field theories}},
	\newblock Riv. Nuovo Cimento {\bf 6}, 1 (1983).
	
	\bibitem{Peccei1988}
	R.~D. Peccei,
	\newblock {\em New phase for an old theory?},
	\newblock Nature {\bf 332}, 492 (1988).
	
	\bibitem{Luican-Mayer2014}
	A.~Luican-Mayer, M.~Kharitonov, G.~Li, C.-P.~Lu, I.~Skachko, A.-M.~B.~Gon\c{c}alves, K.~Watanabe, T.~Taniguchi, and E.~Y.~Andrei,
	\newblock {\em Screening Charged Impurities and Lifting the Orbital Degeneracy
		in Graphene by Populating Landau Levels},
	\newblock Phys. Rev. Lett. {\bf 112}, 036804 (2014).
	
	\bibitem{Mao2016}
	J.~Mao, Y.~Jiang, D.~Moldovan, G.~Li, K.~Watanabe, T.~Taniguchi, M.~R.~Masir, F.~M.~Peeters, and E.~Y.~Andrei,
	\newblock {\em Realization of a tunable artificial atom at a supercritically
		charged vacancy in graphene},
	\newblock Nat Phys {\bf 12}, 545 (2016).
	
	\bibitem{Giovannetti2007}
	G.~Giovannetti, P.~A. Khomyakov, G.~Brocks, P.~J. Kelly, and J.~van~den Brink,
	\newblock {\em Substrate-induced band gap in graphene on hexagonal boron
		nitride: \textit{Ab initio} density functional calculations},
	\newblock Phys. Rev. B {\bf 76}, 073103 (2007).
	
	\bibitem{Son2006}
	Y.-W. Son, M.~L. Cohen, and S.~G. Louie,
	\newblock {\em Energy gaps in graphene nanoribbons},
	\newblock Phys. Rev. Lett. {\bf 97}, 216803 (2006).
	
	\bibitem{Novikov2007}
	D.~S. Novikov,
	\newblock {\em {Elastic scattering theory and transport in graphene}},
	\newblock Phys. Rev. B {\bf 76}, 245435 (2007).
	
	\bibitem{Pereira2008}
	V.~M. Pereira, V.~N. Kotov, and A.~H. Castro~Neto,
	\newblock {\em Supercritical Coulomb impurities in gapped graphene},
	\newblock Phys. Rev. B {\bf 78}, 085101 (2008).
	
	\bibitem{Terekhov2008}
	I.~S. Terekhov, A.~I. Milstein, V.~N. Kotov, and O.~P. Sushkov,
	\newblock {\em {Screening of Coulomb Impurities in Graphene}},
	\newblock Phys. Rev. Lett. {\bf 100}, 076803 (2008).
	
	\bibitem{Shytov2009}
	A.~Shytov, M.~Rudner, N.~Gu, M.~Katsnelson, and L.~Levitov,
	\newblock {\em {Atomic collapse, Lorentz boosts, Klein scattering, and other
			quantum-relativistic phenomena in graphene}},
	\newblock Solid State Comm. {\bf 149}, 1087 (2009).
	
	\bibitem{CastroNeto2009a}
	A.~H.~Castro Neto, V.~N.~Kotov, J.~Nilsson, V.~M.~Pereira, N.~M.~R.~Peres, and B.~Uchoa,
	\newblock {\em {Adatoms in graphene}},
	\newblock Solid State Comm. {\bf 149}, 1094 (2009).
	
	\bibitem{Darwin1913}
	C.~G. Darwin,
	\newblock {\em {On some orbits of an electron}},
	\newblock Philos. Mag. {\bf 25}, 201 (1913).
	
	\bibitem{Gradshtein-book}
	I.~S. Gradshtein and I.~M. Ryzhik,
	\newblock {\em Table of Integrals, Series, and Products},
	\newblock Academic, Orlando, FL, 1980.
	
	\bibitem{Khalilov1998}
	V.~R. Khalilov and C.-L. Ho,
	\newblock {\em {Dirac electron in a Coulomb field in (2+1) dimensions}},
	\newblock Mod. Phys. Lett. A {\bf 13}, 615 (1998).
	
	\bibitem{Fomin1978}
	P.~Fomin, V.~Gusynin, and V.~Miransky,
	\newblock {\em {Vacuum instability of massless electrodynamics and the
			Gell-Mann-Low eigenvalue condition for the bare coupling constant}},
	\newblock Phys. Lett. B {\bf 78}, 136 (1978).
	
	\bibitem{Popov1971}
	V.~S. Popov,
	\newblock {\em Position Production in a Coulomb Field with $Z >137$},
	\newblock Sov. Phys. JETP {\bf 32}, 526 (1971).
	
	\bibitem{FominMiransky1976}
	P.~Fomin and V.~Miransky,
	\newblock {\em {On the dynamical vacuum rearrangement and the problem of
			fermion mass generation}},
	\newblock Phys. Lett. B {\bf 64}, 166 (1976).
	
	\bibitem{Wang2012}
	Y.~Wang, V.~W.~Brar, A.~V.~Shytov, Q.~Wu, W.~Regan, H.-Z.~Tsai, A.~Zettl, L.~S.~Levitov, and M.~F.~Crommie,
	\newblock {\em Mapping Dirac quasiparticles near a single Coulomb impurity on
		graphene},
	\newblock Nat. Phys. {\bf 8}, 653 (2012).
	
	\bibitem{Li1997}
	J.~Li, W.-D. Schneider, and R.~Berndt,
	\newblock {\em {Local density of states from spectroscopic
			scanning-tunneling-microscope images: Ag(111)}},
	\newblock Phys. Rev. B {\bf 56}, 7656 (1997).
	
	\bibitem{Wittneven1998}
	C.~Wittneven, R.~Dombrowski, M.~Morgenstern, and R.~Wiesendanger,
	\newblock {\em {Scattering States of Ionized Dopants Probed by Low Temperature
			Scanning Tunneling Spectroscopy}},
	\newblock Phys. Rev. Lett. {\bf 81}, 5616 (1998).
	
	\bibitem{Krive1992}
	I.~V. Krive and S.~A. Naftulin,
	\newblock {\em Dynamical symmetry breaking and phase transitions in a
		three-dimensional Gross-Neveu model in a strong magnetic field},
	\newblock Phys. Rev. D {\bf 46}, 2737 (1992).
	
	\bibitem{Klimenko1991}
	K.~G. Klimenko,
	\newblock {\em Three-dimensional Gross--Neveu model in an~external magnetic
		field.~I},
	\newblock Theor. Math. Phys. {\bf 89}, 1161 (1991).
	
	\bibitem{Gusynin1994}
	V.~P. Gusynin, V.~A. Miransky, and I.~A. Shovkovy,
	\newblock {\em Catalysis of Dynamical Flavor Symmetry Breaking by a Magnetic
		Field in 2 + 1 Dimensions},
	\newblock Phys. Rev. Lett. {\bf 73}, 3499 (1994).
	
	\bibitem{Gusynin1996}
	V.~Gusynin, V.~Miransky, and I.~Shovkovy,
	\newblock {\em {Dimensional reduction and catalysis of dynamical symmetry
			breaking by a magnetic field}},
	\newblock Nucl. Phys. B {\bf 462}, 249 (1996).
	
	\bibitem{graphene-QHE}
	E.~V. Gorbar, V.~P. Gusynin, V.~A. Miransky, and I.~A. Shovkovy,
	\newblock {\em {Coulomb interaction and magnetic catalysis in the quantum Hall
			effect in graphene}},
	\newblock Phys. Scr. {\bf T146}, 014018 (2012).
	
	\bibitem{Oraevski1977}
	V.~N. Oraevskii, A.~I. Rex, and V.~B. Semikoz,
	\newblock {\em Spontaneous production of positrons by a Coulomb center in a
		homogeneous magnetic field},
	\newblock Sov. Phys. JETP {\bf 45}, 428 (1977).
	
	\bibitem{Schlutter1985}
	P.~Schluter, G.~Soff, K.~H. Wietschorke, and W.~Greiner,
	\newblock {\em Bound electrons in critical magnetic fields},
	\newblock J. Phys. B: At. Molec. Phys. {\bf 18}, 1685 (1985).
	
	\bibitem{Khalilov-attempt}
	C.-L. Ho and V.~R. Khalilov,
	\newblock {\em {Planar Dirac electron in Coulomb and magnetic fields}},
	\newblock Phys. Rev. A {\bf 61}, 32104 (2000).
	
	\bibitem{Gamayun2011}
	O.~V. Gamayun, E.~V. Gorbar, and V.~P. Gusynin,
	\newblock {\em Magnetic field driven instability of a charged center in
		graphene},
	\newblock Phys. Rev. B {\bf 83}, 235104 (2011).
	
	\bibitem{Siedentop2012}
	T.~Maier and H.~Siedentop,
	\newblock {\em Stability of impurities with Coulomb potential in graphene with
		homogeneous magnetic field},
	\newblock J. Math. Phys. {\bf 53}, 095207 (2012).
	
	\bibitem{Yang2014}
	S.~C. Kim and S.-R. {Eric Yang},
	\newblock {\em Coulomb impurity problem of graphene in magnetic fields},
	\newblock Ann. Phys. {\bf 347}, 21 (2014).
	
	\bibitem{Moldovan2017}
	D.~Moldovan, M.~R. Masir, and F.~M. Peeters,
	\newblock {\em Magnetic field dependence of the atomic collapse state in
		graphene},
	\newblock 2D Materials {\bf 5}, 015017 (2018).
	
	\bibitem{Semenoff1999}
	G.~W. Semenoff, I.~A. Shovkovy, and L.~C.~R. Wijewardhana,
	\newblock {\em Universality and the magnetic catalysis of chiral symmetry
		breaking},
	\newblock Phys. Rev. D {\bf 60}, 105024 (1999).
	
	\bibitem{Sobol2016}
	O.~O. Sobol, P.~K. Pyatkovskiy, E.~V. Gorbar, and V.~P. Gusynin,
	\newblock {\em Screening of a charged impurity in graphene in a magnetic
		field},
	\newblock Phys. Rev. B {\bf 94}, 115409 (2016).
	
	\bibitem{Wigner1929}
	J.~von Neumann and E.~P. Wigner,
	\newblock {\em {\"U}ber das Verhalten von Eigenwerten bei adiabatischen
		Prozessen. (On the behavior of eigenvalues in the adiabatic processes.)},
	\newblock Phys. Z. {\bf 30}, 467  (1929).
	
	\bibitem{Muller1972b}
	B.~M{\"u}ller, J.~Rafelski, and W.~Greiner,
	\newblock {\em Electron shells in over-critical external fields},
	\newblock Z. Phys. {\bf 257}, 62 (1972).
	
	\bibitem{Kovner1990}
	A.~Kovner and B.~Rosenstein,
	\newblock {\em {Kosterlitz-Thouless mechanism of two-dimensional
			superconductivity}},
	\newblock Phys. Rev. B {\bf 42}, 4748 (1990).
	
	\bibitem{Dorey1992}
	N.~Dorey and N.~Mavromatos,
	\newblock {\em {QED3 and two-dimensional superconductivity without parity
			violation}},
	\newblock Nucl. Phys. B {\bf 386}, 614 (1992).
	
	\bibitem{Marino1993}
	E.~C. Marino,
	\newblock {\em Quantum electrodynamics of particles on a plane and the
		Chern-Simons theory},
	\newblock Nucl. Phys. B {\bf 408}, 551 (1993).
	
	\bibitem{Gorbar2001}
	E.~V. Gorbar, V.~P. Gusynin, and V.~A. Miransky,
	\newblock {\em Dynamical chiral symmetry breaking on a brane in reduced QED},
	\newblock Phys. Rev.~D {\bf 64}, 105028 (2001).
	
	\bibitem{Pyatkovskiy2011}
	P.~K. Pyatkovskiy and V.~P. Gusynin,
	\newblock {\em Dynamical polarization of graphene in a magnetic field},
	\newblock Phys. Rev. B {\bf 83}, 075422 (2011).
	
	\bibitem{Roldan2009}
	R.~Rold\'{a}n, J.-N. Fuchs, and M.~O. Goerbig,
	\newblock {\em Collective modes of doped graphene and a standard
		two-dimensional electron gas in a strong magnetic field: Linear
		magnetoplasmons versus magnetoexcitons},
	\newblock Phys. Rev. B {\bf 80}, 085408 (2009).
	
	\bibitem{Roldan2010}
	R.~Rold\'{a}n, M.~O. Goerbig, and J.-N. Fuchs,
	\newblock {\em The magnetic field particle?hole excitation spectrum in doped
		graphene and in a standard two-dimensional electron gas},
	\newblock Semicond. Sci. Technol. {\bf 25}, 034005 (2010).
	
	\bibitem{Mizes1989}
	H.~A. Mizes and J.~S. Foster,
	\newblock {\em {Long-Range Electronic Perturbations Caused by Defects Using
			Scanning Tunneling Microscopy}},
	\newblock Science {\bf 244}, 559 (1989).
	
	\bibitem{Zhang2012}
	Y.~Zhang, Y.~Barlas, and K.~Yang,
	\newblock {\em Coulomb impurity under magnetic field in graphene: A
		semiclassical approach},
	\newblock Phys. Rev. B {\bf 85}, 165423 (2012).
	
	\bibitem{DeMartino2017}
	A.~{De Martino} and R.~Egger,
	\newblock {\em {Two-electron bound states near a Coulomb impurity in gapped
			graphene}},
	\newblock Phys. Rev. B {\bf 95}, 085418 (2017).
	
	\bibitem{Lee2016}
	J.~Lee, D.~Wong, J.~Velasco Jr, J.~F.~Rodriguez-Nieva, S.~Kahn, H.-Z.~Tsai, T.~Taniguchi, K.~Watanabe, A.~Zettl, F.~Wang, L.~S.~Levitov, and M.~F.~Crommie,
	\newblock {\em {Imaging electrostatically confined Dirac fermions in graphene
			quantum dots}},
	\newblock Nat. Phys. {\bf 12}, 1032 (2016).
	
	\bibitem{Gutierrez2016}
	C.~Guti\'{e}rrez, L.~Brown, C.-J. Kim, J.~Park, and A.~N. Pasupathy,
	\newblock {\em Klein tunnelling and electron trapping in nanometre-scale
		graphene quantum dots},
	\newblock Nat. Phys. {\bf 12}, 1069 (2016).
	
	\bibitem{Chandrasekhar1944}
	S.~Chandrasekhar,
	\newblock {\em {Some Remarks on the Negative Hydrogen Ion and its Absorption
			Coefficient}},
	\newblock Astrophys. J. {\bf 100}, 176 (1944).
	
	\bibitem{Bethe1957}
	H.~A. Bethe and E.~E. Salpeter,
	\newblock {\em Quantum Mechanics of One- and Two-Electron Atoms},
	\newblock Academic Press, New York, 1957.
	
	\bibitem{Hill1977}
	R.~N. Hill,
	\newblock {\em {Proof that the $H^{-}$ ion has only one bound state. Details
			and extension to finite nuclear mass}},
	\newblock J. Math. Phys. {\bf 18}, 2316 (1977).
	
	\bibitem{Bransden1983}
	B.~H. Bransden and C.~J. Joachain,
	\newblock {\em Physics of Atoms and Molecules},
	\newblock Longmont Group Limited, Hong Kong, 1983.
	
	\bibitem{Andersen2004}
	T.~Andersen,
	\newblock {\em {Atomic negative ions: structure, dynamics and collisions}},
	\newblock Phys. Rep. {\bf 394}, 157 (2004).
	
	\bibitem{Hogaasen2010}
	H.~H{\o}gaasen, J.-M. Richard, and P.~Sorba,
	\newblock {\em {Two-electron atoms, ions, and molecules}},
	\newblock Am. J. Phys. {\bf 78}, 86 (2010).
	
	\bibitem{Phelps1983}
	D.~E. Phelps and K.~K. Bajaj,
	\newblock {\em {Ground-state energy of a $D^{-}$ ion in two-dimensional
			semiconductors}},
	\newblock Phys. Rev. B {\bf 27}, 4883 (1983).
	
	\bibitem{Pang1990}
	T.~Pang and S.~G. Louie,
	\newblock {\em {Negative-donor centers in semiconductors and quantum wells}},
	\newblock Phys. Rev. Lett. {\bf 65}, 1635 (1990).
	
	\bibitem{Larsen1992a}
	D.~M. Larsen and S.~Y. McCann,
	\newblock {\em Excited states of the two-dimensional $D^{-}$ center in magnetic
		fields},
	\newblock Phys. Rev. B {\bf 45}, 3485 (1992).
	
	\bibitem{Larsen1992b}
	D.~M. Larsen and S.~Y. McCann,
	\newblock {\em Variational studies of two- and three-dimensional $D^{-}$
		centers in magnetic fields},
	\newblock Phys. Rev. B {\bf 46}, 3966 (1992).
	
	\bibitem{Sandler1992}
	N.~P. Sandler and C.~R. Proetto,
	\newblock {\em {Negative-donor centers in two dimensions}},
	\newblock Phys. Rev. B {\bf 46}, 7707 (1992).
	
	\bibitem{Ivanov2002}
	M.~V. Ivanov and P.~Schmelcher,
	\newblock {\em Two-dimensional negative donors in magnetic fields},
	\newblock Phys. Rev. B {\bf 65}, 205313 (2002).
	
	\bibitem{Huant1990}
	S.~Huant, S.~P. Najda, and B.~Etienne,
	\newblock {\em Two-dimensional $D^{-}$ centers},
	\newblock Phys. Rev. Lett. {\bf 65}, 1486 (1990).
	
	\bibitem{Shields1995}
	A.~J. Shields, M.~Pepper, M.~Y. Simmons, and D.~A. Ritchie,
	\newblock {\em {Spin-triplet negatively charged excitons in GaAs quantum
			wells}},
	\newblock Phys. Rev. B {\bf 52}, 7841 (1995).
	
	\bibitem{Brown1951}
	G.~E. Brown and D.~G. Ravenhall,
	\newblock {\em On the interaction of two electrons},
	\newblock Proc. Royal Soc. A {\bf 208}, 552 (1951).
	
	\bibitem{Kolakowska1996}
	A.~Kolakowska, J.~D. Talman, and K.~Aashamar,
	\newblock {\em Minimax variational approach to the relativistic two-electron
		problem},
	\newblock Phys. Rev. A {\bf 53}, 168 (1996).
	
	\bibitem{Nakatsuji2005}
	H.~Nakatsuji and H.~Nakashima,
	\newblock {\em {Analytically Solving the Relativistic Dirac-Coulomb Equation
			for Atoms and Molecules}},
	\newblock Phys. Rev. Lett. {\bf 95}, 050407 (2005).
	
	\bibitem{Sucher1980}
	J.~Sucher,
	\newblock {\em {Foundations of the relativistic theory of many-electron
			atoms}},
	\newblock Phys. Rev. A {\bf 22}, 348 (1980).
	
	\bibitem{Sucher1984}
	J.~Sucher,
	\newblock {\em {Foundations of the relativistic theory of many-electron bound
			states}},
	\newblock Int. J. Quantum Chem. {\bf 25}, 3 (1984).
	
	\bibitem{Haeusler2009}
	W.~H{\"{a}}usler and R.~Egger,
	\newblock {\em {Artificial atoms in interacting graphene quantum dots}},
	\newblock Phys. Rev. B {\bf 80}, 161402 (2009).
	
	\bibitem{Egger2010}
	R.~Egger, A.~{De Martino}, H.~Siedentop, and E.~Stockmeyer,
	\newblock {\em {Multiparticle equations for interacting Dirac fermions in
			magnetically confined graphene quantum dots}},
	\newblock J. Phys. A: Math. Theor. {\bf 43}, 215202 (2010).
	
	\bibitem{Marinov1975}
	M.~S. Marinov and V.~S. Popov,
	\newblock {\em Critical distance in collision of heavy ions},
	\newblock Sov. Phys. JETP {\bf 41}, 205 (1975).
	
	\bibitem{Cohen-Tannoudji}
	C.~Cohen-Tannoudji, B.~Diu, and F.~Laloe,
	\newblock {\em Quantum Mechanics}, volume~2,
	\newblock Hermann, Paris, 1977.
	
	\bibitem{Klopfer2014}
	D.~Kl{\"o}pfer, A.~De~Martino, D.~Matrasulov, and R.~Egger,
	\newblock {\em Scattering theory and ground-state energy of Dirac fermions in
		graphene with two Coulomb impurities},
	\newblock Eur. Phys. J. B {\bf 87}, 187 (2014).
	
	\bibitem{Klopfer_thesis}
	D.~Kl{\"o}pfer,
	\newblock {\em Critical Effects and Universality in Graphene Quantum Dots},
	\newblock PhD thesis, D{\"u}sseldorf, 2014.
	
	\bibitem{Matveev2000}
	V.~I. Matveev, D.~U. Matrasulov, and H.~Y. Rakhimov,
	\newblock {\em Two-center problem for the dirac equation},
	\newblock Phys. At. Nucl. {\bf 63}, 318 (2000).
	
	\bibitem{Bondarchuk2007}
	V.~V. Bondarchuk, I.~M. Shvab, D.~I. Bondar, and A.~V. Katernoga,
	\newblock {\em Simple model of scalar-vector interaction for the relativistic
		two-center problem},
	\newblock Phys. Rev. A {\bf 76}, 062507 (2007).
	
	\bibitem{Popov1972b}
	V.~S. Popov,
	\newblock {\em To the problem of the critical charge of nucleus},
	\newblock Sov. J. Nucl. Phys. {\bf 14}, 257 (1972).
	
	\bibitem{Khalilov2013}
	V.~R. Khalilov,
	\newblock {\em Zero-mass fermions in Coulomb and Aharonov-Bohm potentials in
		2+1 dimensions},
	\newblock Theor. Math. Phys. {\bf 175}, 637 (2013).
	
	\bibitem{Chakraborty2013}
	B.~Chakraborty, K.~S. Gupta, and S.~Sen,
	\newblock {\em Effect of topological defects and Coulomb charge on the low
		energy quantum dynamics of gapped graphene},
	\newblock J. Phys. A: Math. Theor. {\bf 46}, 055303 (2013).
	
	\bibitem{Chakraborty2013a}
	B.~Chakraborty, K.~S. Gupta, and S.~Sen,
	\newblock {\em Topology, cosmic strings and quantum dynamics -- a case study
		with graphene},
	\newblock J. Phys. Conf. Series {\bf 442}, 012017 (2013).
	
	\bibitem{Popov1972}
	V.~S. Popov,
	\newblock {\em Critical charge in the two-center problem},
	\newblock JETP Lett. {\bf 16}, 251 (1972).
	
	\bibitem{Popov-review2001}
	V.~S. Popov,
	\newblock {\em Critical charge in quantum electrodynamics},
	\newblock Phys. At. Nucl. {\bf 64}, 367 (2001).
	
	\bibitem{Marinov1975b}
	M.~S. Marinov, V.~S. Popov, and V.~S. Stolin,
	\newblock {\em Variational approach to the relativistic two-center problem:
		Critical internuclear distance},
	\newblock J. Comput. Phys. {\bf 19}, 241 (1975).
	
	\bibitem{Sobol2014UJP}
	O.~O. Sobol,
	\newblock {\em Variational method for the calculation of critical distance
		between two Coulomb centers in graphene},
	\newblock Ukr. Jour. Phys. {\bf 59}, 531 (2014).
	
	\bibitem{Sobol2013}
	O.~O. Sobol, E.~V. Gorbar, and V.~P. Gusynin,
	\newblock {\em Supercritical instability in graphene with two charged
		impurities},
	\newblock Phys. Rev. B {\bf 88}, 205116 (2013).
	
	\bibitem{DeMartino2014}
	A.~De~Martino, D.~Kl{\"o}pfer, D.~Matrasulov, and R.~Egger,
	\newblock {\em Electric-dipole-induced universality for Dirac fermions in
		graphene},
	\newblock Phys. Rev. Lett. {\bf 112}, 186603 (2014).
	
	\bibitem{Matrasulov1999}
	D.~U. Matrasulov, V.~I. Matveev, and M.~M. Musakhanov,
	\newblock {\em Eigenvalue problem for the relativistic electric-dipole system},
	\newblock Phys. Rev. A {\bf 60}, 4140 (1999).
	
	\bibitem{Connolly2007}
	K.~Connolly and D.~J. Griffiths,
	\newblock {\em Critical dipoles in one, two, and three dimensions},
	\newblock Am. J. Phys. {\bf 75}, 524 (2007).
	
	\bibitem{Turner1977}
	J.~E. Turner,
	\newblock {\em Minimum dipole moment required to bind an electron --- molecular
		theorists rediscover phenomenon mentioned in Fermi-Teller paper twenty years
		earlier},
	\newblock Am.~J. Phys {\bf 45}, 758  (1977).
	
	\bibitem{Gorbar2015}
	E.~V. Gorbar, V.~P. Gusynin, and O.~O. Sobol,
	\newblock {\em Supercritical electric dipole and migration of electron wave
		function in gapped graphene},
	\newblock Europhys. Lett. {\bf 111}, 37003 (2015).
	
	\bibitem{Abramov1972}
	D.~I. Abramov and I.~V. Komarov,
	\newblock {\em Weakly bound states of a charged particle in a finite-dipole
		field},
	\newblock Theor. Math. Phys. {\bf 13}, 1090 (1972).
	
	\bibitem{Camblong2001}
	H.~E. Camblong, L.~N. Epele, H.~Fanchiotti, and C.~A. Garc\'{\i}a~Canal,
	\newblock {\em Quantum Anomaly in Molecular Physics},
	\newblock Phys. Rev. Lett. {\bf 87}, 220402 (2001).
	
	\bibitem{Schumayer2010}
	D.~Schumayer, B.~P. van Zyl, R.~K. Bhaduri, and D.~A.~W. Hutchinson,
	\newblock {\em {Geometric scaling in the spectrum of an electron captured by a
			stationary finite dipole}},
	\newblock Europhys. Lett. {\bf 89}, 13001 (2010).
	
	\bibitem{Efimov1970}
	V.~Efimov,
	\newblock {\em Energy levels arising from resonant two-body forces in a
		three-body system},
	\newblock Phys. Lett. B {\bf 33}, 563  (1970).
	
	\bibitem{Braaten2006}
	E.~Braaten and H.-W. Hammer,
	\newblock {\em Universality in few-body systems with large scattering length},
	\newblock Phys. Rep. {\bf 428}, 259 (2006).
	
	\bibitem{Gogolin2008}
	A.~O. Gogolin, C.~Mora, and R.~Egger,
	\newblock {\em {Analytical Solution of the Bosonic Three-Body Problem}},
	\newblock Phys. Rev. Lett. {\bf 100}, 140404 (2008).
	
	\bibitem{AbramowitzStegun}
	M.~Abramowitz and I.~A. Stegun,
	\newblock {\em {Handbook of Mathematical Functions}},
	\newblock Dover, New York, 1964.
	
	\bibitem{Gorbar2015a}
	E.~V. Gorbar, V.~P. Gusynin, and O.~O. Sobol,
	\newblock {\em Supercriticality of novel type induced by electric dipole in
		gapped graphene},
	\newblock Phys. Rev.~B {\bf 92}, 235417 (2015).
	
	\bibitem{McCann2006}
	E.~McCann and V.~I. Fal'ko,
	\newblock {\em Landau-level degeneracy and quantum Hall effect in a graphite
		bilayer},
	\newblock Phys. Rev. Lett. {\bf 96}, 086805 (2006).
	
	\bibitem{Novoselov2006}
	K.~S.~Novoselov, E.~McCann, S.~V.~Morozov, V.~I.~Fal'ko, M.~I.~Katsnelson, U.~Zeitler, D.~Jiang, F.~Schedin, and A.~K.~Geim,
	\newblock {\em Unconventional quantum Hall effect and Berry/'s phase of $2\pi$
		in bilayer graphene},
	\newblock Nat. Phys. {\bf 2}, 177 (2006).
	
	\bibitem{Shytov-arXiv-2015}
	A.~V. Shytov,
	\newblock {\em Cloaked resonant states in bilayer graphene},
	\newblock ArXiv [cond-mat.mes-hall] , 1506.02839 (2015).
	
	\bibitem{Martin2010}
	J.~Martin, B.~E. Feldman, R.~T. Weitz, M.~T. Allen, and A.~Yacoby,
	\newblock {\em Local compressibility measurements of correlated states in
		suspended bilayer graphene},
	\newblock Phys. Rev. Lett. {\bf 105}, 256806 (2010).
	
	\bibitem{Weitz2010}
	R.~T. Weitz, M.~T. Allen, B.~E. Feldman, J.~Martin, and A.~Yacoby,
	\newblock {\em Broken-symmetry states in doubly gated suspended bilayer
		graphene},
	\newblock Science {\bf 330}, 812 (2010).
	
	\bibitem{Freitag2012}
	F.~Freitag, J.~Trbovic, M.~Weiss, and C.~Sch{\"o}nenberger,
	\newblock {\em Spontaneously gapped ground state in suspended bilayer
		graphene},
	\newblock Phys. Rev. Lett. {\bf 108}, 076602 (2012).
	
	\bibitem{Lee2014}
	Y.~Lee,  D.~Tran, K.~Myhro, J.~Velasco, N.~Gillgren, C.~N.~Lau, Y.~Barlas, J.~M.~Poumirol, D.~Smirnov, and F.~Guinea,
	\newblock {\em {Competition between spontaneous symmetry breaking and
			single-particle gaps in trilayer graphene}},
	\newblock Nature Comm. {\bf 5}, 5656 (2014).
	
	\bibitem{Kolomeisky2016}
	E.~B. Kolomeisky, J.~P. Straley, and D.~L. Abrams,
	\newblock {\em {Space charge and screening in bilayer graphene}},
	\newblock J. Phys.: Cond. Mat. {\bf 28}, 47LT01 (2016).
	
	\bibitem{Oriekhov2017}
	D.~O. Oriekhov, O.~O. Sobol, E.~V. Gorbar, and V.~P. Gusynin,
	\newblock {\em Coulomb center instability in bilayer graphene},
	\newblock Phys. Rev. B {\bf 96}, 165403 (2017).
	
	\bibitem{Nandkishore2010}
	R.~Nandkishore and L.~Levitov,
	\newblock {\em Dynamical screening and excitonic instability in bilayer
		graphene},
	\newblock Phys. Rev. Lett. {\bf 104}, 156803 (2010).
	
	\bibitem{Peeters2017}
	R.~{Van~Pottelberge}, M.~Zarenia, P.~Vasilopoulos, and F.~M. Peeters,
	\newblock {\em Graphene quantum dot with a Coulomb impurity: Subcritical and
		supercritical regime},
	\newblock Phys. Rev. B {\bf 95}, 245410 (2017).
	
	\bibitem{Rossi2011}
	S.~{Das~Sarma}, S.~Adam, E.~H. Hwang, and E.~Rossi,
	\newblock {\em Electronic transport in two-dimensional graphene},
	\newblock Rev. Mod. Phys. {\bf 83}, 407 (2011).
	
\end{thebibliography}
\end{document}